Chapter 1

# Light in Slices: How to Enable Precise Laser Processing?

**Stefan Nolte [1,2,*], Jörn Bonse[3,§], and Nadezhda M. Bulgakova[4]**

[1] Friedrich Schiller University Jena, Institute of Applied Physics, Abbe Center of Photonics, Albert-Einstein-Straße 15, 07745 Jena, Germany

[2] Fraunhofer Institute for Applied Optics and Precision Engineering IOF, Center of Excellence in Photonics, Albert-Einstein-Straße 7, 07745 Jena, Germany

[3] Bundesanstalt für Materialforschung und -prüfung (BAM), Unter den Eichen 87, 12205 Berlin, Germany

[4] FZU - Institute of Physics of the Czech Academy of Sciences, Na Slovance 1999/2, 182 00 Praha 8, Czech Republic

[*,§,#]: corresponding author

e-mails: stefan.nolte@uni-jena.de ; joern.bonse@bam.de ; bulgakova@fzu.cz

## Abstract

Ultrashort (femtosecond [fs]) laser pulses have fascinating properties as they allow to confine optical energy on extreme scales in space and time. Such fs-pulsed laser beams can be seen as spatially thin slices of intense light that are radially and axially constrained to the micrometer scale, while simultaneously propagating at the extremely high speed of light. Their high peak intensities and their short duration makes such laser pulses unique tools for materials processing, as their duration is shorter than the time required to transfer absorbed optical energy, via electron-phonon coupling, from the electronic system of the solid to its lattice. Hence, the laser pulse energy remains localized during the interaction and does not spread via diffusion into the area surrounding the irradiated region. As one consequence, the fs-laser thus offers increased precision for material modification or ablation accompanied by a reduced heat-affected zone of only a few hundred nanometers. On the other hand, the high laser peak intensities can enable nonlinear material interactions that render unique material excitation and relaxation pathways possible. In this chapter, we briefly review the reasons for the enormous success of ultrashort pulse lasers in materials processing – both for the processing of the surface or in the bulk of solids. We identify the underlying fundamental processes that can limit the precision or the up-scaling of the laser processing towards large volumes, areas, or processing rates. Strategies to overcome such limitations will be outlined and questions on the ultimate limits of laser material processing will be answered.

**Keywords:** Ultrashort laser pulses; Surface Processing, Volume Processing, Damage thresholds, Plasma shielding, Heat accumulation

# 1. Introduction

# 2. Surface Processing

# 3. Volume Processing

# 4. Outlook: Pushing the Limits through Tailored Ultrashort Laser Processing

# 1 Introduction

Apart from the *monochromaticity*, the most striking property of laser radiation is its *coherence*. The latter manifests in space and time as a direct consequence of its generation via *stimulated emission* processes. Using optical elements such as lenses, mirrors, or objectives for laser beam focusing the coherence enables an extreme *spatial confinement* of the laser radiation to spot sizes somewhat smaller than the laser wavelength itself. This has enabled the enormous success of the laser technology reflected by its ubiquitous presence in daily life applications, e.g., for optical data storage applications in CD- or DVD-technologies for consumer electronics.

Less known but of paramount importance for current materials processing with lasers is the possibility that laser radiation can be emitted in the form of *temporally confined* laser pulses – either as single-pulse events (addressed in the chapter title as slices propagating in space at the speed of light) or even repetitive as trains of such laser pulses. This picture of “light in slices” becomes particularly appealing for extremely short laser pulses – so-called *ultrashort laser pulses* (ULP) with durations in the few-ps- down to the fs-range. For example, a laser pulse of 100 fs in duration that is propagating in free space (air) has a spatial extent of just 30 micrometers in its propagation direction – typically orders of magnitude smaller than the (unfocused) laser beam diameter itself in radial direction (typically a few millimeters).

In this chapter, we will further develop this picture of light in slices and explain why and how this enables precise laser material processing from the macro-, over the micro-, down to the nanoscale. Our introductory text presents some of the most relevant fundamentals of ultrafast laser processing and, thus, provides the ground for many of the following more specialized book chapters.

## 1.1 Laser Processing Geometries: Surface vs. Volume

Laser processing is typically performed in air – preventing complex and costly vacuum technology and enabling an easy integration into industrial in-line manufacturing processes. Due to its low mass density and specific chemical gaseous constituents, air is essentially transparent for most laser wavelengths. Hence, air induces negligible optical losses, shows small nonlinearities, and exhibits a significantly higher “damage threshold” than the processed solids. Thus, laser radiation can be easily guided to the workpiece via free-space optics (mirrors, beam splitters, polarizers, etc.) and focusing elements (lenses, objectives, curved mirrors, etc.), see, for example, Chap. 12 (Schlutow and Fuchs) of this book.

Figure 1.1 shows the two most common scenarios of laser material processing by direct focusing of the laser radiation either onto the surface or into the volume of the sample. For *surface processing* (Figure 1.1a), the surface of the workpiece/target/sample is usually placed near the focal position of the focusing element. The laser radiation is partly reflected, scattered, or absorbed at the surface, leading – at sufficiently high laser intensities (in $W/cm^2$) or fluences (in $J/cm^2$) – to a permanent modification (damage) of the irradiated material. Such modifications may be structural or compositional changes (induced via transient laser-induced phase transitions, such as melting and subsequent re-solidification) or alterations of the surface topography (typically an ablation crater that is formed via material removal from the surface).

In case that the sample material is sufficiently transparent for the laser radiation, *volume processing* in the bulk of the sample (Figure 1.1b) may be possible if the laser radiation is tightly focused by a high numerical optical element. In this processing geometry, the locally laser-induced material modification is spatially confined completely by the surrounding bulk material, thus preventing material removal. Nevertheless, a plethora of other material modifications can be generated, such as structural or compositional changes leading to a reduced (or increased) mass density or refractive index, etc. [Davis, 1996 / Nolte APA, 2003 / Chan APA, 2003/ Eaton, 2005 / Mermillod Chapter, 2023], see for example Chap. 14 (Chambonneau et al.) and Chap. 41 (Zhang et al.) of this book. Even micro-explosions leaving behind a gas-filled void, surrounded by a shell of pressure-compactified material are possible [Juodkazis, 2006].

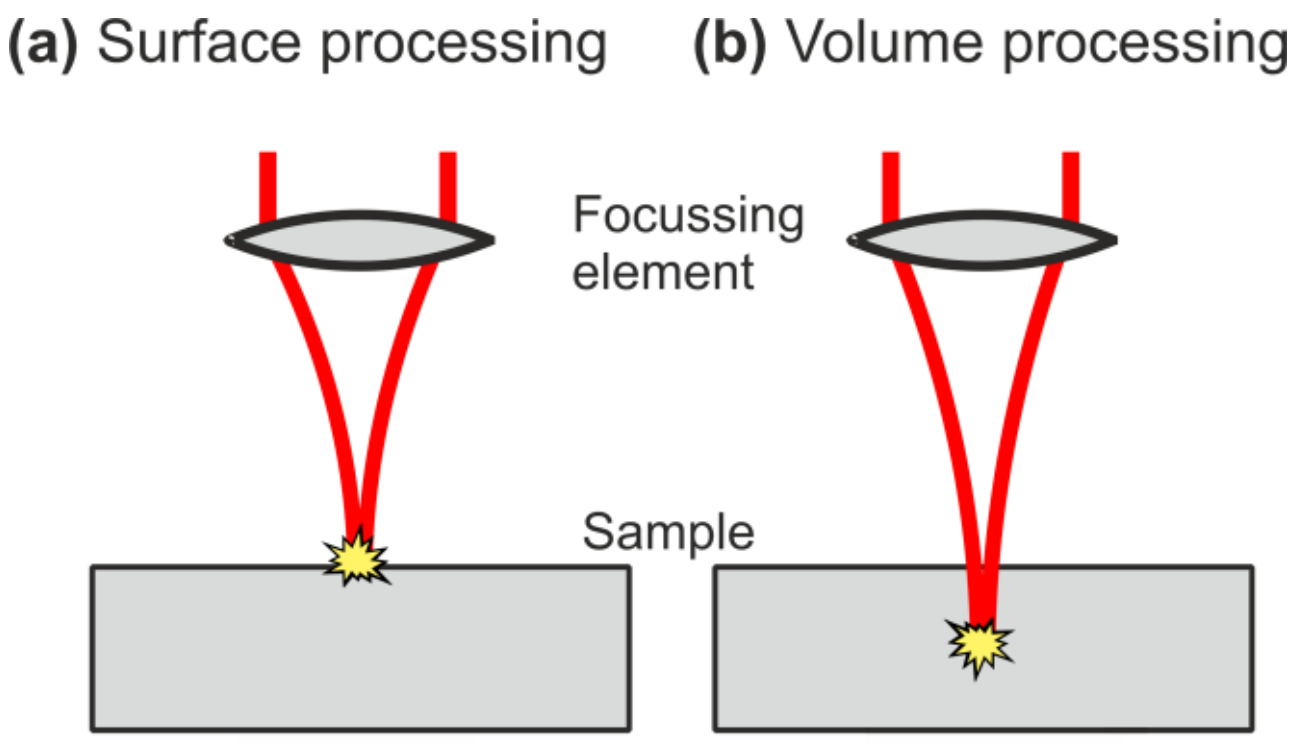

**Fig. 1.1:** Scheme of **(a)** surface and **(b)** volume processing by directly focused laser radiation

Moving the focused laser beam relative to the sample, while simultaneously controlling its focus position, allows the processing of almost arbitrary geometries at the surface (2D) or in the volume (3D) of the material. This can be realized by either (i) moving the sample under a fixed laser beam, or (ii) by deflecting the laser beam in a controlled way, when moving it across a fixed sample. For an up-scaling of the overall processing rates, however, only strategy (ii) is technologically feasible since the mass (inertia) of the sample prevents a sufficiently fast movement of it. For that, advanced scanning technologies, smart laser processing strategies, and laser process optimizations for high repetition rates were developed, see, for example, Chap. 3 (Gräf and Müller), Chap. 4 (Holder et al.), and Chap. 26 (Bulgakov et al.), opening the race to areal laser processing rates towards a square meter per second (see Chap. 15 (Haasler et al.) and Chap. 20 (Schille et al.)).

## 1.2 Advantages of Ultrashort Laser Pulses

For addressing the advantages of laser processing with ultrashort laser pulses, it is instructive to discuss first briefly the most relevant physical processes following the impact of an ultrashort laser pulse on a solid surface. In Figure 1.2, this is exemplified for a semiconducting material, ordering temporally the electron- and lattice-related processes that follow the laser excitation [Sundaram, 2002]. Four main regimes can be identified, i.e., (i) carrier excitation, (ii) thermalization, (iii) carrier removal, and (iv) thermal and structural effects. Each regime involves different fundamental processes and effects on different timescales.

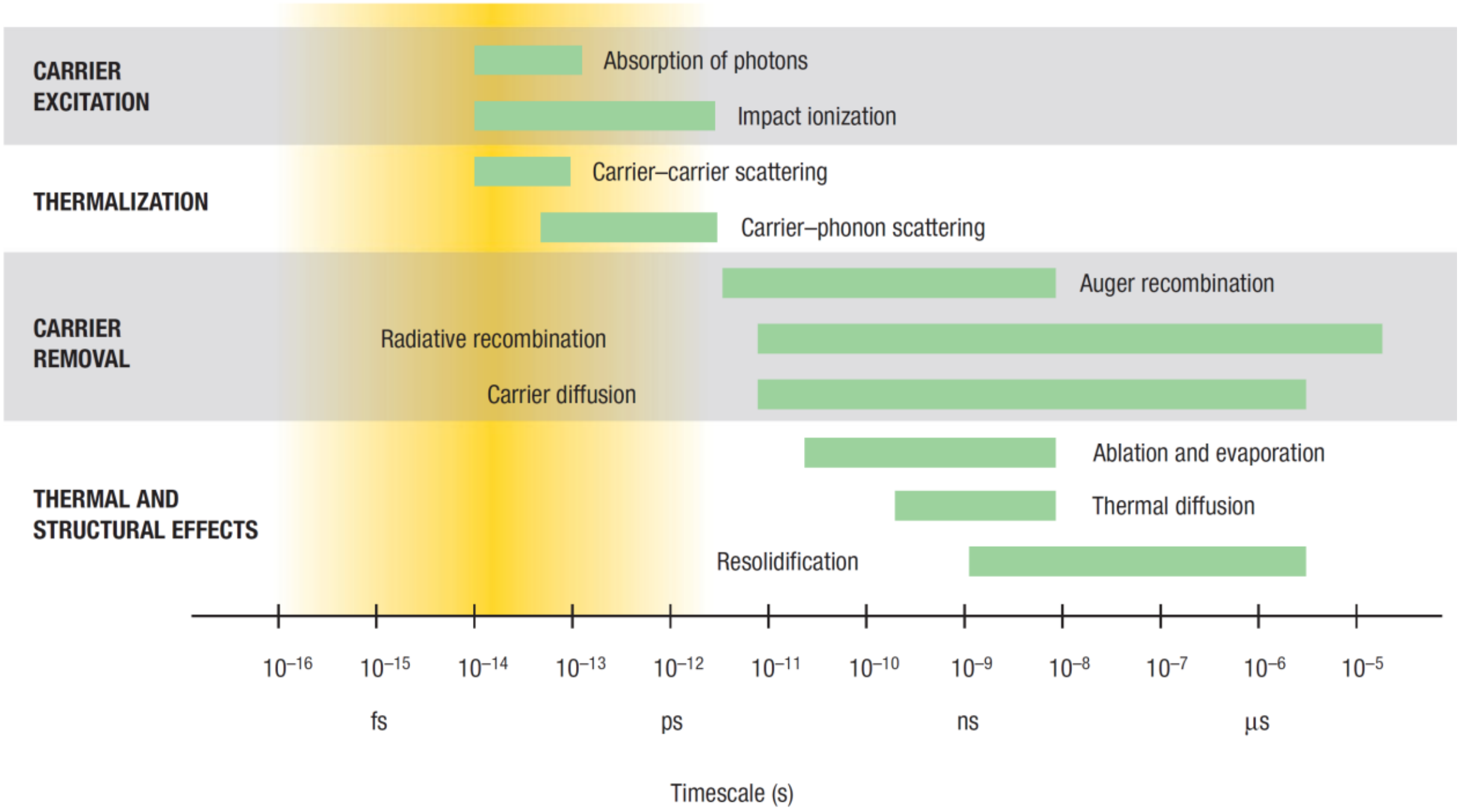

**Fig. 1.2:** Timescales of various electron- and lattice-related processes in laser-excited semiconductors [Sundaram, 2002]. Each horizontal green bar represents an approximate range of characteristic times over a range of carrier densities from $10^{17}$ to $10^{22}$ $cm^{-3}$. The yellow shaded region marks the typical range of accessible ultrashort laser pulse durations (Reprinted from [Sundaram, 2002], S. Sundaram et al., Inducing and probing non-thermal transitions in semiconductors using femtosecond laser pulses, Nature Mater. **1**, 217 – 224, 2002, Springer Nature)

Laser processing with ultrashort (fs – ps) pulses has numerous consequences and advantages that are enabled either by their extremely short optical pulse duration, or by their extremely high peak intensity, as briefly summarized in the following [Bonse APA, 2023]:

- By definition of the term "ultrashort", as used in the context of laser technology, the duration ($\tau_p$) of such laser pulses is shorter than the typical material-dependent *electron-phonon coupling time* ($\tau_{e\text{-}ph}$). This implies that heat diffusion into the surrounding of the laser irradiation spot is insignificant during the absorption of the laser pulse itself.
- The extremely fast energy deposition leads to a local confinement of the laser pulse energy. This results in reduced fluence thresholds of thermal melting and ablation when compared to longer pulse durations in the ns-range or even longer.
- The strong local confinement of energy manifests in a small *heat-affected zone* (HAZ) of typically less than 100 nm in extent. As a consequence, high lateral and vertical machining precision can be obtained by processing with ultrashort laser pulses (see Chap. 6 (Lenzner and Bonse)). Thus, features with sizes even below the focused laser beam diameter can be generated.

- The surface of almost any type of material (metals, semiconductors, polymers, ceramics, dielectrics, or composites) can be processed successfully by employing ultrashort laser pulses with peak intensities ranging between $10^{12}$ and $10^{14}$ W/cm$^2$.
- For all materials, a very sharply defined damage threshold ($\phi_{th}$) exists, above which a permanent material modification (structural, chemical, or ablative) can be induced by laser irradiation. Compared to longer pulse durations ($\tau_p \geq$ ns), the material damage (surface or volume) occurs more deterministic for ultrashort laser pulses.
- Since $\tau_p < \tau_{e\text{-}ph}$, the absorption of the optical radiation through the electronic system of the solid (for times $t < \tau_p$) and the material response following thermal energy relaxation pathways (for $t > \tau_{e\text{-}ph}$) are temporally well separated. This has important consequences, particularly for the ablative surface processing with ultrashort pulsed lasers, as illustrated in Figure 1.3.

Figure 1.3 schematically compares the surface ablation by ultrashort (fs) laser pulse with that of a longer (ns) laser pulses. In the case of fs-laser pulse irradiation ($\tau_p < \tau_{e\text{-}ph}$ , Figure 1.3a), material removal (ablation) from the surface generally begins after the irradiation of the workpiece is completed. The energy input of the laser radiation into the solid and the subsequent ablation process are temporally decoupled. While at early delay times (tens of ps until a few ns) the ablated material may consist of a thermo-mechanically expulsed spallation layer of near-solid-state-density (see Chap. 20 (Bonse)), at later delay times the ablated material forms an ablation plume consisting of nanoparticles, clusters, atoms, ions, and free electrons.

In contrast, for ns-pulse durations ($\tau_p >> \tau_{e\text{-}ph}$, Figure 1.3b), the ablation starts already during the arrival of the laser-pulse to the surface. It forms a plasma plume that is still expanding during the laser pulse absorption by the sample material. This ablation plasma plume then shields the target being processed from the incoming laser beam, a process that is referred to as *plasma shielding*. The effect is strongly affected by the presence of an atmosphere that can prevent the free expansion of the plasma plume and, thus, strongly depends on the ambient pressure.

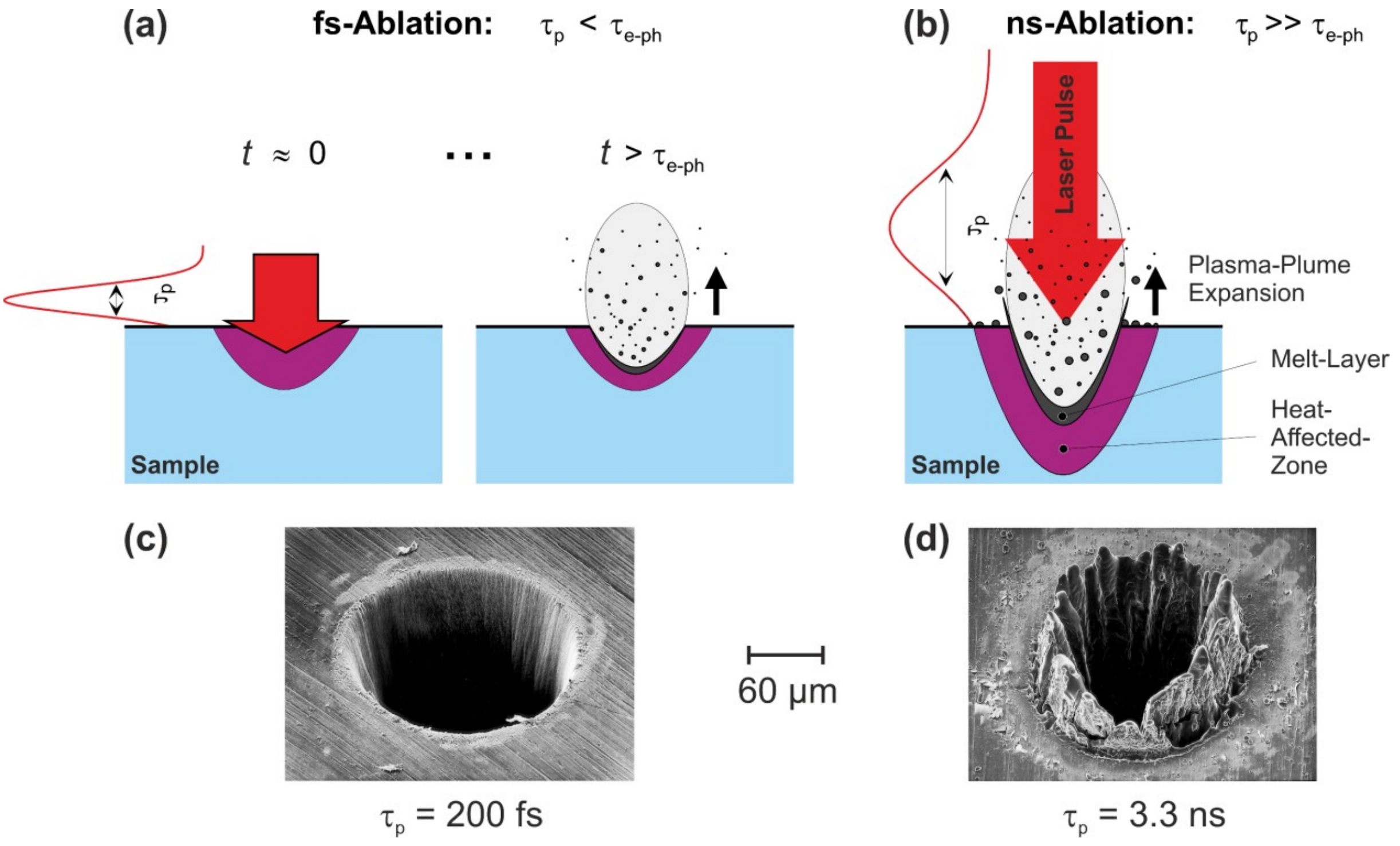

**Fig. 1.3:** Schematic comparison between surface ablation by fs-laser pulses **(a)** and by ns-laser pulses **(b)**, illustrating the reduced heat-affected zone and the effect of *plasma shielding*. Tilted-view scanning electron micrographs of a 100 µm thick steel foil ablated by **(c)** multiple fs-laser pulses [200 fs, 780 nm, 0.5 J/cm$^2$, $10^4$ pulses] or **(d)** by multiple ns-laser pulses [3.3 ns, 780 nm, 4.2 J/cm$^2$, $10^4$ pulses]; both spots were ablated in a rough vacuum environment ($10^{-4}$ mbar). (Figures (a and b) adapted from [Bonse PhD, 2001]. Figures (c and d) reprinted from [Chichkov, 1996], B.N. Chichkov et al., Femtosecond, picosecond and nanosecond laser ablation of solids, Appl. Phys. A **63**, 109 – 115, 1996, Springer Nature)

The incident ns-laser radiation can be defocused, scattered, or absorbed by the plasma cloud, which is ablated from the surface at velocities typically ranging between $10^3$ and $10^5$ m/s. Absorption of the laser radiation is mostly caused by the free charge carriers in the plasma plume via inverse Bremsstrahlung. As a consequence, the energy coupling into the workpiece is reduced, i.e., only a part of the radiation (laser pulse energy) provided can be used for the laser machining process. Studies with ns-pulse durations have shown that at peak intensities of $2\times10^{10}$ W/cm$^2$ only about 20% of the laser pulse energy reaches the workpiece [Yoo, 2000].

Figures 1.3c and 1.3d visualize through tilted-view scanning electron micrographs the superior laser processing quality of a steel foil upon irradiation with multiple fs-laser pulses when compared to that with multiple ns-laser pulses [Chichkov, 1996]. The improved quality manifests via an increased machining precision, a missing burr around the ablation crater, and less material redeposition (debris).

As another consequence of plasma shielding, the surface ablation rates depend on the focusing conditions of the laser beam (focal length, size of the laser spot). This was confirmed for various materials by ablation tests in vacuum with ps- or fs-laser pulses by direct comparison with ns-pulse durations [Beuermann, 1990 / Wolff-Rottke, 1995]. Moreover, in the case of ns-laser irradiation, the ablation rates (depth/pulse) increase with decreasing laser spot diameters at the surface. For fs-pulse durations, on the other hand, the ablation rates are independent of the spot size of the focused laser beam.

The “lifetime” of the ablation plasma plume typically manifesting in the nanosecond to millisecond regime usually imposes an upper limit of the pulse repetition rate for both short (ns) and ultrashort (fs) laser pulses during laser processing: it must generally be avoided that a laser pulse can still interact with the radiation absorbing plasma generated by the preceding pulse. Irrespective of the availability of appropriate laser radiation sources, pulse repetition frequencies in the range between kHz and MHz are, therefore, most suitable for material processing.

The pulse repetition frequency also has an influence on laser material processing if the time interval between two successive individual pulses is shorter than the lifetime of optically excited states in the material (e.g., self-trapped excitons (STEs) or color centers in dielectrics) or the thermal dissipation time of the locally introduced residual heat. In the latter case, if the sample area is not cooled down to room temperature, *heat accumulation* can occur at the workpiece surface, i.e., the successive incremental heating of the irradiation region leads to macroscopic thermal side effects, which then can negate the specific advantages of the ultrashort laser pulses.

This effect of successive heat accumulation at the surface of the irradiated sample (workpiece) is illustrated in Figure 1.4. The effect was theoretically and experimentally explored by Gamaly and co-workers in the context of laser-induced evaporation for pulsed laser deposition (PLD) of thin carbon film with ultrashort laser pulses [Gamaly, 1999]. Later, heat accumulation was analyzed analytically and numerically in the context of ultrashort pulsed laser scan-processing of metals [Bauer OPEX, 2015] and in more detail on the basis of energy balance considerations [Weber OPEX, 2017]. Heat accumulation is also occurring in volume processing [Schaffer, 2003]. Here, it is used e.g. for inscribing optical waveguides [Eaton OPEX, 2008] or for welding, when the laser radiation is focused at the interface between two samples [Miyamoto, 2007], see Sect. 3.3.

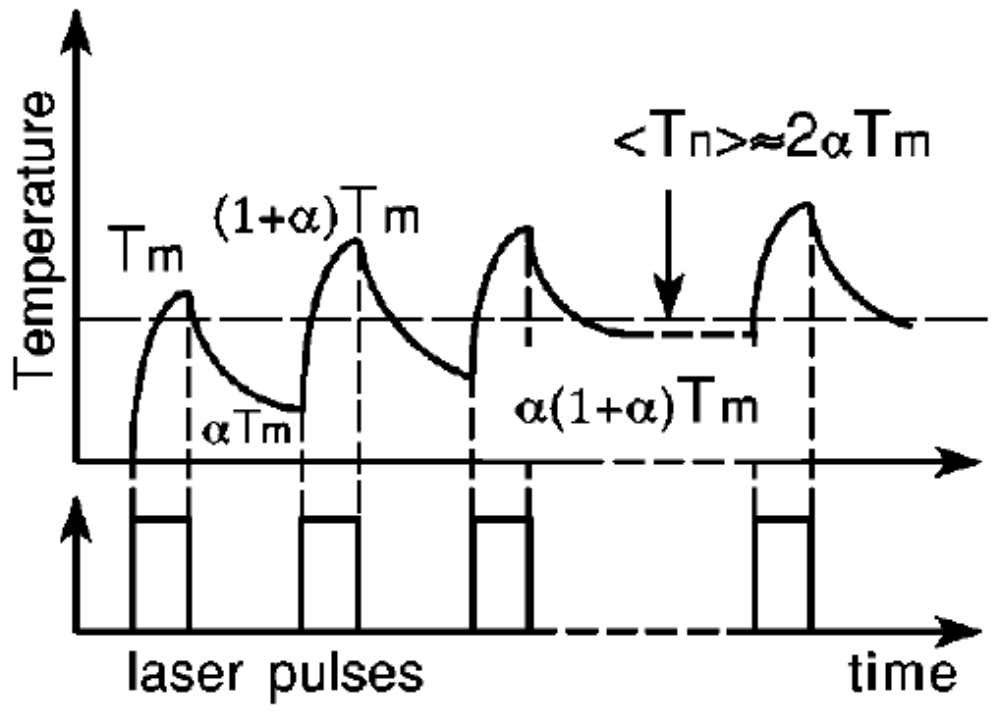

**Fig. 1.4:** Scheme of the evolution of the temperature of the workpiece surface heated by a series of successive laser pulses, illustrating the effect of *heat accumulation* [Gamaly, 1999]. $T_m$ denotes the melting temperature of the sample material and $\alpha = (t_p/t_{pp})^{1/2}$ is defined via the ratio of the pulse duration $t_p$ and the inter-pulse separation time ($t_{pp}$). (Reprinted from [Gamaly, 1999], E.G. Gamaly et al., Ultrafast ablation with high-pulse-rate lasers. Part I: Theoretical considerations, J. Appl. Phys. **85**, 4213 – 4221 (1999), with the permission of AIP Publishing)

The effects of plasma shielding and heat accumulation must be carefully checked for scaling-up the laser processing as they both can seriously affect the processing efficiency and the quality of the obtained results. For more details and for strategies preventing these effects, the reader is referred here to Chap. 4 (Holder et al.) of this book and to Sect. 2.4.

## 1.3 Material Responses

The laser processing with ultrashort laser pulses features additional material-specific aspects, as briefly outlined in this section. These aspects can either directly relate to the energy deposition required for laser material processing, or they indirectly affect the laser beam management and the selection of suitable optical elements.

### Linear and Nonlinear Effects

Most obviously, the high peak intensities ($I_0$) of ultrashort laser pulses render nonlinear interactions in the sample material possible. This enables, for example, that dielectric materials (e.g., optical glasses or crystals) being transparent for low intensity laser radiation start to absorb photons from the ultrashort laser pulse via *multi-photon absorption* (MPA) – a process, which probability scales with $I_0^m$, when $m$ is the number of photons being simultaneously absorbed. Through this nonlinear absorption, optical energy can be deposited in the focal region of a tightly focused ultrashort pulsed laser beam, eventually leading to the desired localized material modification (see Sect. 2.1 for surface processing and Sect. 3.1 for volume processing. In that physical picture, ultrashort laser pulses often feature an intra-pulse intensity-driven transient interplay of the order of nonlinear interaction with the sample material. While the low-intensity wing of the ultrashort laser pulses interacts linearly ($m = 1$) with the irradiated material, the high-intensity peak enables multi-photon absorption processes ($m > 1$). Simultaneously, the energy deposition depth may also transiently vary (see Sect. 2.1).

On the other hand, the *intensity-dependent refractive index* $n_I$ can transiently lead to focusing of the spatially propagating pulsed laser beam via the optical *Kerr-effect* – either in the air environment or even in the sample material itself.

$$n_I(x, y, z, t) = n_0 + n_2 \cdot I(x, y, z, t) \tag{1.1}$$

Here, $n_0$ is the usual (real-valued) refractive index, $n_2$ is the *second-order nonlinear refractive index*, and $I$ is the local laser beam intensity. For a first assessment, the calculation of the so-called $B$-integral ($B = 2\pi/\lambda \int_0^l n_2 I(z) dz$) along the optical beam path ($l$) may help addressing the question of whether or not the focusing of the ultrashort laser pulses is affected by the Kerr-effect [Momma, 1998]. The critical value for the intensity-induced phase delay, i.e., when self-focusing and/or filamentation of the laser beam occur, is at $B \approx \pi$.

Similarly, such beam distortion effects can prevent the use of the *mask projection technique* with high energetic ultrashort laser pulses in air. In such a classical demagnification of an amplitude mask placed in the laser beam, while demagnifying its contour onto the sample surface, an intermediate focus region manifests shortly above the surface. Here, the high laser beam intensities can lead to the Kerr effect (self-focusing), beam filamentation, or even plasma formation (optical breakdown) in the air. The impact of these effects on the laser microdrilling quality can be seen in Figure 1.5, where the laser treatment results with multiple fs-laser pulses (780 nm, 150 fs, 350 µJ, 1.35 J/cm$^2$) are provided for the projection of a square-shaped projection mask onto a stainless steel surface. Here, the laser processing was performed with identical conditions, except that the pressure of the air environment was reduced from ambient conditions (**a**, 1000 mbar) over 100 mbar (**b**) and 10 mbar (**c**) down to 0.1 mbar (**d**). The effect can be widely prevented when using tailored *diffractive optical elements* (DOEs) for beam delivery and confinement [Momma, 1998 / Tönshoff, 2000].

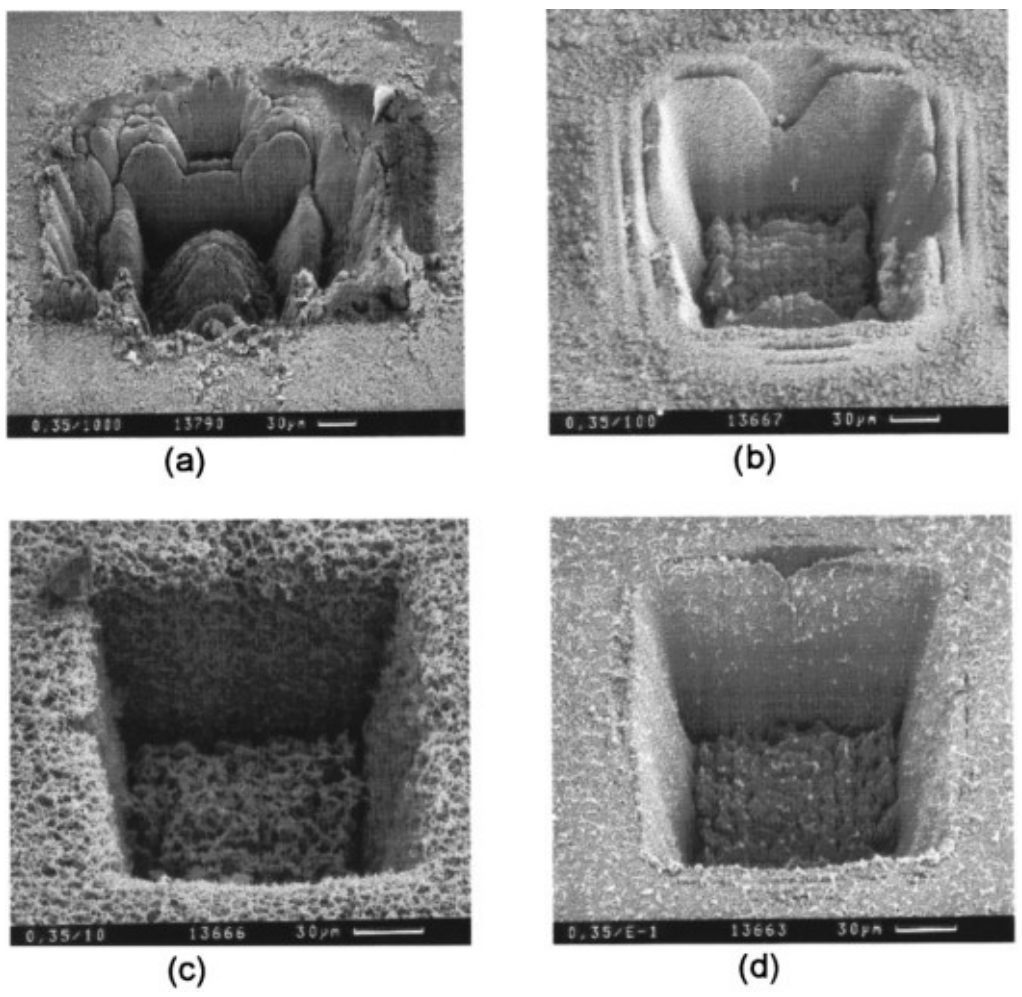

**Fig. 1.5:** Influence of atmospheric pressure when using mask projection for microdrilling with ultrashort laser pulses ($\lambda$ = 780 nm, $\tau_p$ = 150 fs, $E_p$ =350 mJ, $\phi$ = 1.35 J/cm$^2$]. (**a**) 1000 mbar, (**b**) 100 mbar, (**c**) 10 mbar; (**d**) 0.1 mbar. (Reproduced from [Tönshoff, 2000], H.K. Tönshoff et al., Microdrilling of metals with ultrashort laser pulses. J. Laser Appl. **12**, 23 – 27 (2000), with the permission of the Laser Institute of America)

## Nonthermal Effects

At high laser peak intensities, respectively strong material excitation, other material-specific effects may come into play. Several laser-driven nonthermal effects were reported. "Non-thermal" refers here just to the fact that the effect manifests on timescales shorter than $\tau_{e\text{-}ph}$. Note that it does not mean that thermal effects are not involved at later times. In this context, it is important to point to a common misconception: often ultrashort pulses ablation is referred to as "cold ablation". This phrase is a simple but simply wrong oversimplification of the physical processes. Even in laser processing with ultrashort pulses, electrons and ions of the lattice reach extremely high temperatures that can transiently exceed the temperatures for thermal melting or evaporation. Often, it is simply meant that the HAZ is reduced when compared to longer pulse durations (see Sect. 1.2).

One prominent example that has continuously gained attraction over several decades already is the phenomenon of *nonthermal melting* (NTM) of solids. NTM was observed first in time-resolved optical pump-probe experiments monitoring the laser-induced reflectivity change upon irradiation of semiconductors by ultrashort laser pulses [Shank, 1983 / Tom, 1988]. A strong sub-ps increase of the surface reflectivity change was found for silicon, germanium, and other compound semiconductors, indicating the ultrafast transition into a molten (metal-like) liquid state. NTN occurs when a large number of electrons is promoted via laser-induced inter-band transitions from bonding states in the valence band (VB) into anti-bonding states of the conduction band (CB). A few years later, Stampfli and Benemann predicted theoretically that if more than ~10% to 15% of the VB electrons are excited, the materials can be considered as nonthermally molten [Stampfli & Benemann PRB 1992 / Stampfli & Benemann PRB 1994]. NTM is a purely electronic effect that changes the electronic bonds between the atoms of the irradiated solid (see Figure 1.6). Thus, in all-optical experiments on semiconductors, NTM is typically identified via three criteria that must be simultaneously fulfilled: (i) the laser-induced reflectivity increase occurs on a sub-picosecond time scale smaller than $\tau_{e\text{-}ph}$, (ii) the rapidly reached reflectivity level of the NTM state is fluence independent, and (iii) the rise-time required to reach the high reflective state does depend on the laser fluence (carrier generation rate). Later, around the turn of the millennium, pump-probe experiments using ultrafast X-rays to directly probe the lattice structure and its dynamics, along with theoretical modelling confirmed and extended the view on NTM [Siders, 1999 / Lindenberg, 2000 / Rousse, 2001 / Zier, 2015].

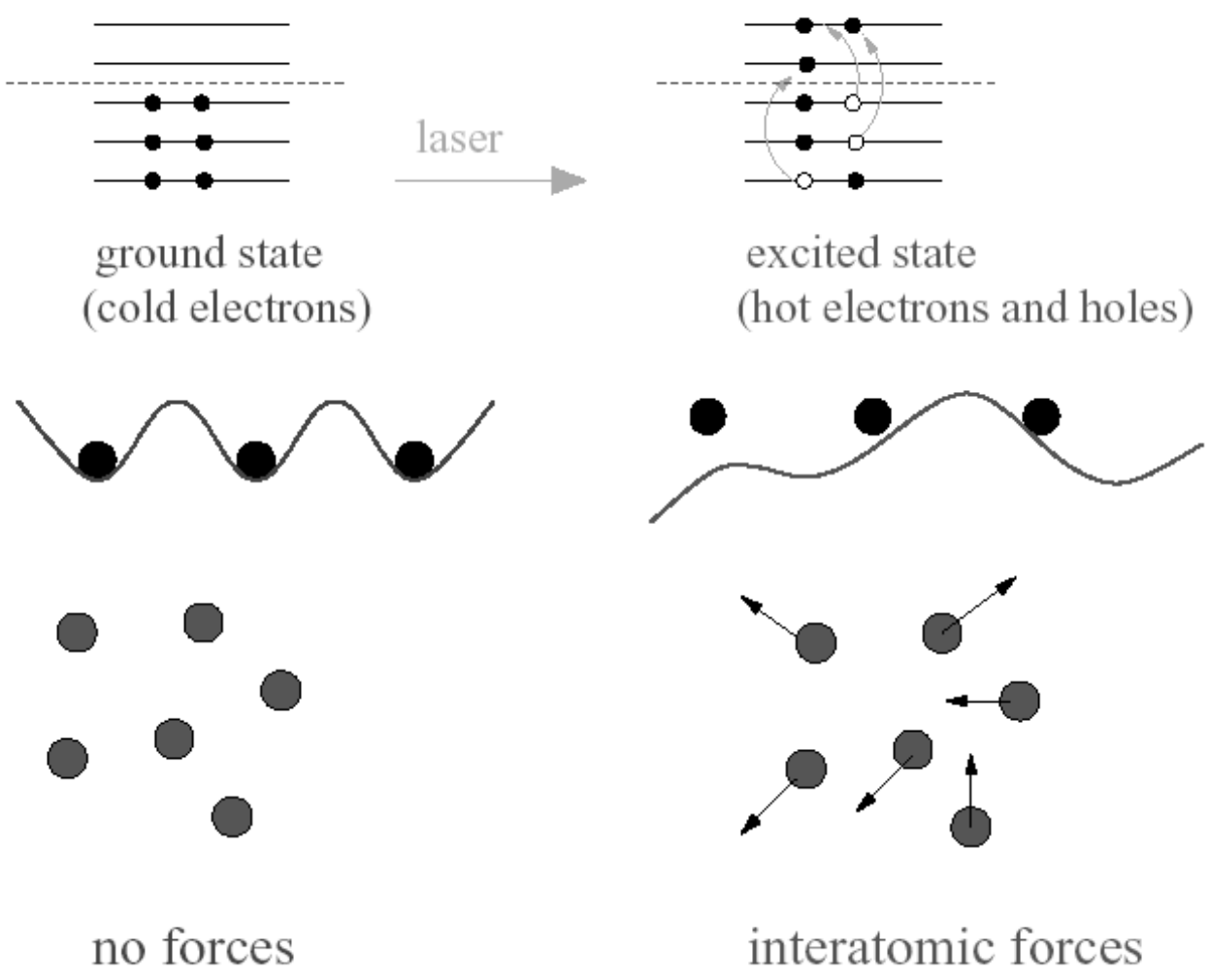

**Fig. 1.6:** Illustration of laser-induced *nonthermal melting* [Jeschke Bookchapter, 2003]. (Left) Solid before the laser irradiation in thermodynamic equilibrium. The electrons fill the ground state and the atoms are at the equilibrium position of the ground state potential energy surface. (Right) The laser pulse changes the electronic occupations, leading to a rapid change of the potential energy landscape. As consequence, the lattice is destabilized and interatomic forces are driving a structural change. (Reprinted from [Jeschke Bookchapter, 2003], H.O. Jeschke et al. in *Nonlinear Optics*, pp. 175 – 214, 2003, Springer Nature)

Another nonthermal effect manifesting in laser processing of dielectrics and semiconductors, is referred to as *Coulomb-explosion* (CE) [Stoian, 2000 / Stoian, 2002 / Bulgakova, 2008]. Here, upon laser irradiation, electrons are removed via the photo-effect from the solid, leaving behind a near-surface layer consisting of positively charged ions. In dielectrics and semiconductors, the moderately low reflow rate of electrons from the surrounding gives the laser-induced ions enough time to be driven explosively apart from the surface via the Coulomb forces between the positively charged ions. In this way, a smooth, few nanometer deep crater is formed at the surface. In contrast, in laser-irradiated metals, the reflow rate of electrons from the surrounding is large enough to rapidly neutralize the charge-separated near-surface layer.

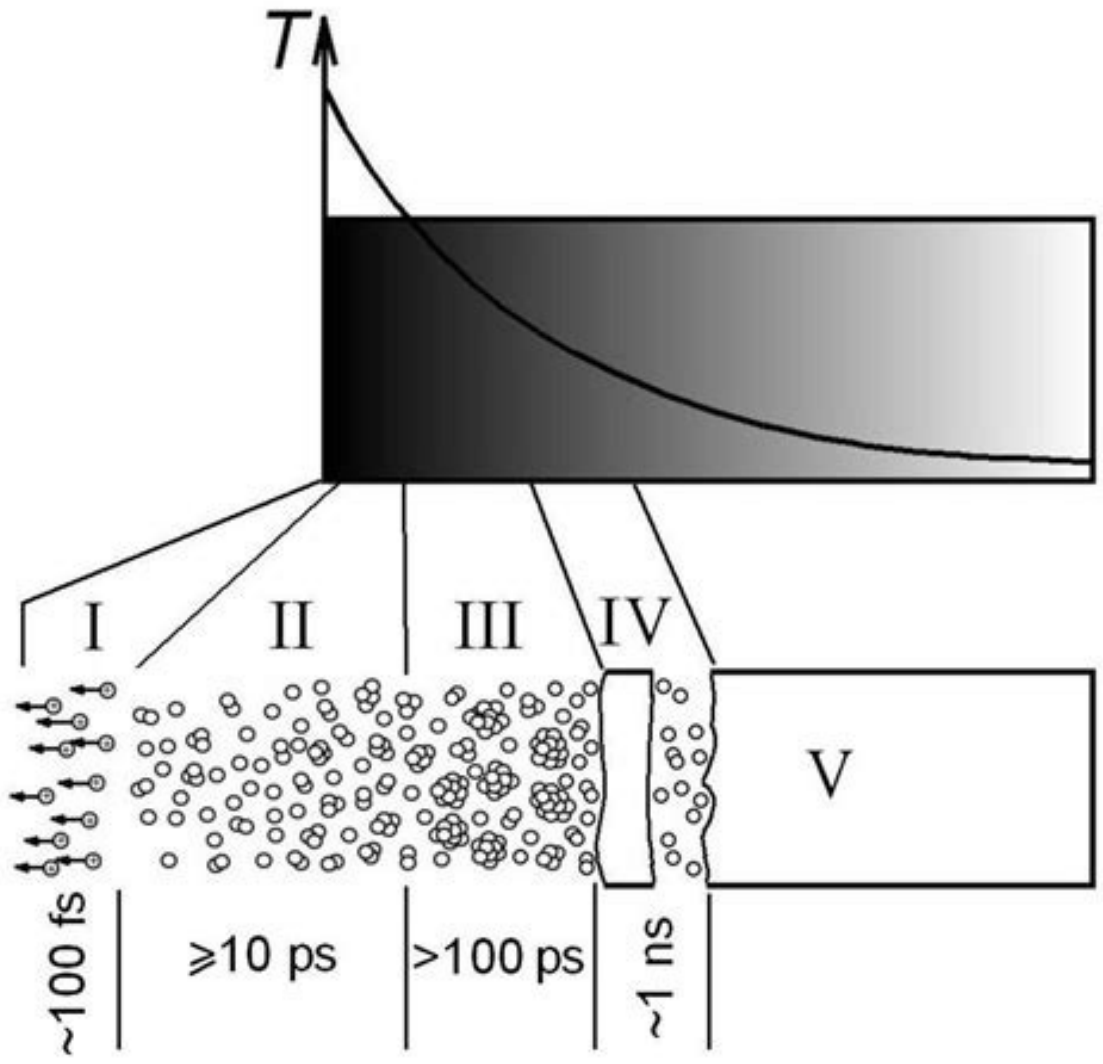

**Fig. 1.7:** Schematic of the main ablation mechanisms (in case of their coexistence) upon fs-laser irradiation of solids [Bulgakova, 2008]. Regions: I – Coulomb explosion; II – fragmentation into plasma, atoms, and small clusters; III – phase explosion; IV – spallation. V represents the non-ablated target. Typical time scales of the mechanisms are indicated at the bottom. (Reproduced from [Bulgakova, 2008], with permission from N.M. Bulgakova et al., Charging and plasma effects under ultrashort pulsed laser ablation, Proc. SPIE **7005**, 70050C (2008). Copyright 2008, SPIE)

Figure 1.7 schematically illustrates the situation of a fs-laser heated surface, visualizing different spatial domains dominated by different ablative mechanisms (I – V), each associated with typical time scales: I – removal of electrons from the solid (within ~ 100 fs), II – Coulomb-explosion of positively charged ions (within > 10 ps), III – *Phase-explosion* of material rapidly driven towards the thermodynamically critical point (within > 100 ps), IV – Spallation of a near-solid-density layer by thermo-mechanical expulsion (100 ps to few ns), and V: Remaining solid. Note that the ablative mechanisms I – IV do not necessarily manifest at the same time during the laser processing with ultrashort laser pulses. Their emergence depends crucially on the material and specific laser irradiation conditions.

A secondary, typically unwanted, laser intensity-driven effect can manifest through the optically excited electrons that form a dense plasma of quasi-free electrons in the solid. Via (intra-pulse) resonance absorption of the laser radiation with that free-electron plasma, high energetic electrons can be generated that interact and collide with inner-shell electrons of the atoms of the solid [Luther-Davies, 1978]. Via inverse Bremsstrahlung and inner shell ionization, a continuous spectrum of soft X-rays superimposed with the characteristic X-ray lines of the constituents of the material can be emitted, featuring photon energies up to a few tens of keV. This secondary emission occurs at laser peak intensities exceeding ~$10^{13}$ W/cm$^2$. Particularly for modern laser systems, providing moderately high laser pulse energies at high laser pulse repetition rates, this can already cause the emission of hazardous dose rates of secondary X-rays and may represent a serious working safety issue [Legall APA, 2018], see Chap. 11 (Schille and Krüger)].

### Near-field Effects

The ultrashort pulse durations enable that intra-pulse *near-field effects* can become effective since they are not screened by materials responses that occur on time scales larger than the laser pulse duration (e.g., phase transitions, melt displacement effects, etc.). Optical near-fields can exploit laser-matter interaction at a nanometer scale, where the electromagnetic field is evanescent rather than propagating. Hence, near-field optics is not subject of the classical far-field diffraction limit, enabling outstanding spatial resolution due to the evanescent field characteristics that is decaying spatially typically on the few tens of nanometer scale.

Most relevant in this context is the fact that the presence of sharp conductive topographic structures exposed to an external optical field can lead to a significant enhancement of the local electric field intensity by several orders of magnitude. This enables a coherently driven energy localization effect that does not rely on classical far-field optics. As one consequence, the self-ordered formation of sub-wavelength periodic nanostructures with spatial periods between tens and hundreds of nanometers can form and manifest as the surface as specific type of *high spatial frequency laser-induced periodic surface structures* (HSFL) or in the volume of transparent materials as *nanogratings* [Buividas, 2014 / Bonse, 2017 / Rudenko, 2017 / Rudenko AM, 2020 / Bonse & Gräf LPR, 2020], see the example in Sect. 2.3 and Chap. 9 (Ancona et al.).

Apart from such self-organized effects, the optical near-field can be tailored and technologically used for site-controlled surface nanostructuring [Plech LPR, 2009 / Huang, 2025], for example combining ultrashort laser pulses with *scanning near field microscope* (SNOM) or *atomic force microscope* (AFM) technology [Lieberman, 1999 / Nolte PB, 1999 / Chimalgi, 2003], see the example in Sect. 2.3.

However, such near-field effects can also impose undesired effects in USP laser processing. This may become relevant when samples are laser treated that already exhibit a surface texture with sharp topographic features on the micrometer to nanometer scale, as for example used for silicon-based solar cells in photovoltaics. Then, the optical near-field enhancement can prevent a homogeneous material removal of an additionally overcoated anti-reflection layer since preferential ablation occurs at locations with strong near-field enhancement [Bonse TSF, 2013].

# 2 Surface Processing

In this section, we discuss some fundamentals of laser surface processing and point to the peculiarities of ultrashort laser pulses. Wherever it is possible, we provide simple models that allow to estimate analytical scaling laws of the laser processing, e.g., of the damage/ablation threshold fluence or modification/ablation depths with regard to the incident laser intensity/fluence. Sect. 2.1 addresses aspects of surface processing with Gaussian laser beams and related material-specific laser-induced carrier generation at the surface, Sect. 2.2 describes the subsequent transfer of optical energy to the materials lattice and subsequent energy relaxation processes. In Sect. 2.3 selected applications are presented, before in Sect. 2.4 the most relevant factors are discussed that can limit the up-scaling of laser surface processing.

## 2.1 Optical Energy Deposition: Loose Focusing

Laser surface processing usually relies either on the tailored precise sculpturing of very small topographic surface features in a direct focusing geometry with tightly focusing optics (high NA) or it relies on the areal processing with larger focused beam sizes using loose focusing conditions (low NA) for obtaining high areal processing rates. In this section, we will address the optical energy deposition of loosely focused ultrashort pulsed laser radiation. This situation manifests in the constraint that the focused laser beam diameter at the surface is much larger than the energy deposition depth. As a consequence, residual heat transport predominantly occurs one-dimensionally (1D) into the depth of the workpiece.

### Irradiation Spots

The laser irradiation of solids with spatially Gaussian beams can be analyzed analytically and allows to render important insights for laser processing. Figure 1.8 displays the situation, where a fixed spot at the sample surface is irradiated by a radially symmetric Gaussian beam having the fluence profile

$$\phi(r) = \phi_0 \cdot e^{-2\left(\frac{r}{w_0}\right)^2} \quad . \qquad (1.2)$$

Herein, $\phi_0$ is the peak fluence in the center of the beam in front of the sample, $r$ represents the radial coordinate, and $w_0$ is the Gaussian beam waist radius ($1/e^2$-decay). For a sample spot irradiated at a peak fluence $\phi_0$ exceeding the materials ablation threshold fluence $\phi_{th}$, a crater with a diameter $D = 2 \cdot r_{th}$, a radial profile $d(r)$, and a maximum depth value $d_{max}$ is formed. Note that the residual optical energy $E_{res} = 2\pi \int_{r_{th}}^{\infty} \phi(r) dr$ contained integrated over the tail of the Gaussian fluence profile for $r > r_{th}$ only contributes to the heating of the sample and cannot be used for material removal (shaded region in Figure 1.8).

Note that the following analysis is not restricted to the physical process of ablation. The formalism can similarly be applied to any kind of permanent material modification that exhibits a defined threshold fluence and can be unambiguously identified at the laser-irradiated surfaces, e.g., laser-induced melting that is followed by superficial amorphization or re-crystallization [Bonse APA, 2002 / Bonse APA, 2006, Florian, 2021], laser-induced oxidation [Bonse APA, 2000], etc.

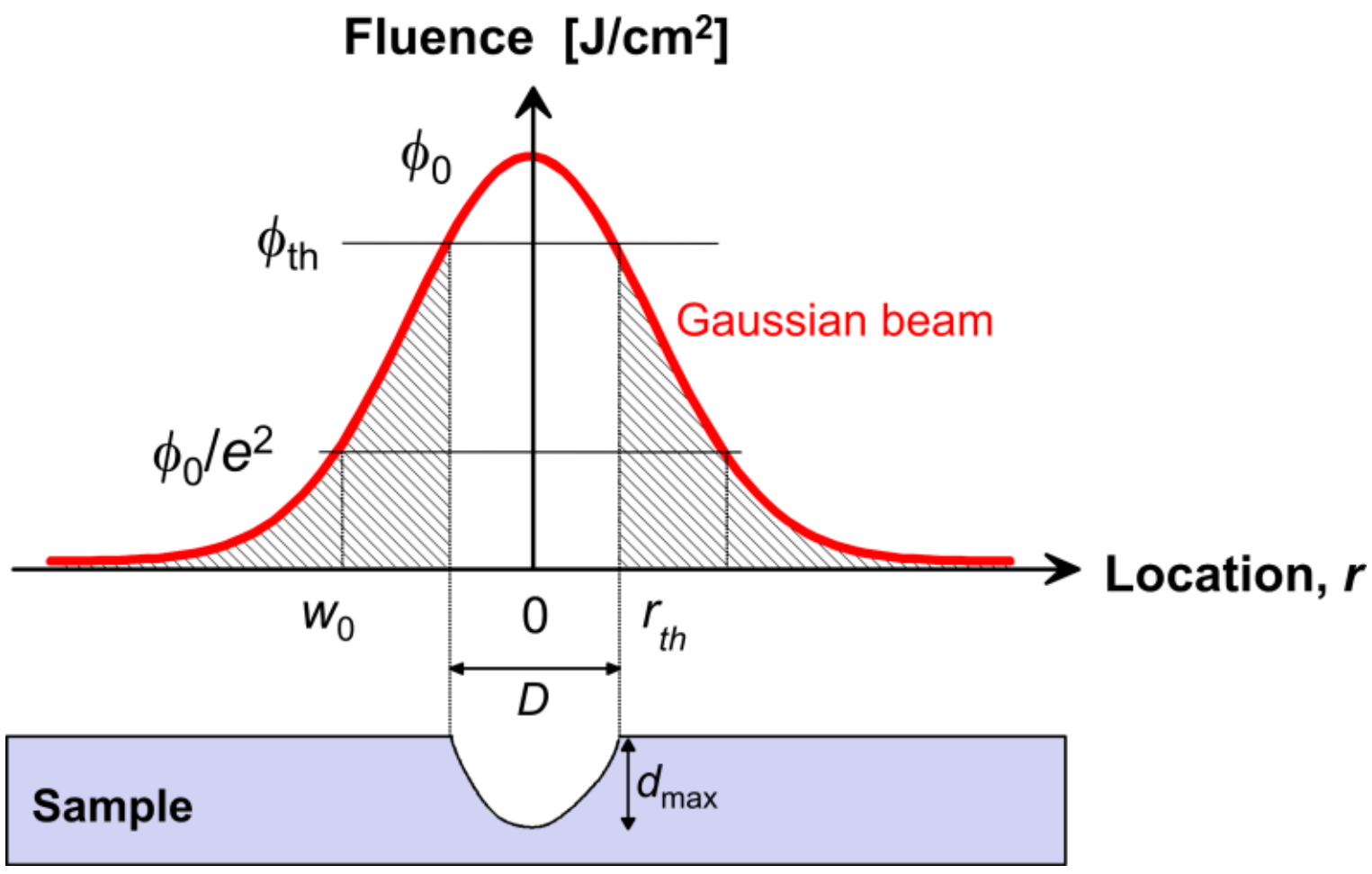

**Fig. 1.8:** Scheme of laser spot processing with a spatially Gaussian laser beam. For local fluences $\phi$ above the ablation threshold $\phi_{th}$, a crater with a diameter $D$ and a maximum depth $d_{max}$ is formed at the surface

From the boundary condition that the ablation threshold fluence $\phi_{th}$ is locally reached at the ablation crater radius ($\phi(r_{th}) = \phi_{th}$), Equation (1.2) allows to derive an expression for the squared crater diameter obtained for a radially Gaussian laser beam:

$$D^2 = 2w_0^2 \cdot \ln\left(\frac{\phi_0}{\phi_{th}}\right) \tag{1.3}$$

The fluence profile $\phi(x,y)$ and the laser pulse energy $E_p$ are generally related via

$$E_p = \iint_{-\infty}^{\infty} \phi(x,y)\, dx\, dy. \tag{1.4}$$

For a spatially Gaussian beam profile as defined via Equation (1.2), this integral can be computed analytically, resulting in a linear relation between the peak fluence $\phi_0$ and the laser pulse energy $E_p$:

$$\phi_0 = \frac{2\,E_p}{\pi w_0^2} \tag{1.5}$$

The combination of Equation (1.3) with Equation (1.5) enables a simple experimental method for the determination of the ablation threshold fluence, as proposed first by J.M. Liu in 1982 [Liu, 1982]. In brief, in a first step, the measured ablation crater diameters are squared and are plotted semi-logarithmically as a function of the laser pulse energy. The slope of the least-squares-fit to the experimental data provides the Gaussian beam waist radius $w_0$ according to Equation (1.3). In a following step, the corresponding peak fluences $\phi_0$ can be calculated from the laser pulse energies via Equation (1.5). The threshold pulse energies $E_{th}$ and threshold fluences $\phi_{th}$ are obtained from the data (fit) extrapolation $D^2 \rightarrow 0$. By its principle, the approach provides the value of $w_0$ precisely in the laser processed surface plane and does not require an extra measurement.

Figure 1.9 shows an example of a threshold determination according to this "$D^2$-method" (sometimes referred to as "Liu-plot") for two different silicate-based multicomponent glasses (commercial Borofloat® (Schott AG) and in-house-made $LiGe_{50}Si_{25}$) irradiated with single Ti:sapphire fs-laser pulses (800 nm, 120 fs) [Grehn, 2014].

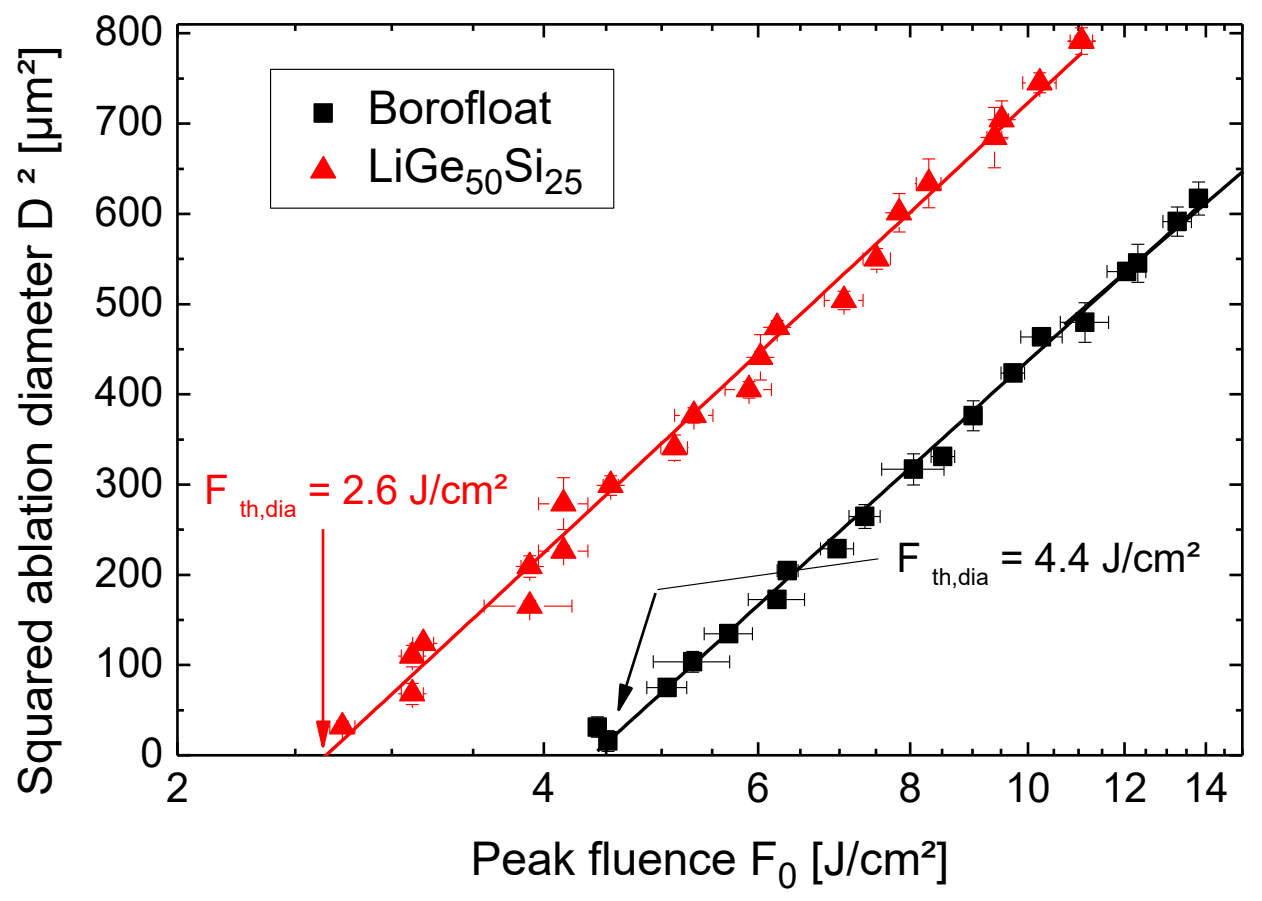

**Fig. 1.9:** Liu-plot of the squared ablation diameter vs. the incident peak fluence for two different silicate-based multicomponent glasses (Borofloat® and $LiGe_{50}Si_{25}$) upon irradiation with single Ti:sapphire fs-laser pulses [Grehn, 2014]. Note the semilogarithmic data representation. The solid lines are least-squares-fits of the corresponding data set to Equation (1.3). The two arrows indicate the ablation threshold fluences obtained from the fits. (Reprinted with permission from [Grehn, 2014] © Optical Society of America)

From the least-squares-fits to the data, a Gaussian beam waist radius $w_0 = (16.2 \pm 0.3)$ µm was obtained, along with single-pulse ablation threshold fluences of $\phi_{th}$ of (2.6 ± 0.1) J/cm$^2$ ($LiGe_{50}Si_{25}$) and (4.4 ± 0.1) J/cm$^2$ (Borofloat®), respectively [Grehn, 2014].

The $D^2$-method is very robust and does not rely on specific laser-matter interaction mechanisms – just the existence of a sharply-defined threshold, an unambiguously determined spot diameter measurement, along with the use of a spatially Gaussian beam are required for establishing the formalism. The formalism presented for radially symmetric Gaussian beams here can straightforwardly be applied also to elliptical Gaussian beams. In 2021, Garcia-Lechuga and Grojo presented another extension of the formalism to Airy-disk-like laser beams, as they arise from truncated Gaussian beams [Garcia-Lechuga, 2021]. The authors demonstrated that although the associated Airy-disk pattern is, in principle, inappropriate to directly use the Liu's method, the extended formalism leads to a superior reliability in threshold determination, provided that additional correction factors are applied.

## Optical Absorption Depth

Using a generalized "Lambert-Beer" law, the maximum ablation crater depth $d_{max}$ and its shape $d(r)$ can be calculated from the optical absorption processes occurring in the irradiated material. In this model, the differential equation describing the depth- ($z$-)dependent intra-sample laser beam intensity change via linear and nonlinear (multi-photon) absorption can be written as

$$\frac{dI(z)}{dz} = -\alpha_1 \cdot I(z) - \alpha_2 \cdot I^2(z) - \alpha_3 \cdot I^3(z) - O(\alpha_4). \tag{1.6}$$

The different coefficients $\alpha_m$ are related to the number $m$ of simultaneously absorbed photons lead to a reduction of the beam intensity upon propagation into depth.

Without additional simplifications, Equation (1.6) must be solved numerically. However, for the cases of purely linear absorption (only $\alpha_1 \neq 0$), the well-known Lambert-Beer law follows as an analytical solution featuring an exponential intensity decay. Also, in the case of simultaneous one- and two-photon absorption (only $\alpha_1 \neq 0$ and $\alpha_2 \neq 0$) – a situation that can be valid, for example, for semiconducting or low bandgap materials – the differential equation can be solved analytically via separation of variables.

However, for that it must be considered that a part of the incident laser radiation (fluence according to Equation (1.2)) is being reflected at the sample-air interface with a Fresnel-reflectivity $R$ that is assumed to be constant here during the ultrashort pulsed laser irradiation. Assuming additionally for simplicity a temporal square-shaped laser pulse, the intra-sample intensity $I$ can then be substituted by the fluence $\phi$ incident in front of the sample via $I = \phi \cdot (1 - R)/\tau_p$ (for other temporal pulse shapes, an additional numerical factor close to one may have to be introduced, arising from the temporal integration over the dedicated pulse profile). Along with the boundary condition that the ablation threshold fluence is precisely reached at the maximum ablation crater depth $d_{max}$, the solution of the differential equation can be written as [Krüger, 1999 / Florian, 2021]

$$d_{\max} = \frac{1}{\alpha_1} \cdot \ln\left(\frac{\left[\frac{\alpha_1}{\phi_{\mathrm{th}}} + \frac{\alpha_2(1-R)}{\tau_{\mathrm{p}}}\right] \cdot \phi}{\alpha_1 + \frac{\alpha_2(1-R)}{\tau_{\mathrm{p}}} \cdot \phi}\right). \tag{1.7}$$

For purely linear absorption in the material ($m = 1$), as usually valid for metals, Equation (1.7) simplifies to the relation

$$d_{\max} = \frac{1}{\alpha_1} \cdot \ln\left(\frac{\phi}{\phi_{\mathrm{th}}}\right). \tag{1.8}$$

The maximum ablation depth then shows a logarithmic saturation behavior with the incident laser fluence. This characteristic scaling of the ablation depth per pulse was verified for single- and multi-pulse ablation in vacuum and in air at fluences moderately above the ablation threshold [Preuss APA, 1995 / Nolte JOSAB, 1997].

For bandgap materials such as transparent glasses or crystals, often the transient optical absorption upon irradiation with high-intensity ultrashort laser pulses is dominated by a single multi-photon absorption coefficient $\alpha_m$, i.e., the nonlinear order $m$, where $m$ represents the lowest number of photons that is required to bridge the energy band gap ($m \cdot h\nu \geq E_g$). Assuming solely $\alpha_m \neq 0$ and $m > 1$, the differential equation Equation (1.6) can be rewritten as [Preuss APL, 1993 / Puerto, 2010 / Grehn, 2014]

$$\frac{d\phi(z)}{dz} = \underbrace{-\frac{\alpha_m}{\tau_{\mathrm{p}}^{m-1}} \cdot (1-R)^{m-1} \phi^m(z)}_{\text{MPA}}. \tag{1.9}$$

Using the above boundary conditions, the maximum crater depth $d_{max}$ for dominant $m$-photon absorption is obtained as

$$d_{\max} = \frac{\tau_{\mathrm{p}}^{m-1}}{(m-1)\alpha_m(1-R)^{m-1}} \left[\frac{1}{\phi_{\mathrm{th}}^{m-1}} - \frac{1}{\phi^{m-1}}\right]. \tag{1.10}$$

Hence, for band gap materials ($m > 1$), the maximum crater depth $d_{max}$ scales with a characteristic power law proportional to $\phi^{1-m}$. Moreover, it is interesting to note that the maximum crater depth saturates at a constant level that is affected by the laser pulse duration and by material-specific parameters, such as the order of dominant absorption, the corresponding multi-photon absorption coefficient, the surface reflectivity, and the ablation threshold fluence.

Figure 1.10 exemplifies some experimental results of the maximum crater depth measured for two different silicate-based multicomponent glasses (commercial Borofloat® (Schott AG) and in-house-made $LiGe_{50}Si_{25}$) as a function of the peak fluence of single Ti:sapphire laser pulses (800 nm, 120 fs) [Grehn, 2014].

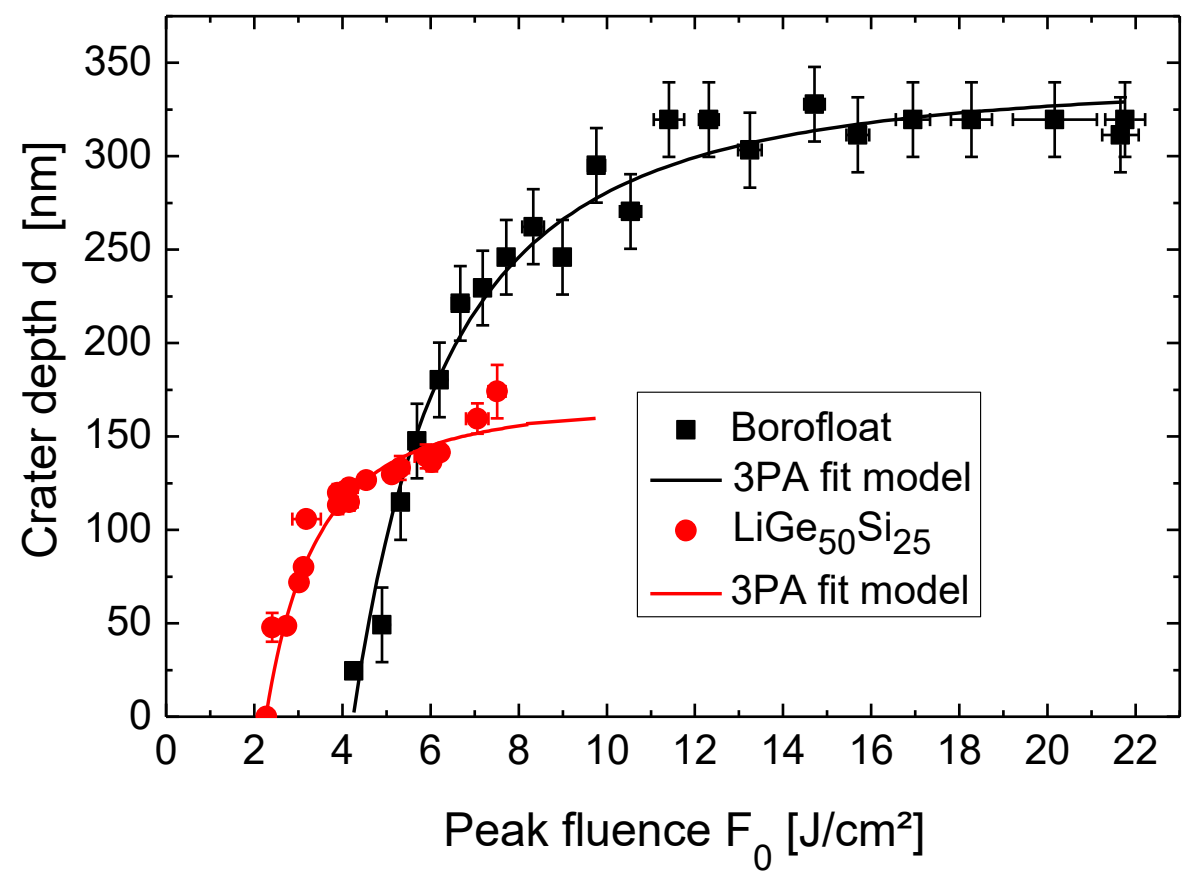

**Fig. 1.10:** Maximum depth of the ablation craters vs. the incident peak fluence for two different silicate-based multicomponent glasses (Borofloat® and $LiGe_{50}Si_{25}$) upon irradiation with single Ti:sapphire fs-laser pulses [Grehn, 2014]. The solid lines are least-squares-fits of the corresponding data set to Equation (1.10) using $m = 3$. (Reprinted with permission from [Grehn, 2014] © Optical Society of America)

The two selected glasses exhibit band gap energies of $E_g = 4.4$ eV ($LiGe_{50}Si_{25}$) and $E_g = 4.7$ eV (Borofloat®) [Grehn, 2014]. Thus, for a photon energy of hν = 1.5 eV (800 nm wavelength), three photons must be simultaneuosly absorbed from the fs-laser beam to bridge the band gap of the glasses. Hence, the crater depths were evaluated using Equation (1.10) with $m = 3$. Reasonably good agreement is obtained for the three-photon absorption (3-PA) coefficient used as a fit-parameter along with the ablation threshold fluence. Values of $\alpha_3 = (9.3 \pm 0.1)\times 10^{-23}$ ($LiGe_{50}Si_{25}$) and $\alpha_3 = (1.3 \pm 0.1) \times 10^{-23}$ $cm^3/W^2$ (Borofloat®) were obtained, respectively. These values are consistent with the ones independently measured for similar silicate-based glasses by the *z*-scan method [Grehn, 2013]. The corresponding damage threshold fluences obtained from the fit to the crater maximum depth are $\phi_{th} = (2.4 \pm 0.1)$ J/cm² ($LiGe_{50}Si_{25}$) and $\phi_{th} = (4.3 \pm 0.1)$ J/cm² (Borofloat®), in excellent agreement with the threshold values obtained from the fit to the ablation crater diameters (see Figures 1.9 and 1.10).

For finally obtaining radial crater profiles $d(r)$, the incident fluence level $\phi$ in Equations (1.7), (1.8), and (1.10) can be straightforwardly substituted by Equation (1.2) for obtaining the radial scaling. Specific examples for the two above-mentioned glasses can be found in Fig. 3 of Grehn et al. [Grehn, 2014] (not shown for brevity here).

## Carrier Excitation

Following the pioneering work of Bloembergen [Bloembergen, 1974], in 1996 Stuart et al. published a very appealing and simple analytical model of the optical absorption of ultrashort laser pulses in bandgap materials, i.e., for semiconductors and dielectrics [Stuart JOSAB, 1996 / Stuart PRB, 1996].

In such materials, free charge carriers (electrons) in the conduction band are only present in small concentrations under normal conditions (typically $10^8$ - $10^{10}$ $cm^{-3}$) but can be generated by intense photoexcitation. This occurs for intensities in the range between $10^{11}$ and $10^{14}$ $W/cm^2$ either by (1) linear or nonlinear multi-photon interband absorption or by (2) linear intraband absorption of the radiation by electrons already present in the CB (*free-carrier absorption*, FCA) with subsequent *collisional ionization* (CI, see below). The combined sequence of FCA and CI is called *avalanche ionization* (AI). In this context, "ionization" means the generation of an electron-hole pair in the intact solid. Repeated as a cascade, the mechanism (2) can lead to an avalanche-like increase in the number of electrons in the CB during the laser pulse excitation (see below).

These processes are illustrated in more detail in the energy scheme of a laser-excited semiconductor (band gap energy $E_g$) presented in Figure 1.11 (inspired and extended from van Allmen and Blatter [van Allmen, 1988]). The optical excitation is displayed here via a single-photon (linear) *interband absorption* (red vertical arrow), generating an electron in the CB. Such a carrier can induce *electron-phonon collisions* (generating heat), or it can absorb another photon from the laser beam via *free-carrier absorption*. When a sufficiently large number of electrons is present in the CB, the electrons are additionally starting to interact with each other via *electron-electron collisions*, leading to a redistribution of the electron momentum and the relaxation of their kinetic energies into a Fermi-Dirac distribution. For CB electron densities above a critical density, a *free-electron plasma* is formed, where the electrons interact collectively and exhibit characteristic resonances, such as *Plasmons*. Electrons of large kinetic energy can relax to the lower energy band edge (dashed vertical arrow), while the released energy is used to create another electron-hole pair (solid back arrow). This combination is called *collisional ionization*. In semiconductors and at sufficiently large electron densities in the CB, such electrons can undergo *Auger recombination* processes. Then, an electron from the CB recombines with a hole from the VB, and the released potential band gap energy is transferred as kinetic energy to another electron in the CB. Since three carriers are involved in the process, its probability scales with $N_e^3$. Auger recombination becomes effective only for strong laser excitation and exhibits minimum Auger recombination times of $\tau_{Auger} \sim 6$ ps at high carrier densities [Yoffa, 1980].

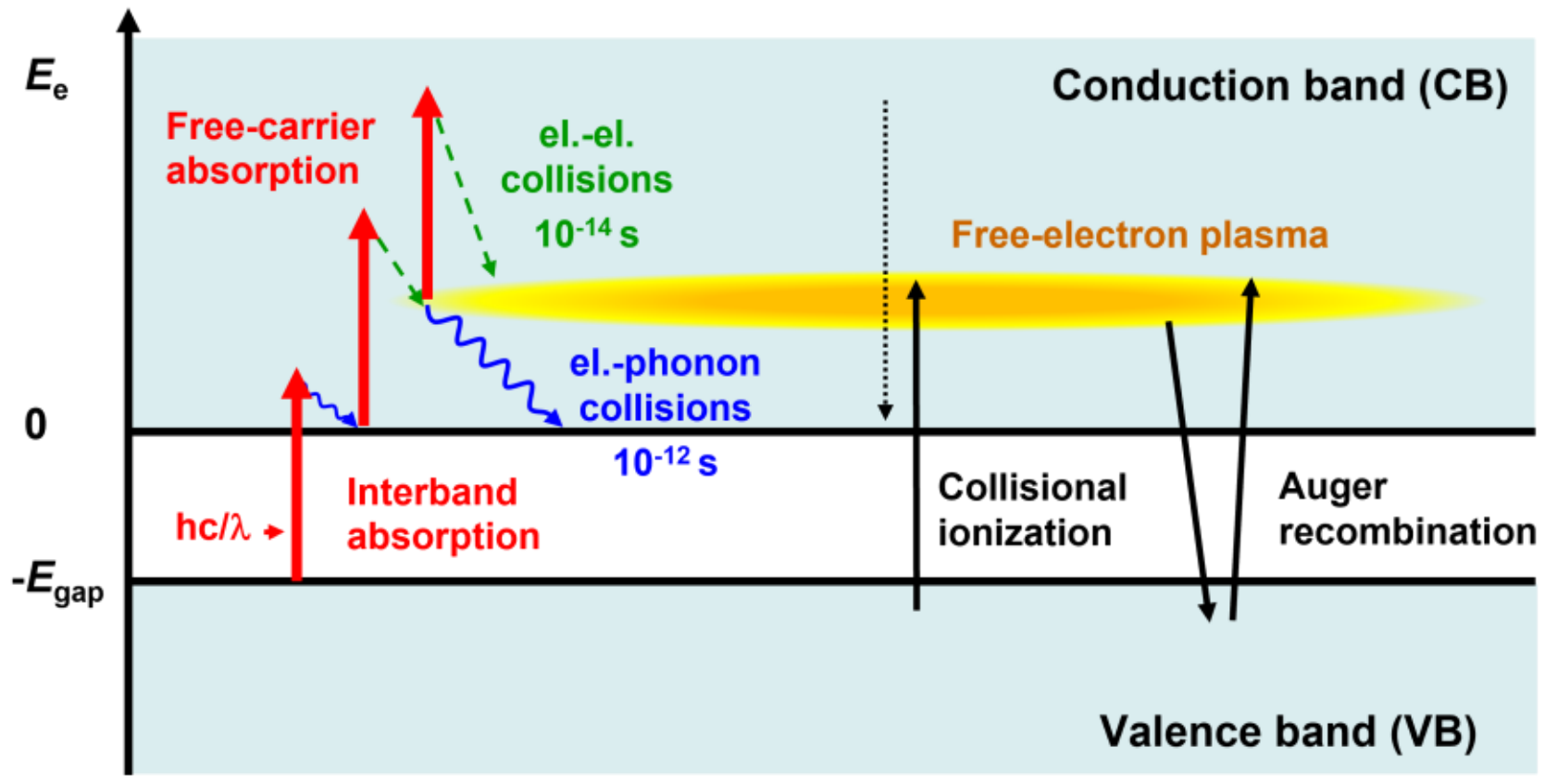

**Fig. 1.11:** Scheme of electron excitation and relaxation in a laser-irradiated semiconductor, visualizing a free-electron plasma exchanging energy with the laser beam and the lattice via various fundamental physical processes. (Inspired by van Allmen and Blatter [van Allmen, 1988])

One refers to an *optical breakdown* (*laser-induced breakdown*) when a critical electron density $N_e^{(cr)}$ is exceeded in the conduction band of the solid as a result of the irradiation, which then ultimately leads to material damage [Bloembergen, 1974]. This characteristic charge carrier density depends on the material and typically ranges between $10^{19}$ cm$^{-3}$ and $10^{21}$ cm$^{-3}$ [Bloembergen, 1974 / Bäuerle, 2011].

For a quantitative understanding of these effects, the temporal development of the charge carrier density $N_e$ of the electrons in the conduction band as a result of the irradiation of the band gap material by a laser pulse with an intensity curve $I(t)$ can be described by the rate equation (1.11) that takes into account both the charge carrier generation in the conduction band by avalanche ionization (AI) and by multi-photon absorption (MPA) [Stuart PRB, 1996]:

$$\frac{dN_e(t)}{dt} \cong \underbrace{\epsilon_{av} \cdot I(t) \cdot N_e(t)}_{\text{AI}} + \underbrace{\gamma_m \cdot I^m(t)}_{\text{MPA}} \tag{1.11}$$

Here, $\varepsilon_{av}$ denotes the avalanche coefficient, and $\gamma_m$ is the coefficient for the simultaneous absorption of $m$ photons. Here, $m$ is the minimum number of photons required for a transition between the valence and conduction band ($m \cdot h\nu \geq E_g$). Loss mechanisms (e.g., spatial carrier diffusion or recombination) were neglected in the rate equation (1.11), which is particularly justified for ultrashort laser pulses.

At high radiation intensities, however, the two effects (AI and MPA) often occur simultaneously, so Equation (1.11) must then be solved numerically. Nevertheless, it is very instructive to solve the equation analytically in the two limiting cases of pure AI or pure MPA by separation of variables. In both cases, this procedure allows to gain useful insights about the scaling behavior of the damage threshold ($\phi_{th}$) with the laser pulse duration $\tau_p$.

## Damage Threshold Fluence

Case 1 (Only AI; $\epsilon_{av} \neq 0$, $\gamma_m = 0$): In the limiting case of pure AI, immediately after laser pulse excitation, the solution of Equation (1.11) gives the expression $N_e(\tau_p) = N_0 \cdot \exp\{\epsilon_{av} \int_0^{\tau_p} I(t)dt\}$, i.e., there is an avalanche-like (exponential) increase in the electron density. The quantity $N_0$ is the number of electrons present in the CB at the start of the laser pulse. The expression $\int_0^{\tau_p} I(t)dt$ represents the fluence $\phi$ of the laser pulse. The critical electron density $N_e^{(cr)}$ is, therefore, just reached at the threshold fluence

$$\phi_{th}(\tau_p) \cong \frac{1}{\epsilon_{av}} \cdot \ln\left(\frac{N_e^{(cr)}}{N_0}\right) \cong \text{Const.} \tag{1.12}$$

In the case of pure AI (representing a linear absorption process), the damage threshold fluence $\phi_{th}$ is independent of the laser pulse duration $\tau_p$.

Case 2 (Only MPA; $\epsilon_{av} = 0$, $\gamma_m \neq 0$): In the limiting case of pure $m$-photon absorption, the solution $N_e(\tau_p) = N_0 + \gamma_m \cdot \int_0^{\tau_p} I^m(t)dt$ results for rate equation (1.11). If we assume a temporal rectangular pulse of intensity $I_0$ to simplify the mathematical analysis, there is a linear relationship between fluence $\phi$ and intensity $I_0$ of the pulse ($\phi = I_0 \cdot \tau_p$). This allows the time dependence of the damage threshold to be derived:

$$\phi_{\text{th}}(\tau_{\text{p}}) = \gamma_m^{-1/m} \cdot \left[N_{\text{e}}^{(\text{cr})} - \text{N}_0\right]^{1/m} \cdot \tau_{\text{p}}^{\frac{m-1}{m}} \quad (1.13)$$

This relation reveals that the damage threshold $\phi_{\text{th}}$ scales with $\tau_{\text{p}}^{\frac{m-1}{m}}$. For $m = 1$ (linear absorption) it is (again) independent of the pulse duration $\tau_{\text{p}}$. For pure two-photon absorption ($m = 2$), it leads to a $\sqrt{\tau_{\text{p}}}$-law, and for $m = 3$ and higher, it represents a power law.

In summary, linear absorption mechanisms result in a constant damage threshold, while multi-photon absorption leads to decreasing damage thresholds for shorter laser pulse durations. An experimental determination of $\phi_{\text{th}}(\tau_{\text{p}})$, therefore, allows conclusions to be drawn about the specific optical absorption mechanisms acting in the sample during the ultrashort pulse irradiation featuring pulse duration shorter than the electron-phonon relaxation time ($\tau_{\text{p}} < \tau_{\text{e-ph}}$).

## Material Incubation

One particularity of ultrashort laser pulses is related to their high peak intensities, which can drive the irradiated material into highly excited electronic states, which subsequently can relax along complex pathways to a new equilibrium state. Hereby, the laser-irradiated but non-ablated residual sample material may undergo transient or permanent material modifications that can be reinforced successively upon exposure to multiple laser pulses. Such material modifications can be the formation of laser-induced electronic defect states, frequently denoted as *incubation centers*. Among them are vacancies and interstitials in crystalline semiconductors, self-trapped excitons (STEs) and color centers in ionic crystals or glasses, broken chemical bonds, molecular fragments, etc.

Such incubation centers are of particular importance for the laser processing of wide bandgap materials at photon energies $h\nu < E_{\text{g}}$. Multiple laser pulses may successively increase the material's absorption at the irradiation wavelength (see Figure 1.12), thus lowering its damage threshold fluence $\phi_{\text{th}}$. As consequence, several "incubation pulses" may have to be applied to the surface of a material before ablation begins.

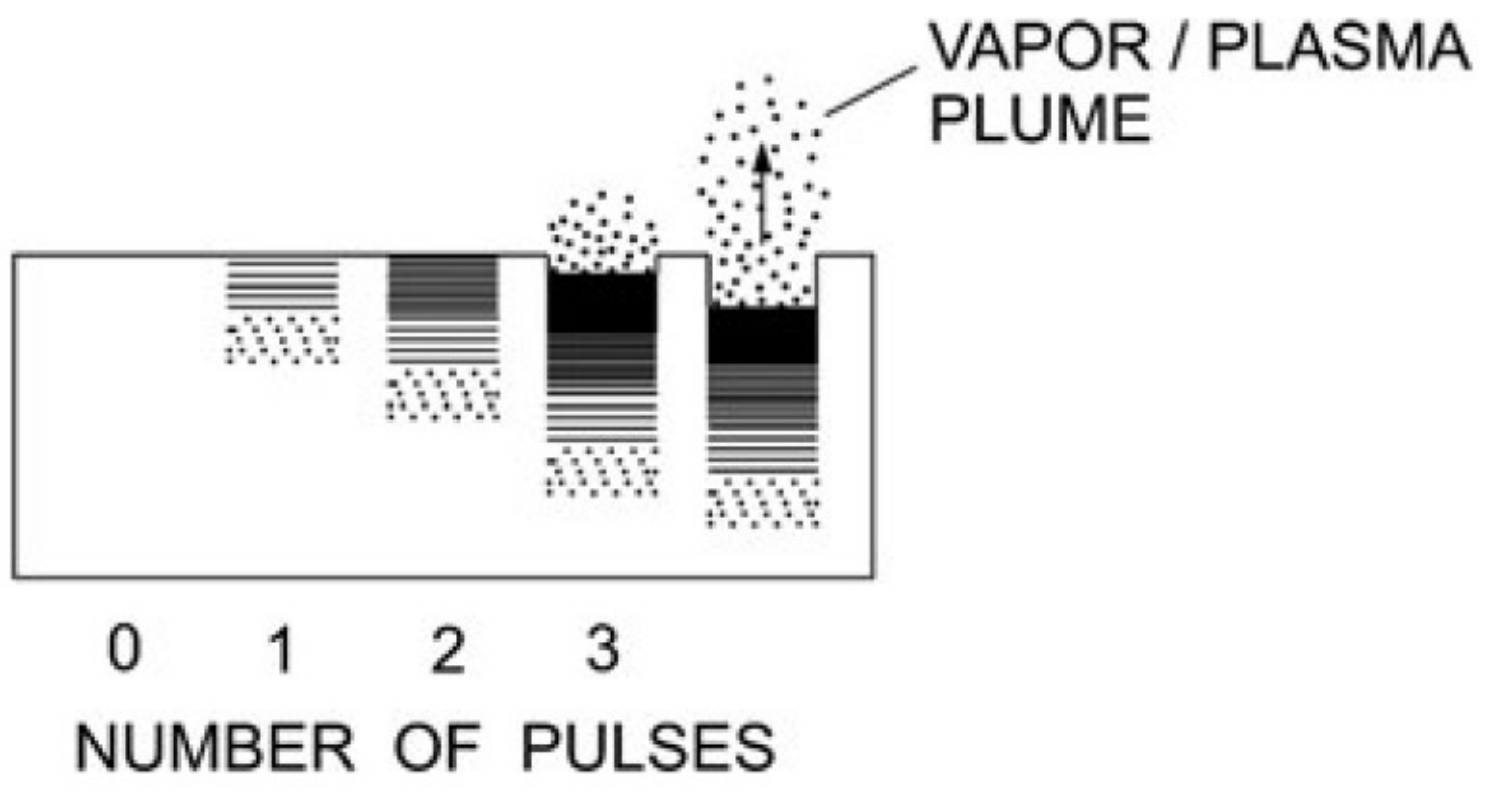

**Fig. 1.12:** Scheme illustrating the effect of incubation through an increase of optical absorption with successive laser pulses [Bäuerle, 2011]. The gray tone/shading reflects the amount of laser-induced defect states increasing the effective optical absorption coefficient. (Reprinted from [Bäuerle, 2011], D. Bäuerle, *Laser Processing and Chemistry*, 4th edn., pp. 237 – 278, 2011, Springer Nature)

The incubation effect can be seen as a pulse number dependent change of the *effective optical absorption coefficient* [Bäuerle, 2011]

$$\alpha_{\text{eff}} = \alpha_1 + \sum_{m=2}^{\infty} \alpha_m \cdot I^m + \alpha_{\text{D}}(N_p)\,. \qquad (1.14)$$

The first two terms on the right-hand side represent the linear absorption coefficient and the nonlinear absorption coefficients according to Equation (1.6). The third term $\alpha_D$ considers the pulse-number ($N_p$) dependent change of the absorption due to laser-generated defect states. Via $1/\alpha_{eff}$, an *effective optical penetration depth* can be defined. Its values inversely decrease with $\alpha_{eff}$ when incubation turn the irradiated material more absorbent at the laser wavelength.

At very high intensities where self-induced transparency or avalanche ionization take place, Equation (1.14) exceeds its validity range. For more details regarding laser-induced incubation, the reader is referred to Chap. 6 (Lenzner and Bonse) of this book.

## 2.2 Energy Relaxation

After the laser excitation of charge carriers at the surface, the deposited optical energy starts relaxing toward the thermodynamic equilibrium. The possible pathways depend crucially on the degree of material excitation that rule the number and nature of subsequent physical processes.

### Two-Temperature Model

In strongly absorbing materials and at low to moderate degrees of laser-induced material excitation, the *two-temperature model* (TTM) has proven to be suitable for describing the light-matter interaction for ultrashort laser pulses. It was first developed in 1974 by Anisimov et al. to describe the irradiation of metals [Anisimov, 1974] but has also been used successfully for semiconductors [Pronko, 1996] or dielectrics [Bulgakova Chapter, 2010 / Derrien Chapter, 2023].

In metals, the energy of the electromagnetic radiation is mainly absorbed by the conduction electrons of the solid. When the laser pulse duration $\tau_p$ is significantly shorter than the time of energy transfer between the electron system and the solid-state lattice, a non-equilibrium state is created between these two systems, and they must be described separately by an electron temperature $T_e$ and a lattice temperature $T_l$.

Immediately after the absorption of the ultrashort optical laser pulse by the electrons, a non-equilibrium state is formed within the electron system (a temperature cannot yet be defined at this time), and the electrons move ballistically with velocities up to the order of $10^6$ m/s [Wellershoff, 1999]. The excited electron system relaxes mainly by electron-electron collisions within 10 fs to 1 ps ($\tau_e$) to form a *Fermi-Dirac distribution* (“hot electron gas”). When the electrons are in thermal equilibrium with each other, they penetrate the solid by diffusion at velocities < $10^4$ m/s. At $T_e >> T_l$, an energy transfer between the excited electron system and the solid lattice takes place – depending on the strength of the coupling between electrons and lattice (phonons) – through a large number of electron-phonon interactions on a typical time scale ($\tau_{e\text{-}ph}$) between 1 and 100 ps, depending on the specific metal. The energy transfer continues until electrons and lattice reach the same temperature ($T_e = T_l$). From this point onwards, the excitation energy dissipates into the surrounding material through ordinary heat diffusion driven by phonons. At this point, the electron velocities are < $10^2$ m/s.

For a quantitative mathematical analysis, the time- ($t$) and space-dependent ($\vec{x}$) temperature fields can be described by a set of two coupled heat conduction equations [Anisimov, 1974 / Wellershoff, 1999 / Bulgakova Chapter, 2010 / Bäuerle, 2011].

$$C_e \frac{\partial T_e}{\partial t} = \overbrace{\nabla(\kappa_e \nabla T_e)}^{\text{Electron diffusion}} - \overbrace{\Gamma_{e-ph}(T_e - T_l)}^{\text{e-ph coupling}} + \overbrace{Q_{Laser}(\vec{x},t)}^{\text{Source term}} \quad (1.15)$$

$$C_l \frac{\partial T_l}{\partial t} = \underbrace{\nabla(\kappa_l \nabla T_l)}_{\text{Heat diffusion}} + \underbrace{\Gamma_{e-ph}(T_e - T_l)}_{\text{e-ph coupling}} \quad (1.16)$$

Here, $C_e$ and $C_l$ denote the heat capacities of the electrons or the solid lattice (per unit volume), $\kappa_e$ and $\kappa_l$ are the corresponding thermal conductivities, and $Q_{Laser}(\vec{x},t)$ is the heat source term due to the absorbed laser radiation. The *electron-phonon coupling constant* $\Gamma_{e\text{-}ph}$ is the critical quantity that determines the strength of electron-lattice coupling and varies significantly among different metals [Lin PRB, 2008]. It is related to the electron-phonon coupling time ($\tau_{e\text{-}ph}$) via [Bäuerle, 2011]

$$\tau_{e-ph} = \frac{C_e \cdot C_l}{(C_e + C_l) \cdot \Gamma_{e-ph}} \,. \quad (1.17)$$

Thus, the TTM consists of a set of two partial differential equations for $T_e$ and $T_l$, coupled via $\Gamma_{e\text{-}ph}$ multiplied by the temperature difference ($T_e - T_l$). Hence, any temperature difference between the electrons and the lattice drives an energy exchange among both sub-systems until a thermal equilibrium is reached.

In case of loose focusing, i.e., when the focused beam diameter of the exciting laser pulse is larger than the optical penetration depth of the laser radiation (usually fulfilled for metals), a one-dimensional simplification of the differential equation system is justified ($\vec{x} \rightarrow z$), and the source term can be described in a first approximation by $Q_{Laser}(\vec{x},t) = \alpha_1 \, (1-R) \, I(t) \, e^{-\alpha_1 z}$ [Nolte PhD, 1999]. The function $I(t)$ represents the time-dependent laser intensity in front of the sample surface, $\alpha_l$ is the linear absorption coefficient of the metal (see Eq. (1.6) and the discussion of the optical absorption depth), and $R$ is its reflectivity at the irradiation wavelength. Together with specific boundary conditions (sample stratigraphy, material properties, laser irradiation conditions, etc.) Equations (1.15) and (1.16) can be numerically solved together at different depths ($z$) to obtain temporal solutions of the temperature fields of the electrons and the lattice, respectively.

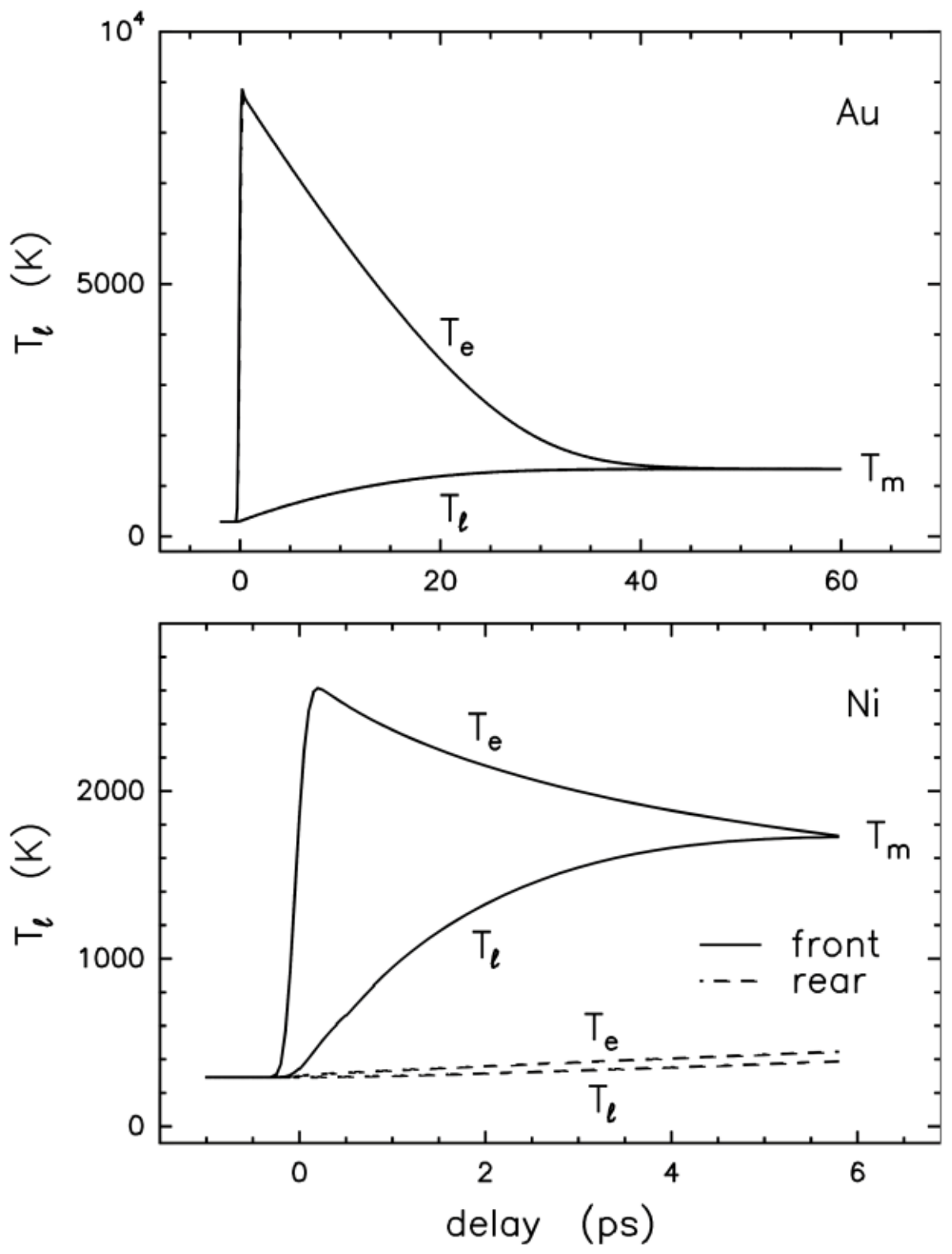

**Fig. 1.13:** Time dependence of the electron temperature ($T_e$) and the lattice temperatures ($T_l$) of 100 nm thick Au (top) and Ni (bottom) films on an $SiO_2$ substrate upon irradiation with single fs-laser pulses (400 nm, 200 fs, 0.023 J/cm$^2$) as numerically calculated via solving Equations (1.15) and (1.16) [Wellershoff, 1999]. The fs-laser pulse arrives to the surface at delay time zero. The solid lines indicate the temperatures at the air/metal interface (sample surface). The dashed lines represent the values at the underlying metal/silica interface. Note the different scales of the abscissa. (Reprinted from [Wellershoff, 1999], S.-S. Wellershoff et al., The role of electron–phonon coupling in femtosecond laser damage of metals, Appl. Phys. A **69** [Suppl.], S99 – S107, 1999, Springer Nature)

Figure 1.13 exemplifies some selected results of the TTM by displaying the time dependence of the electron and the lattice temperatures at the front surface (solid lines) and the rear interface (dashed lines) of a gold film (top graph) and a nickel film (bottom graph). In both cases, the 100 nm thick metal films were deposited on a fused silica substrate. [Wellershoff, 1999]. The absorption of a single laser pulse with a duration of 200 fs at 400 nm wavelength and with a laser fluence of 0.023 J/cm$^2$ is suitable to reach the melting temperature ($T_{melt}$) of the lattices of both materials.

Due to the much stronger electron-phonon coupling in nickel ($\Gamma_{e\text{-}ph}$(Ni) = $3.6\times10^{17}$ W/(m$^3$K) [Bäuerle, 2011]), the relaxation of energy to the lattice occurs about eight times faster for nickel when compared to gold ($\Gamma_{e\text{-}ph}$(Au) = $2.1\times10^{16}$ W/(m$^3$K) [Bäuerle, 2011]), as manifesting in the surface temperatures of $T_e$ and $T_l$ merging after ~6 ps for Ni and after ~45 ps for gold, respectively. However, no significant heat transport takes place through the Ni-film to the rear surface, where nearly no significant temperature difference to the initial sample temperatures of $T_e$ and $T_l$ is computed, even when the front surface reaches the melting temperature $T_m$ after approx. 6 ps delay (see Figure 1.12, bottom graph). In contrast, for gold (Figure 1.12, top graph), front and rear surface temperatures are not to distinguish for $T_e$ and $T_l$. Obviously, for gold, the absorbed optical energy can spread through the ballistic transport and diffusion of electrons through the entire 100 nm thick film, before the energy of the excited electrons converts via electron-phonon scattering into heat.

Some additional comments regarding the validity range of the TTM should be made. The model, as discussed above, assumes that the heat conduction equation is also valid for the femtosecond time regime. It ignores the energy transport in the solid by ballistic charge carriers, which can initially penetrate the solid almost unhindered due to their high kinetic energies and directed momentum. However, this deficiency can be remedied by a modified heat source term $Q$ [Wellershoff, 1999].

Moreover, thermal emission of electrons, photoelectric effect, lattice deformations, thermodynamic phase transitions, and ablation are also not taken into account here. In order to integrate the ablation process into the model in addition to melting, the hydrodynamic *Navier-Stokes equations* (NSE), the *continuity equation* (conservation of mass), and the *Equations of State* (EOS) can be implemented in extended two-temperature models as additional equations, while latent heats (melting and evaporation) must be considered as additional boundary conditions. In Equations (1.15) and (1.16), pressure- and velocity-dependent additional terms must then also be added [Korte, 1999].

Within the frame of these extended models, lattice deformations are also taken into account. The high transient electron temperatures ($T_e \sim 10^5$ K) due to the ultrashort pulse irradiation transiently may lead to a large electron pressure within the solid, which results in internal tensile stresses of up to several Mbar with subsequent material expansion. The hot electron gas thus generates localized internal stresses, which can lead to fracture of the solid-state material and subsequent ablation (macroscopic picture of the microscopic electron-phonon interaction). Moreover, when repetitively being applied, such localized stress may contribute to incubation effects via *laser-induced fatigue* effects (see Chap. 6 (Lenzner and Bonse)).

The two-temperature model also takes into account the case of interaction with long laser pulse durations (e.g., ns-laser pulses). If $\tau_p >> \tau_{e\text{-}ph}$ applies, the lattice of the solid then already heats up during the laser pulse. The electron and lattice system can then be described by a common temperature $T = T_e = T_l$ and the TTM differential equation system reduces to the *parabolic heat conduction equation*, when neglecting the heat capacity of the electrons ($C_e << C_l$) and assuming that heat conduction in metals is essentially due to electrons ($\kappa_l << \kappa_e$).

$$C_l \frac{\partial T}{\partial t} = \kappa_e \nabla^2 T + Q_{\mathrm{Laser}}(\vec{x}, t) \tag{1.18}$$

The solution of this equation must generally be computed by numerical integration along with the specific boundary conditions.

For a δ-function laser pulse in time (as an approximation of an ultrashort laser pulse with $\tau_p << \tau_{e\text{-}ph}$) and an absorbed fluence, $\phi_{abs} = (1 - R)\cdot\phi_0$ deposited at $z = 0$, one finds for times $t > 0$ a temperature increase $\Delta T = T - T_0$ of

$$\Delta T_\delta(z, t) = \frac{(1-R)\cdot\phi_0}{C_l\cdot\sqrt{\pi\cdot D_{\mathrm{th}}\cdot t}} \cdot \exp\left(\frac{-z^2}{4\cdot D_{\mathrm{th}}\cdot t}\right), \tag{1.19}$$

where $T_0$ is the sample temperature prior to the laser irradiation and $D_{\mathrm{th}} = \frac{\kappa_e}{C_l}$ is the *thermal diffusivity* of the material [Wellershoff, 1999].

This temperature distribution decays rapidly in time and spreads into the material with a *thermal diffusion length* $L_{\mathrm{th},\delta} = \sqrt{\pi \cdot D_{\mathrm{th}} \cdot t}$. For a temporally Gaussian laser pulse, Equation (1.18) must be solved numerically. The thermal diffusion length then accounts to $L_{\mathrm{th,Tmax}} = \sqrt{D_{\mathrm{th}} \cdot \tau_{\mathrm{p}}}$ at the moment when the lattice temperature has reached its maximum at the surface, i.e., typically when surface damage occurs. For times larger than the pulse duration ($t > \tau_{\mathrm{p}}$), the thermal diffusion length approaches the value [Wellershoff, 1999]

$$L_{\mathrm{th}} = \sqrt{\frac{\pi}{2} \cdot D_{\mathrm{th}} \cdot \tau_{\mathrm{p}}} \,. \tag{1.20}$$

As an example, for the irradiation of metals (Au, Ni) with laser pulses of a duration of $\tau_{\mathrm{p}} = 14$ ns, values of $L_{\mathrm{th}} \approx 1$ µm were determined [Matthias, 1994]. However, it is important to note here that the definition of $L_{\mathrm{th}}$ according to Equation (1.20) breaks down if $\tau_{\mathrm{p}} < \tau_{\mathrm{e\text{-}ph}}$, i.e., for ultrashort laser pulses, since then the temperature field of the lattice is usually not established during the laser pulse.

## Surface Melting and Ablation: Scaling of $\phi_{\mathrm{th}}$ with Pulse Duration

Detailed investigations on the (numerical) solution of the complete TTM (Equations (1.15) and (1.16)) have shown that there is a *critical laser pulse duration* $\tau_{\mathrm{cr,1D}}$, above which the melting threshold of the irradiated metal scales with the square root of the pulse duration, i.e., $\phi_{\mathrm{melt}} \propto \sqrt{\tau_{\mathrm{p}}}$ [Corkum, 1988]. This characteristic scaling behavior for $\tau_{\mathrm{p}} > \tau_{\mathrm{cr,1D}}$ is a direct consequence of the 1D heat dissipation into the depth ($z$) during the heating of the sample by the incident laser pulse itself, as it can be seen already from the following simple energy balance analysis:

The optical energy (per unit volume) required to heat the irradiated surface area by the laser to the melting temperature $T_{\mathrm{melt}}$ and to induce the solid-liquid phase transition of the metal can be written as

$$\frac{(1-R) \cdot \phi_{\mathrm{melt}}}{L_{\mathrm{th}}} = C_{\mathrm{l}} \cdot T_{\mathrm{melt}} + \Delta H_{\mathrm{melt}} \tag{1.21}$$

The left-hand side of Equation (1.21) addresses absorbed laser fluences (the $(1-R)$ that is not reflected) and considers the laser-induced heating and melting of a sample layer of $L_{\mathrm{th}}$ in depth by experimentally providing the laser fluence $\phi_{\mathrm{melt}}$ just being sufficient to induce its melting. The right-hand side of Equation (1.21) considers that this sample volume is energetically heated up in the solid state to the melting temperature $T_{\mathrm{melt}}$ and that the volumetric latent heat of melting $\Delta H_{\mathrm{melt}}$ is additionally provided to transform the solid metal into a liquid state. Resolving Equation (1.21) to the melting threshold fluence and substituting $L_{\mathrm{th}} = \sqrt{\frac{\pi}{2} \cdot D_{\mathrm{th}} \cdot \tau_{\mathrm{p}}}$ then reveals the characteristic square-root-scaling of $\phi_{\mathrm{melt}}$with $\tau_{\mathrm{p}}$. Note that a similar analysis can be performed for the ablation threshold fluence ($\phi_{\mathrm{abl}}$) when substituting the enthalpy of melting by the enthalpy of sublimation $\Delta H_{\mathrm{sub}}$, leading to the same functional dependence $\phi_{\mathrm{abl}} \propto \sqrt{\tau_{\mathrm{p}}}$.

For laser pulse durations smaller than the critical pulse duration ($\tau_{\mathrm{p}} < \tau_{\mathrm{cr,1D}}$), however, the melting threshold $\phi_{\mathrm{melt}}$ is almost constant – again the implication of a lack of heat dissipation into the depth of the sample. (This becomes clear since in Equation (1.21) the heated layer thickness $1/L_{th}$ is then substituted by the optically excited layer thickness $1/\alpha_l$, which does not depend on $\tau_p$, fully in line with Equation 1.13).

The value of the critical pulse duration $\tau_{cr,1D}$ depends strongly on the electron-phonon coupling constant $\Gamma_{e\text{-}ph}$ and can be written as [Corkum, 1988]

$$\tau_{cr,1D} = \left(\frac{8}{\pi}\right)^{1/4} \cdot \sqrt{\frac{C_l^3}{C'_e \cdot T_{melt}}} \cdot \frac{1}{\Gamma_{e-ph}} \quad . \tag{1.22}$$

The parameter $C'_e$ is defined by the relationship $C_e = C'_e \cdot T_e$. An experimental determination of $\tau_{cr,1D}$ by measuring the pulse duration dependence of the melting threshold fluence therefore enables the quantification of the material-specific electron-coupling constant $\Gamma_{e\text{-}ph}$.

If the thermal diffusion length $L_{th}(\tau_p)$ exceeds the lateral extent of the laser spot ($L_{th}(\tau_p) > 2w_0$), the assumption of loose focusing is not valid anymore, and a one-dimensional simplification of the two-temperature model is no longer justified. Then, a three-dimensional (3D) heat transport must be considered in the laser irradiated sample material. This happens for long laser pulses ($\tau_p > 10$ µs), and the melting threshold fluence of the metal then increases linearly with the pulse duration ($\phi_{melt} \propto \tau_p$) [Bäuerle, 2011 / Endo, 2023]. Thus, the above given limit rules another critical pulse duration that can be written as

$$\tau_{cr,3D} = \frac{2 \cdot w_0^2}{D_{th}} \quad . \tag{1.23}$$

Figure 1.14 compiles recent measurements of the melting or ablation threshold as a function of the laser pulse duration (reproduced from Endo et al. [Endo, 2023]). This includes different laser wavelength (marked via differently shaped symbols) and single- or multi-pulse irradiation conditions (open vs full symbols). One data set (orange diamond-shaped symbols) includes measurements of the multi-pulse (10,000 pulses per spot) ablation threshold at 1 µm laser wavelength and covers pulse durations varying by eight orders of magnitude, ranging from tens of femtoseconds to microseconds. It is complemented by a fit to a thermal model (for details see [Endo, 2023]). The three different regimes of 0D (no heat diffusion), 1D, and 3D heat transport can be clearly distinguished. The decrease of the ablation threshold for sub-ps pulse duration is very likely related to the multi-pulse irradiation conditions.

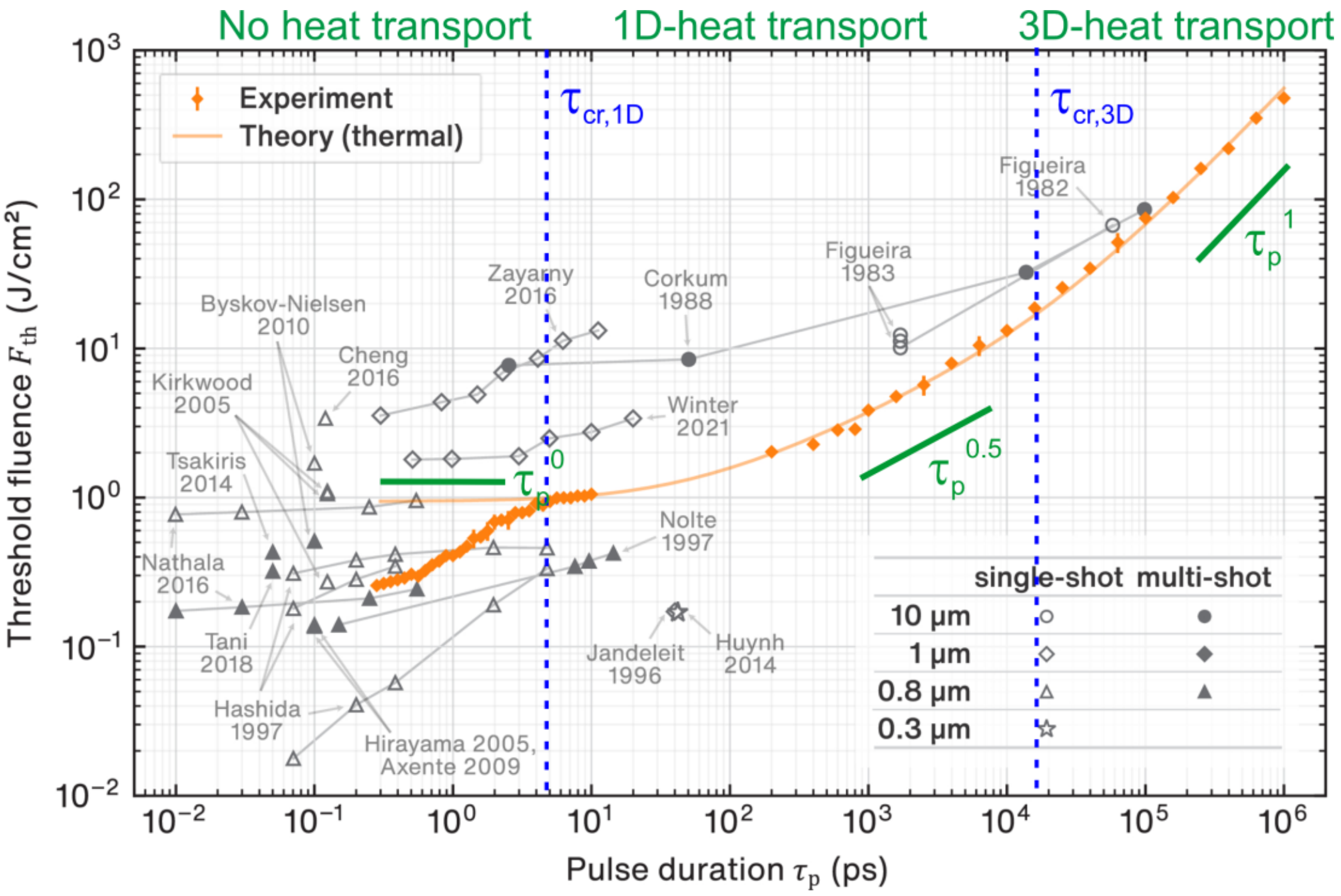

**Fig. 1.14:** Comparison of melting or ablation threshold measurements of copper. Laser wavelengths and single- or multi-pulse experiments are distinguished by open or full data points. The solid orange line relates to a thermal model presented in [Endo, 2023]. Three different $\tau_p$-scaling regimes can be distinguished (see the explanatory schemes above, visualizing 0D, 1D, and 3D heat transport), separated by the critical pulse durations $\tau_{cr,1D/3D}$. The three green solid line segments indicate the characteristic scaling laws (slopes) related to each individual regime. (Reprinted (adapted) with permission from [Endo, 2023] © Optica Publishing Group)

## Melting Dynamics ($\phi < \phi_{abl}$)

It is very instructive to follow the melt dynamics into depth, i.e., to analyze the melting and solidification process initiated by laser radiation at the surface. For crystalline materials irradiated at laser fluences slightly above the melting threshold, the melt-in process occurs interfacially (heterogeneous melting), i.e., a melt front (liquid-solid interface) moves into the depth of the sample, while "consuming" the latent heat of melting from the overheated surrounding. For loose focusing conditions, this process occurs essentially one-dimensionally into the depth ($z$) of the sample material. Simultaneously, heat flow into depth cools both the melt pool and the heated region underneath. If the lattice temperature falls below the melting temperature, the melting terminates. Continuous heat flow in the surrounding further lowers the temperatures of the melt. It then initiates and drives the interfacial re-solidification process, i.e., a solid-liquid interface that now moves again back towards the surface, while releasing the latent heat of melting again at the interface. Depending on the material and the depth gradient created by the specific laser irradiation conditions, the velocity of the re-solidification front may speed up at the end of the process. If a critical velocity $v_{cr}$ of the solid-liquid interface is exceeded, the material may turn into a disordered (amorphous) state if the atoms do not have enough time to solidify in an ordered state of minimal potential energy.

While this melt-dynamics is simple to describe and line out the physical processes involved, mathematically it cannot be implemented in an easy way. Here, the heat flow equation must be numerically solved, while considering the boundary condition of moving interfaces consuming or releasing latent heat. Thus, for details, the reader is referred to the books of Wood et al. [Wood, 1984], van Allmen and Blatter [van Allmen, 1988], and Bäuerle [Bäuerle, 2011].

Figure 1.15 provides an example for irradiation of a crystalline silicon wafer surface by a single ns-laser pulse ($\lambda$ = 532 nm, $\tau_p$ = 18 ns) at fluences ranging between 0.4 and 1.8 J/cm$^2$, reprinted here from Chap. 4 in the book of Wood et al. [Wood, 1984]. The left part (a) of the figure shows the numerically calculated time dependence of the surface temperature, while the right part (b) visualizes the position of liquid-solid interface during the corresponding melt-in and re-solidification cycles. The numerical calculations assumed that undercooling of the melt pool cannot occur here.

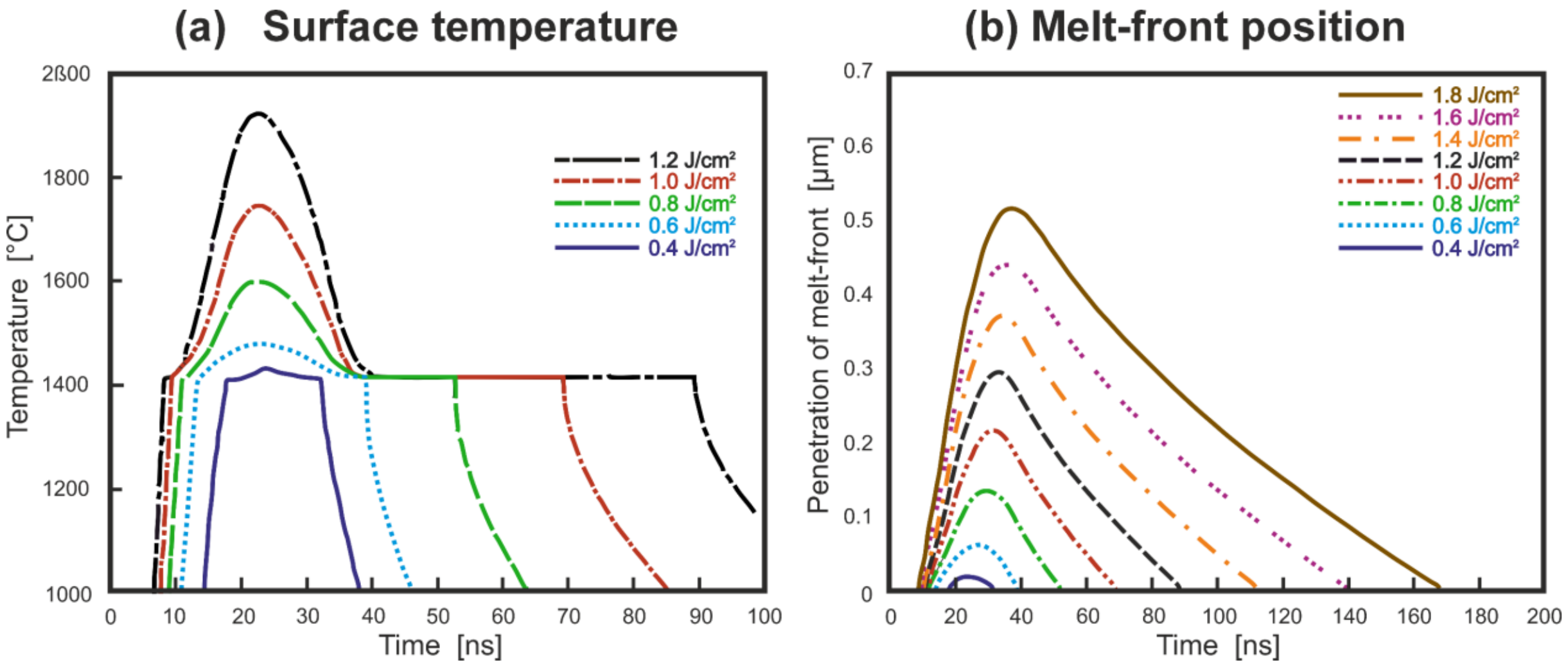

**Fig. 1.15:** Rapid melting and solidification of crystalline silicon upon ns-laser irradiation ($\lambda$ = 532 nm, $\tau_p$ = 18 ns) for eight different laser fluences ranging between 0.4 and 1.8 J/cm$^2$, respectively. **(a)** Numerical computation of the surface temperature as a function of time. (**b**) Numerical computation of the position of the liquid-solid interface as a function of time. Data extracted and redrawn from [Wood, 1984].

The laser-induced surface temperature very rapidly rises until it reaches the melting temperature of silicon at $T_m$ = 1410°C already during or very shortly after the laser pulse (Figure 1.15a). There, it pauses momentarily until the latent heat of melting is absorbed, and then the temperature begins to increase again to some fluence dependent maximum value. Upon cooling, the process is reversed, with the exception that the surface temperature drops quickly to $T_m$, where it remains until the latent heat of melting is completely released again during solidification. Subsequently, the still hot but solidified surface cools down to room temperature again.

The melt-front penetrates very rapidly into the solid, before reaching its maximum extent between ~0.02 and 0.52 µm, depending on the laser fluence, see Figure 1.15b. The melt-in velocity is similar for all melt-front position curves. Near the maximum penetration depth, the melt-front velocity first drops sharply and then changes its sign, before the melt-front recedes back to the surface. The significantly different maximum melt depths can be explained by the varying amounts of the laser pulse energy being absorbed and consumed as latent heat for the solid-liquid phase transition. The set of curves also demonstrates an impact of the released latent heat on the interfacial velocities (being the local slopes to these curves and ranging at several meters per second here) during re-solidification as well as on the overall lifetime of the melt (*melt duration* $\tau_{melt}$), ranging up to 165 ns here. Recent advanced numerical modelling of He and Zhigilei of the melting and solidification of crystalline silicon irradiated by 30-ps-laser pulses at the same wavelength (532 nm) revealed a lower melting threshold fluence along with much smaller melt durations [He, 2024]. Moreover, superficial amorphization may occur for pulse durations shorter than ~90 ps. Particularly for single-crystalline semiconductor samples (wafers), it is straightforward to experimentally monitor this melt and solidification dynamics via time-resolved optical techniques as they show large reflectivity changes upon melting and amorphization. Time-resolved measurements probing the reflectivity of fs-laser irradiated semiconductors have shown that the lifetime of the melt ($\tau_{melt}$) can last up to several tens of nanoseconds tentatively forming an amorphous surface layer [Bonse JAP, 2004 / Bonse APA, 2005].

Such effects are demonstrated in Figure 1.16. It is compiling the results of measurements of the normalized surface reflectivity change $\Delta R/R = (R - R_c)/R_c$ (normalized with the surface reflectivity $R_c$ measured for the pristine single-crystalline wafer material) upon irradiation of (100)-InP with single Ti:sapphire fs-laser pulses (150 fs, 800 nm, normal incidence). The left column (Figure 1.16a) is related to the melt-in dynamics measured by a point-probing *fs-pump-probe setup* (probe beam: 800 nm, angle of incidence $\theta = 12°$, p-polarized) [Bonse APA, 2005]. The upper panel plots $\Delta R/R$ as a function of the pump-probe delay time $\Delta t$, ranging between 0.1 and 1,000 ps. The peak fluence was chosen as $\phi_0 = 1.26 \cdot \phi_{melt}$ below the ablation threshold. At this fluence level, a surface reflectivity change of more than 100% occurs within less than 2 ps. In the following, the reflectivity moderately decreases to a constant level of $\Delta R/R \approx 0.75$ after ≈ 18 ps that is lasting for more than 1 ns.

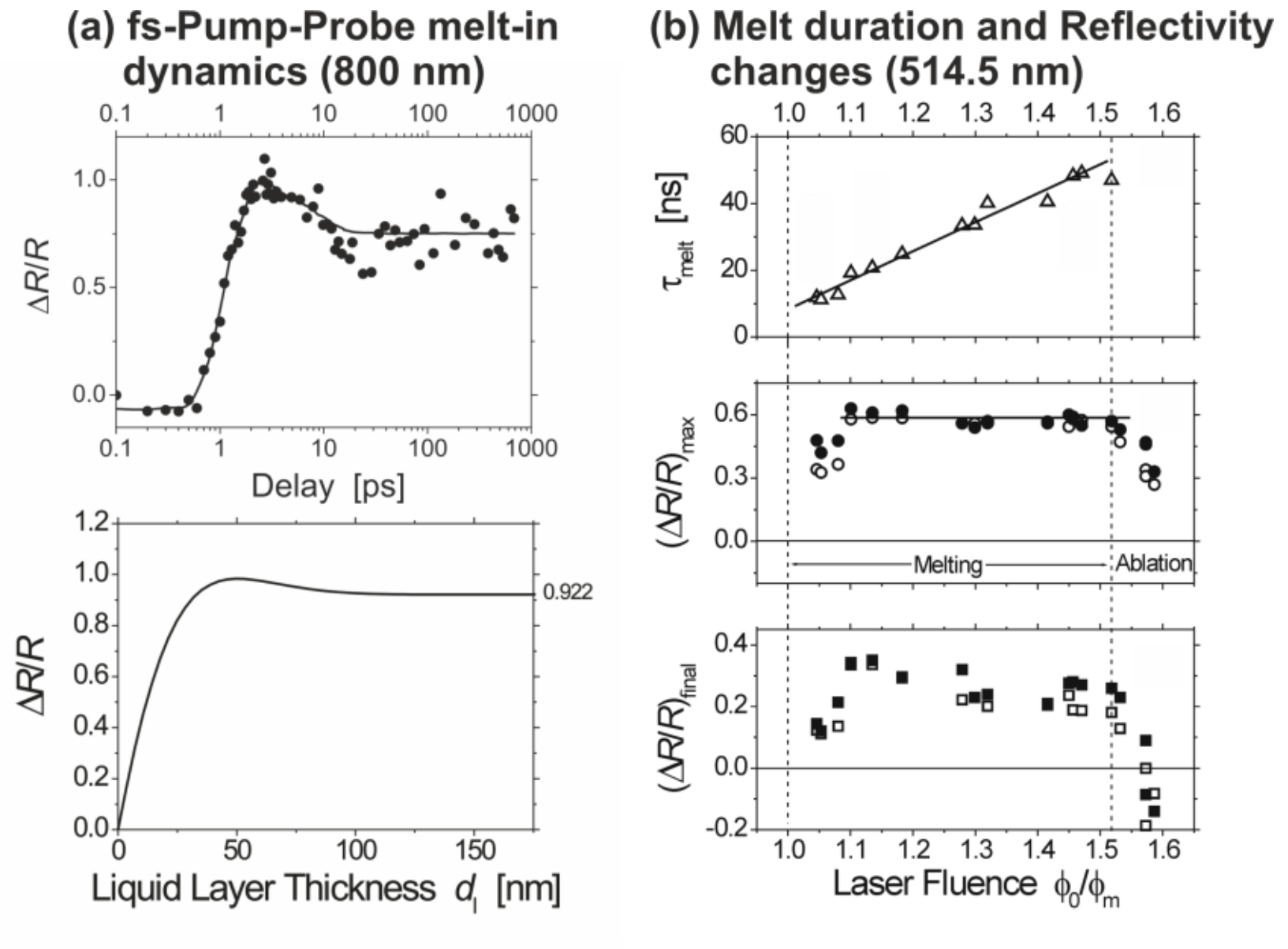

**Fig. 1.16: (a)** Normalized surface reflectivity change $\Delta R/R$ (800 nm) vs. the pump-probe delay time $\Delta t$ (top panel), recorded for a pump laser fluence sufficient for melting ($\phi_0 = 1.26\ \phi_{melt} < \phi_{abl}$) [Bonse APA, 2005]. The solid line guides the eye. In the bottom panel, the corresponding $\Delta R/R$ obtained from thin film optical simulations for a liquid layer (thickness $d_l$) on a crystalline substrate (c-InP) is shown. (Reprinted from [Bonse APA, 2005], J. Bonse et al., Dynamics of phase transitions induced by femtosecond laser pulse irradiation of indium phosphide, Appl. Phys. A **80**, 243 – 248, 2005, Springer Nature); **(b)** Melt duration ($\tau_{melt}$, top panel), transient maximum reflectivity change ($(\Delta R/R)_{max}$, middle panel), and final reflectivity change ((($\Delta R/R)_{final}$),

bottom panel) as a function of the incident laser fluence $\phi_0$ normalized, normalized by the melting threshold fluence $\phi_{melt}$. [Bonse JAP, 2004]. Full symbols represent streak camera measurements, while open symbols refer to photodiode measurements. The melting and ablation thresholds are by vertical lines. (Reprinted from [Bonse JAP, 2004], J. Bonse et al., Dynamics of femtosecond laser-induced melting and amorphization of indium phosphide, J. Appl. Phys. **96**, 2352 – 2358 (2004), with the permission of AIP Publishing)

This behavior can be explained by the formation of a thin liquid layer on the surface via thermal melting that propagates into depth, leading to the formation of an optically thick melt layer after 18 ps. The evolution of the surface reflectivity change in such a scenario has been mathematically modelled with a thin film of molten material with variable thickness $d_l$ that is covering the single-crystalline InP substrate, as seen in the bottom panel [Bonse APA, 2005]. The similarity among the two reflectivity curves allows to relate the delay time (top) with the molten layer thickness (bottom), indicating that the melt-in occurs here at interfacial velocities close to the longitudinal speed of sound in the solid (~5,000 m/s). At higher peak fluences in the ablative regime, InP undergoes nonthermal melting (see Sect. 1.3) within less than 400 fs (data not shown here) [Bonse JAP, 2004].

These fs-pump probe measurements were complemented by so-called *real-time-resolved reflectivity* measurements (RTR) acquiring the surface reflectivity change covering longer times scales, by probing the reflectivity of a continuous wave (cw) $Ar^+$-ion laser beam (probe beam: 514.5 nm, angle of incidence $\theta$ = 18°) in the center of the fs-laser excited spot simultaneously by a streak-camera and a photodiode detection system [Bonse JAP, 2004]. From the reflectivity transients $\Delta R/R(t)$ three characteristic entities were determined, i.e. (i) the maximum transient reflectivity change $(\Delta R/R)_{max}$, (ii) the final reflectivity change $(\Delta R/R)_{final}$ at the end of the temporal detection windows (streak camera: $t$ = 35 ns; photodiode: $t$ = 400 ns), and (iii) the melt duration $\tau_{melt}$. Figure 1.16b (right column) summarizes these three entities as a function of the normalized fluence ratio of the peak fluence $\phi_0$ and the melting threshold fluence $\phi_{melt}$.

The melt duration $\tau_{melt}$ scales linearly with the laser peak fluence in the entire fluence window of melting and ranges between ~10 ns and 50 ns. The linear dependence of the melt duration on the laser fluence is in good agreement with the 1D-thermal heat transport model for loose focusing with strong surface absorption at fluences slightly above $\phi_{melt}$ [Bäuerle, 2011]. The maximum reflectivity change $(\Delta R/R)_{max}$ exhibits a wide plateau at a level of ~0.60, which coincides with the value of optically thick molten InP (indicated by the horizontal solid line). The final reflectivity values $(\Delta R/R)_{final}$ are generally quite similar among the streak camera (full symbols) and photodiode (open symbols) measurements and are very sensitive to the onset of ablation at peak fluences exceeding $\sim 1.5 \cdot \phi_{melt}$. In that plot, the positive permanent relative reflectivity changes between ~0.15 and ~0.35 indicate the formation of an amorphous (a-InP) surface layer with a thickness of a few tens of nanometers. Moreover, the measured reflectivity transients $\Delta R/R(t)$ along with more detailed thin-film optical calculation (both not shown here for brevity) allow to estimate the mean interfacial velocity at the end of the solidification process, ranging between 1.3 and 4 m/s here [Bonse JAP, 2004]. Since amorphization occurs here, that range can be considered as a lower limit of the critical interface velocity $v_{cr}$. Thus, upon fs-laser-induced interfacial (thermal) melting of semiconductors, the typical melt-in velocities are more than three orders of magnitude higher than the ones during re-solidification.

These results for InP indicate that ultrashort laser pulses have the potential to create site-selectively very flat and microscopically smooth amorphous surface layers of several tens of nanometer thickness atop single-crystalline wafer material. This was further explored upon irradiation of single-crystalline silicon wafers with single 30-fs Ti:sapphire laser pulses and subsequent all-optical surface characterization by spectroscopic imaging ellipsometry (SIE) and complementary transmission electron microscopy (TEM) for (destructive) verification [Florian, 2021].

Figure 1.17 provides a thickness map of the laser-induced amorphous (a-Si) surface layer created on a (111)-Si wafer, along with the corresponding map of the covering oxide layer after irradiation at a single ultrashort laser pulse at a peak fluence $\phi_0 = 0.24$ J/cm$^2$ below the ablation threshold. The maps were obtained from the SIE data through point-by-point fitting of the ellipsometric entities $\Psi$ and $\Delta$ to a thin-film optical model represented by the stratigraphy c-Si / a-Si ($d_{am}$) / $SiO_2$ ($d_{ox}$) [Florian, 2021] and smooth interfaces. The a-Si surface layer (Figure 1.17a) exhibits an almost perfect parabolic thickness profile with a maximum value around 60 nm. In contrast, the covering oxide layer (Figure 1.17b) exhibits a rather constant thickness around 3 - 4 nm in the amorphized region and a native oxide layer thickness around 2 nm.

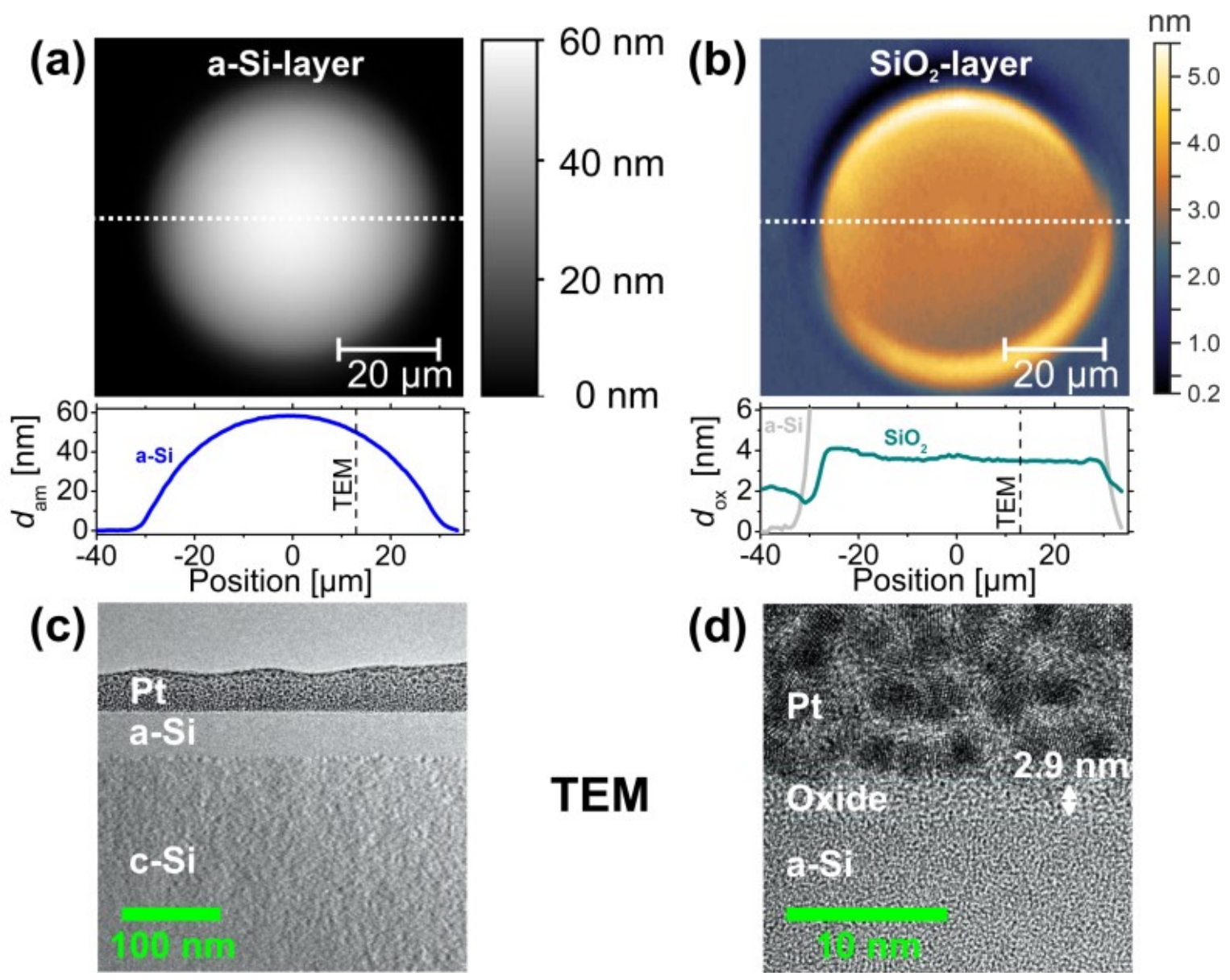

**Fig. 1.17:** Maps of the amorphous silicon layer thickness ($d_{am}$, **(a)**) and the corresponding covering silicon oxide layer thickness ($d_{ox}$, **(b)**), as determined by spectroscopic imaging ellipsometry (SIE) on a 30-fs-laser-irradiated spot on (111)-Si at a peak fluence of $\phi_0 = 0.24$ J/cm$^2$. Note the different thickness scales. **(c)** and **(d)** are corresponding brightfield transmission electron microscopy (TEM) images taken at the position marked by the vertical dashed lines in (a,c). Note the different magnifications. (Reprinted (adapted) from [Florian, 2021], C. Florian et al., Single femtosecond laser-pulse-induced superficial amorphization and re-crystallization of silicon, Materials (Basel, Switzerland) **14**, 1651 (2021), Copyright 2021 under Creative Commons BY 4.0 license. Retrieved from https://doi.org/10.3390/ma14071651 )

These results obtained from an all-optical and non-destructive measurement were subsequently confirmed by preparing a cross-sectional Pt-covered lamella by focused ion beam etching (FIB) and inspecting it by TEM at the location indicated by vertical dashed lines in the cross-sectional thickness profiles (Figure 1.17c and 1.17d). Reasonably good agreement was found between the SIE and TEM measurements.

## Spallative Ablation ($\phi > \phi_{abl}$)

At laser fluences exceeding the ablation threshold, the processing with ultrashort pulsed laser radiation exhibits a peculiarity. For strong absorbing materials, the single-pulse ablation craters usually follow the crater profiles predicted by the solution of Equation (1.6) only in the center of the ablation spots. At its edges, typically steep crater walls are observed [von der Linde, 1997 / Rethfeld SPIE, 2002 / Bonse JAP, 2009].

This effect was explained first by Sokolowski-Tinten, von der Linde et al. through *fs-time resolved microscopy* (fs-TRM) experiments studying the ablation of metals and semiconductors [von der Linde, 1997 /von der Linde, 2000]. In these optical pump-probe measurements, the authors revealed the occurrence of transient Newton rings caused by a semi-transparent layer ejected in the ablated region. This ring pattern emerges after a few tens of picoseconds and lasts for a few nanoseconds (for an example see Chap. 13 (Bonse)). As an explanation, the authors proposed the following hydrodynamic (spallative) ablation mechanism at laser fluences close to the ablation threshold that is visualized in the left part of the Figure 1.18 [Inogamov, 1999 / von der Linde, 2000] and summarized in the following.

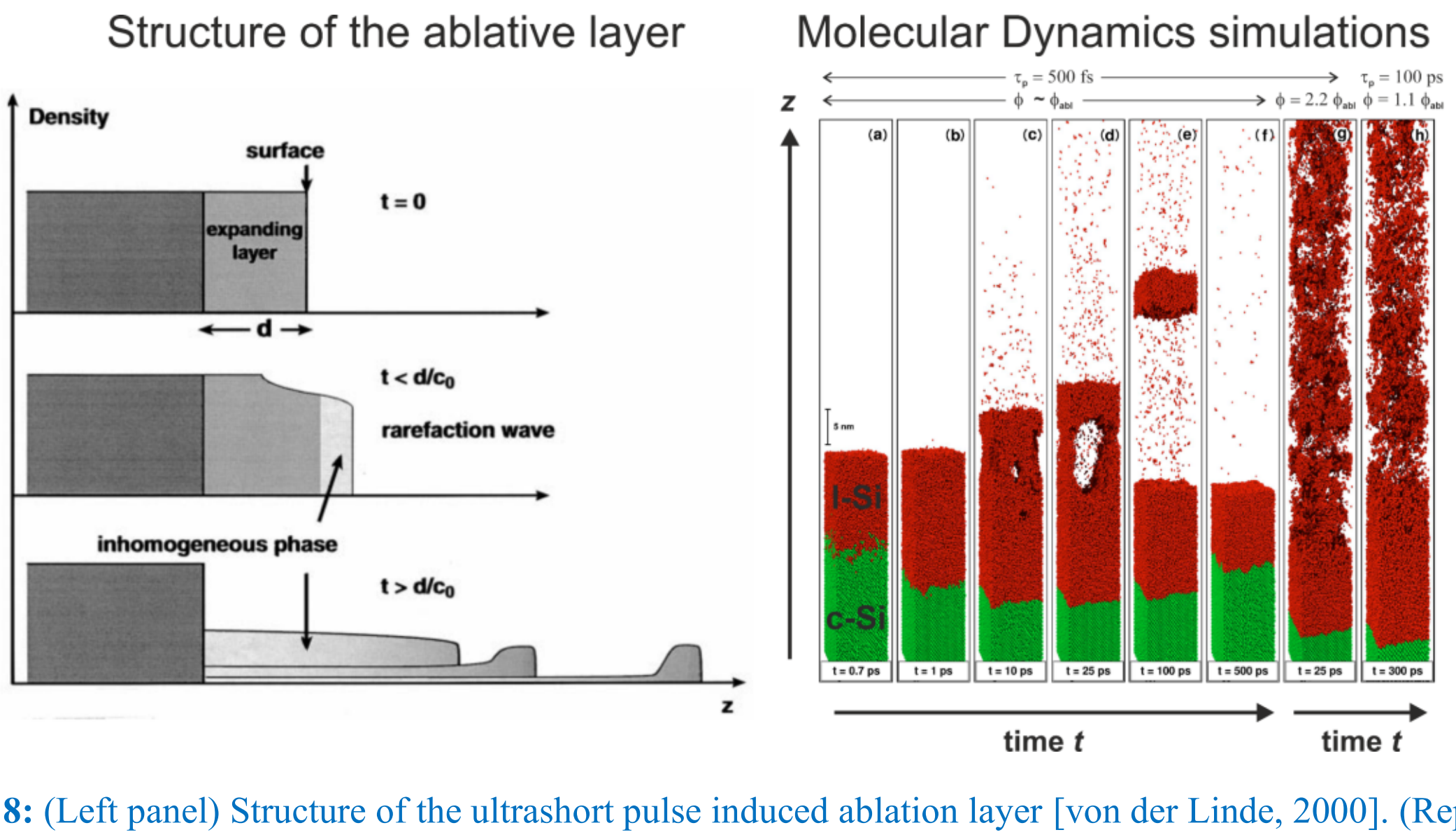

**Fig. 1.18:** (Left panel) Structure of the ultrashort pulse induced ablation layer [von der Linde, 2000]. (Reprinted from [von der Linde, 2000], Appl. Surf. Sci., **154–155**, D. von der Linde et al., The physical mechanisms of short-pulse laser ablation, 1 – 10, Copyright (2000), with permission from Elsevier); (Right panel) Molecular Dynamics simulations of the ultrashort pulse ablation of silicon by a single laser pulse [Lorazo, 2006]. The frames visualize different delay times. Molten material is colored in red, while crystalline material is displayed in green color. (Reprinted figure with permission from [Lorazo, 2006], P. Lorazo et al., Phys. Rev. B **73**, 134108, 2006. Copyright (2006) by the American Physical Society)

The figure represents a qualitative scheme that shows the mass density profile $\rho$ along the $z$-direction at three different times $t$ after the impact of the laser pulse. The top panel sketches the situation just after the energy relaxation ($t \approx 0$), i.e., before any significant material expansion has occurred, i.e., the material is still at the mass density of the solid ($\rho_{solid}$). The two differently shaded areas indicate a laser-excited surface layer (thickness $d_{abl}$) in above-ablation-threshold condition (medium gray) that is separated from sub-threshold bulk material, which will not undergo ablation (dark gray).

In the center panel (for times $t < d_{abl}/c_{sound}$), the contour line indicates the density profile of a so-called *rarefaction wave* (RW). The inner head of the RW is propagating into the solid at the speed of sound $c_{sound}$. Starting from a moderately reduced mass density of the expanded tail of surface layer, there is still a sharp drop of the density of the RW towards vacuum (front interface) that is covering a region of thermodynamic two-phase regime (shaded in light gray), where the liquid phase and the gas phase of the material both coexist.

When the inner head of the RW reaches the rigid interface to the solid at $t_0 = d_{abl}/c_{sound}$ (where the flow velocity must vanish), it is back-reflected towards the sample surface, resulting in a drop of the mass density at the boundary and creating a density discontinuity on its backside. For a typical value of the sound velocity $c_{sound}$ of 2,500 m/s and an ablation layer thickness of $d_{abl}$ = 100 nm, the time for sub-surface material separation through this discontinuity is estimated to be $t_0$ ~40 ps [von der Linde, 2000].

The situation following at times $t > d_{abl}/c_{sound}$ is sketched in the bottom panel of the scheme. Now, a layer of material expands that is confined by two sharp interfaces, i.e., the inner interface to the solid bulk material and the outer interface moving away from the sample at the maximum liquid flow velocity $v_{max,l} = c_{sound} \cdot \ln(\rho_l/\rho_{solid})$, with $\rho_l$ being the liquid density at the binodal in the phase diagram. Typical values of $v_{max,l}$ range between ~100 and ~1,500 m/s and depend on the applied laser fluence [Inogamov, 1999 / von der Linde, 2000 / Bonse PRB, 2006]. The material confined between the interfaces consists mostly of a low average density inhomogeneous two-phase mixture. Inside this expanding layer, the mass density further drops with elapsing time, while in the vicinity of the out-moving ablation front, the density remains rather constant at $\rho = \rho_l$. Hence, as $t$ further elapses, the expanding ablation layer keeps its structure of a thin outer shell of approximately liquid density surrounding a low-density gas inside.

In other words, through the rarefaction wave, a several tens of nanometer thick layer of material spalls from the solid after some tens of picoseconds and develops within a few nanoseconds into a liquid-confined gas bubble. After several nanoseconds, the gas-confining shell loses its sharp interfaces [Bonse PRB, 2006] and disintegrates into smaller fragments and nanoparticles. Similarly, for high laser fluences the sharp shell interfaces may smoothen at too high thermodynamic entropies [Rethfeld SPIE, 2002].

The above qualitatively sketched scenario was later confirmed and impressively visualized by *Molecular Dynamics* simulations of the ultrafast laser matter interaction processes [Zhigilei, 2004], as demonstrated in the right part of Figure 1.18 [Lorazo, 2006]. Here, a silicon crystal surface irradiated by a single ultrashort laser pulse ($\tau_p$ = 500 fs or 100 ps, 266 nm laser wavelength) is modelled using an empirical Stillinger-Weber potential for the microscopic interaction of the silicon atoms. Green-colored atoms belong to the crystalline (semiconducting) solid and red-colored atoms represent silicon in a molten (metallic) state. Sub-figures a - f show snapshots of the material expansion at different delay times of 0.7 ps (a), 1.0 ps (b), 10 ps (c), 25 ps (d), 100 ps (e), and 500 ps (f), respectively. Here, the silicon is irradiated at an average fluence of 0.225 J/cm$^2$, very close to the ablation threshold of the material. A superficial melt layer with a liquid-solid interface moving into the crystal at early times (0.7 ps (a) and 1.0 ps (b)) and moving back at later times ($t$ > 100 ps, (e) – (f)) is evident. The expanding liquid layer forms a sub-surface void emerging for times $t$ > 10 ps, eventually ejecting a thin layer of 5 to 10 nm thickness here (c - f).

For a laser pulse of the same duration but at a more than doubled average fluence of 0.50 J/cm$^2$ (g), however, the situation of the expanding matter shifts from a subcritical liquid to a supercritical fluid. After 25 ps, the removal of molten silicon does not occur via the growth of a localized sub-surface cavity and the subsequent ejection of a liquid shell. Instead, the expanding material has a cluster-like, diffuse structure with a reduced average mass density and without a well-defined ablation front. Moreover, the ejected molten material expands significantly faster.

When irradiating silicon at a significantly longer pulse duration of $\tau_p$ = 100 ps at an average laser fluence of 0.45 J/cm$^2$ close to the ablation threshold, after 300 ps delay the ablating fluid does not exhibit a bubble-like character but exhibits again a rather diffuse, cluster-like structure (h).

## Phase-Explosion

The spallative mechanism of ablation, when a surface layer is swiftly molten by ultrashort laser pulses and ejected from the material target, clearly manifests itself in a relatively narrow range of laser fluences straight above the ablation threshold [Zhigilei, 2009 / Shugaev, 2020]. With increasing laser fluence, the mechanism connected to the disintegration of the liquid phase into a mixture of vapor and liquid droplets becomes dominant [Zhigilei, 2009]. This mechanism, known in thermodynamics as explosive boiling, is usually referred to as *phase explosion* in the fields of laser-matter interaction and the electrical explosion of metallic wires [Skripov, 1974 / Martynyuk, 1977 / Miotello & Kelly, 1999]. The physical origin of phase explosion lies in the overheating of liquid material in its volume (a thin surface layer in the case of laser action) toward its metastable state, where a homogeneous nucleation of vapor phase develops [Skripov, 1974 / Martynyuk, 1977 / Miotello & Kelly, 1999 / Bulgakova & Bulgakov, 2001]. A typical thermodynamic *p-T* diagram is given in Figure 1.19 (left panel), where the ways of liquid superheating metastable state are indicated. The metastable region M lies between the binodal (line 1) and the spinodal (line 2), which join in a region called the thermodynamic critical point (CP). We recall that the binodal is a line corresponding to a sequence of equilibrium states between a liquid surface and an ambient gas just above the surface at certain pressures and temperatures, while the spinodal determines the limit of liquid superheating. In other words, a liquid immediately decays into atoms or molecules upon reaching a state in the vicinity of the spinodal. A metastable state (e.g., point *d*, left panel in Figure 1.19) can be reached in two obvious ways, isothermal pressure reduction (*ad*) and isobaric heating (*cd*). Note that, due to the low compressibility of the liquid, the line of adiabatic expansion *bd* is close to *ad* (the effect of adiabatic cooling is small) [Bulgakova & Bourakov, 2002]. Applied to short and ultrashort pulse laser heating, *cd* corresponds to the case when the ablation threshold is not reached. However, reaching the metastable states from which the phase explosion can be realized upon dynamic laser heating is more complicated and strongly depends on different parameters of material and irradiation, among which the pulse duration plays the most important role.

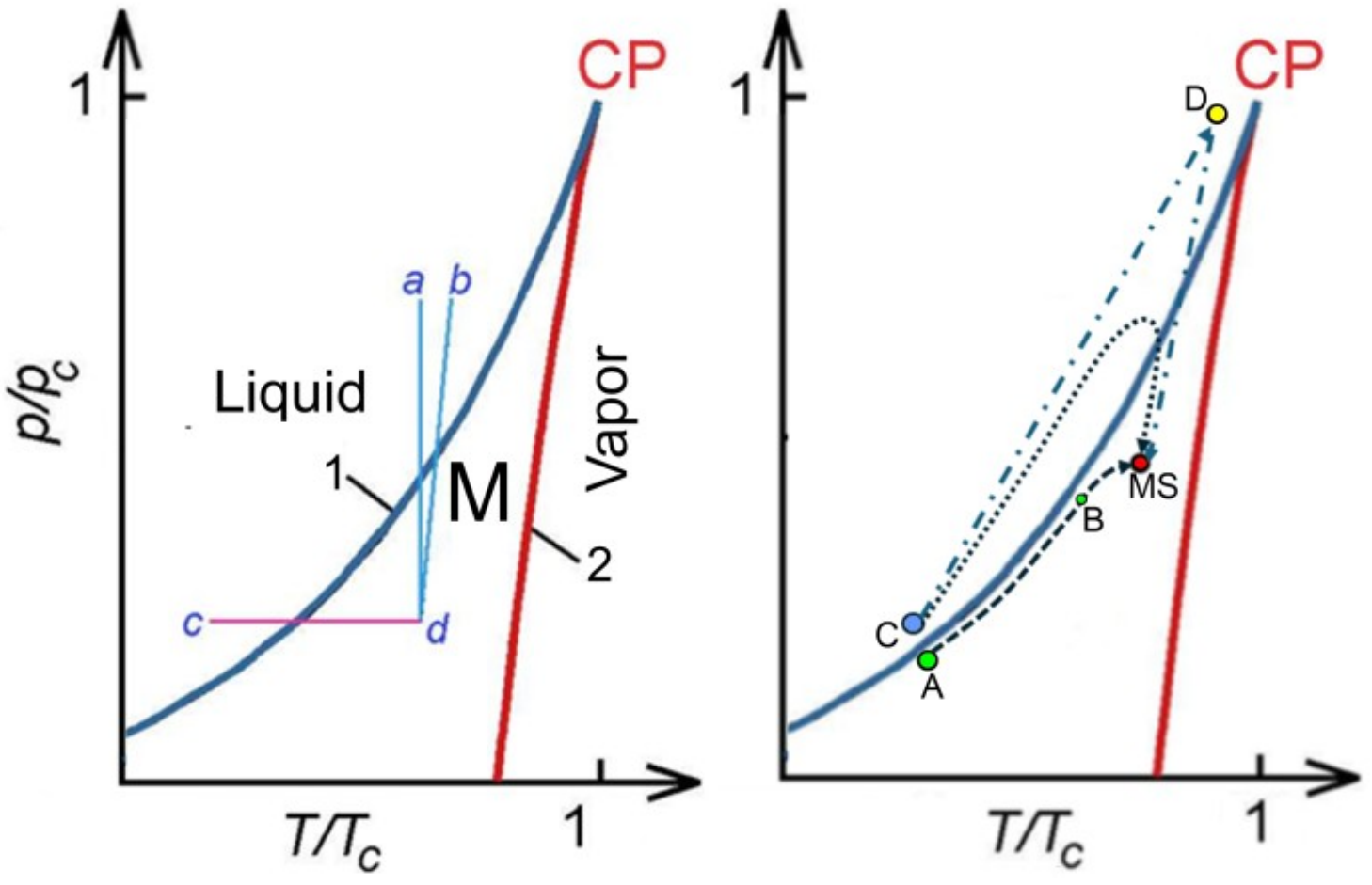

**Fig. 1.19:** (Left panel) Schematics of the thermodynamic *p-T* diagram. The lines *ad*, *bd*, and *cd* indicate ways to reach a metastable state *d* in the region M between binodal 1 and spinodal 2. CP denotes the critical point. (Right panel) Schematics illustrating the material evolution upon swift laser heating with reaching a metastable state MS culminating with phase explosion. A-B-MS and C-D-MS correspond to ns and very short (between fs and a few ps) laser pulses, respectively. For intermediate pulse durations, the material evolves upon laser heating along a line between the two extreme cases as shown exemplary by the dotted curve. Line A-B corresponds to the normal vaporization regime (without phase explosion)

The phase explosion mechanism has specific features for ultrashort laser pulses as compared to longer ones, which is conditioned by the generation of high laser-induced stresses. For nanosecond laser pulse irradiation in the ablation regime of "normal vaporization" (evaporation from the surface) [Miotello & Kelly, 1999 / Bulgakova & Bulgakov, 2001], heating of the material surface layer follows the binodal line (the route A-B in Figure 1.18, right panel), however with some minor superheating otherwise vaporization would be balanced by the backflow of atoms or molecules recondensing on the surface from the vapor phase. When the laser fluence exceeds a certain value specific for materials and irradiation conditions, an abrupt increase in the ablation rate with the appearance of liquid droplets in the ablation products happens [Yoo, 2000 / Bulgakova & Bulgakov, 2021] (line A-B-MS in Figure 1.19).

By contrast, in the case of ultrashort laser pulses, a stress confinement condition is the governing factor of material spallation as shown by Leveugle et al. [Leveugle, 2004]. The stress confinement regime is realized for pulse durations $\tau_{\mathrm{p}}$ at which laser-irradiated material is heated nearly isochorically, thus occurring under high compressive stress [Leveugle, 2004] before being able to expand for releasing stress (shown schematically by line CD in Figure 1.19, right panel). Upon the following expansion, the material swiftly enters a state within the metastable region (D-MS). The condition for stress confinement can be expressed as $\tau_{\mathrm{p}} \leq \tau_{\mathrm{s}} \sim L/c_{\mathrm{sound}}$, where $\tau_{\mathrm{s}}$ is a characteristic time of stress release, which is evaluated as the ratio of the laser-heated layer thickness $L$ to the speed of sound of material $c_{\mathrm{sound}}$ [Leveugle, 2004]. Taking into account a thickness of the laser molten layer of several dozens of nanometers, one can evaluate the characteristic time of existence of the stress confinement regime (or manifestation of photomechanical effects) for metals to be of the order of 10 ps. Indeed, MD simulations [Zhigilei, 2009] demonstrated a massive phase explosion in the absence of spallation for 50-ps laser pulses. It should be noted that the contribution of "normal vaporization" to material ablation at such laser pulses is still negligibly small, keeping phase explosion as the dominant ablation mechanism at high fluences.

The transition between stress confinement and its absence cannot be expected to be abrupt. For intermediate pulse durations, the material starts to gradually expand at the very surface layer already during the laser pulse, so the stress somewhat releases in dynamics. Thus, it can be foreseen that the material evolves upon laser heating along a line between the two extreme cases, with and without stress confinement, as shown exemplary by the dotted curve in Figure 1.19, right panel.

According to classical thermodynamics [Anisimov, 1974], when a liquid enters a metastable state, its parameters (density, entropy, temperature, and pressure) undergo fluctuations. The matter acquires a fine-grained structure which, in a simplified form, may be called a 'gas of droplets' whose characteristic size $r_c$ (or a correlation length) increases when approaching the CP. However, upon matter heating with ultrashort laser pulses, the time is insufficient for developing fluctuations and, in the initial stages of expansion, the highly heated/stressed liquid layer expands homogeneously. Upon expansion, density fluctuations are initiated and grow, leading to a gradual decomposition of expanding matter into gas- and liquid-phase regions [Zhigilei, 2003]. Growth of the density fluctuations leads to the formation of a foamy transient structure. A snapshot of such a part of the expanding ablation plume is shown in Figure 1.20, left panel [Zhigilei, 2003]. In the right panel, the evolution of the ablation plume ejected in the phase explosion regime is presented [Wu & Zhigilei, 2014].

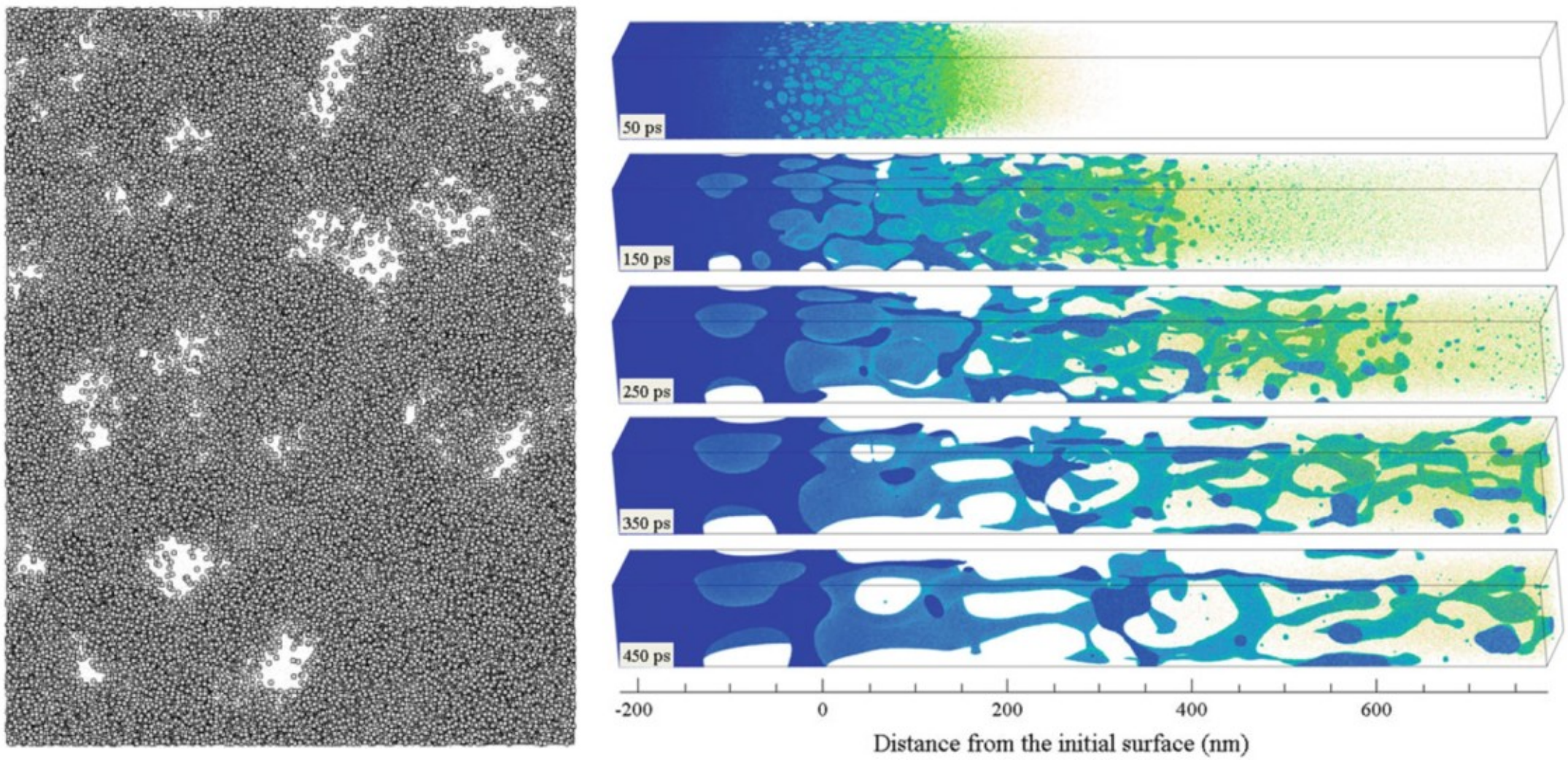

**Fig. 1.20:** (Left panel) A snapshot of the foamy transient structure of interconnected liquid clusters and individual molecules formed in the process of phase explosion of overheated material [Zhigilei, 2003]. (Reprinted from [Zhigilei, 2003], L.V. Zhigilei et al., Dynamics of the plume formation and parameters of the ejected clusters in short-pulse laser ablation, Appl. Phys. A **76**, 339 – 350, 2003, Springer Nature); (Right panel) Snapshots of phase explosion upon irradiation of a bulk Al target by 100-fs laser pulse at the absorbed fluence of 0.2 J/cm$^2$ as predicted by MD simulations [Wu & Zhigilei, 2014]. (Reprinted from [Wu & Zhigilei, 2014], C. Wu et al., Microscopic mechanisms of laser spallation and ablation of metal targets from large-scale molecular dynamics simulations, Appl. Phys. A **114**, 11 – 32, 2014, Springer Nature)

From Figure 1.20, right panel, it is seen that the front part of the ablation plume represents a mixture of vapor phase and fine clusters, while deeper layers of the expanding material are decomposing into larger particles [Wu & Zhigilei, 2014]. It is natural due to the inhomogeneous heating of the absorbing layer toward the target depth. In the deeply extending laser-affected layers, the pores/voids can be observed after target solidification [Savolainen, 2011 / Ashitkov, 2012], which is one of the features of the phase explosion mechanism. Furthermore, across the laser irradiation spot, the phase explosion regime can be realized in the middle part of the spot while, in the spot periphery, the spallation mechanism can be dominating as shown by Wu and Zhigilei [Wu & Zhigilei, 2014]. It should be underlined that the characteristics of ablation of different target materials by ultrashort laser pulses have substantial similarities in both, MD simulations and experimental observations that point to generally the same nature of this complicated phenomenon governed by thermodynamic and photomechanical effects.

## Multi-Pulse Effects ($N > 1$)

Apart from energy relaxation that occurs during each individual laser pulse that is hitting the workpiece, multi-pulse feedback effects must also be discussed. Often the assumption that all pulses are leading to the same ablation effects is not fulfilled, or only becomes valid after a certain number of "conditioning pulses" driving the system towards steady state conditions at the surface.

This becomes obvious when analyzing the volumetric energy density that is deposited at the surface and that must exceed specific threshold values to melt or ablate the material (see Equation (1.21)). In the case of linear optical absorption of an ultrashort laser pulse (e.g. in a metal), the absorbed fluence deposited in the optical penetration depth is given by $(1\text{-}R)\phi_0/\alpha_1$. However, from the effects discussed above it is already clear that the surface reflectivity $R$ will change with the laser ablated surface topography and that the absorption coefficient $\alpha_1$ may be modified if the material changes its structural state, e.g., by superficial amorphization.

Moreover, the surface absorption may be driven pulse-by-pulse into absorptive resonance, e.g., via the excitation of specific surface waves, such as *Surface Plasmon Polariton*s (SPPs), enabled by increasing surface roughness or an increasing local angle of incidence ($\theta$), provided that the dielectric permittivity $\varepsilon^*$ of the laser excited material supports SPP excitation ($\mathrm{Re}(\varepsilon^*) < -1$ [Raether book, 1988]). Even if SPPs are not excited, upon drilling a deep hole into the sample surface, the local Fresnel-reflectivity may change the energy deposition across the spatial laser beam profile at the gradually steeper crater walls [Nolte APA, 1999].

Additionally, the intrinsic ablation threshold fluence $\phi_{\mathrm{th,intr}}$ can be affected by structural and chemical material changes or mechanical stress/strain. Thus, all three physical parameters become dependent on the number of laser pulses $N$ applied to the same spot and the laser processing strategy ($R \to R(N)$ and $\alpha_1 \to \alpha_{\mathrm{eff}}(N)$, $\phi_{\mathrm{th,intr}} \to \phi_{\mathrm{th}}(N)$, see Equation (1.14)).

Figure 1.21 sketches such multi-pulse effects that can manifest during the laser processing of surfaces by ultrashort laser pulses into (i) changes of the surface reflectivity (affecting the optical energy deposited into the processed material), (ii) changes of the surface absorption (affecting the energy deposition depth), and (iii) alterations of the material's intrinsic threshold fluences. Depending on the "lifetime" of the transient states involved or on the time-scales of material modification and relaxation processes — compared to the time between successive laser pulses — these effects may even affect the interaction with the next laser pulse and, thus, create a *feedback* for the subsequent (repetitive) pulsed irradiations.

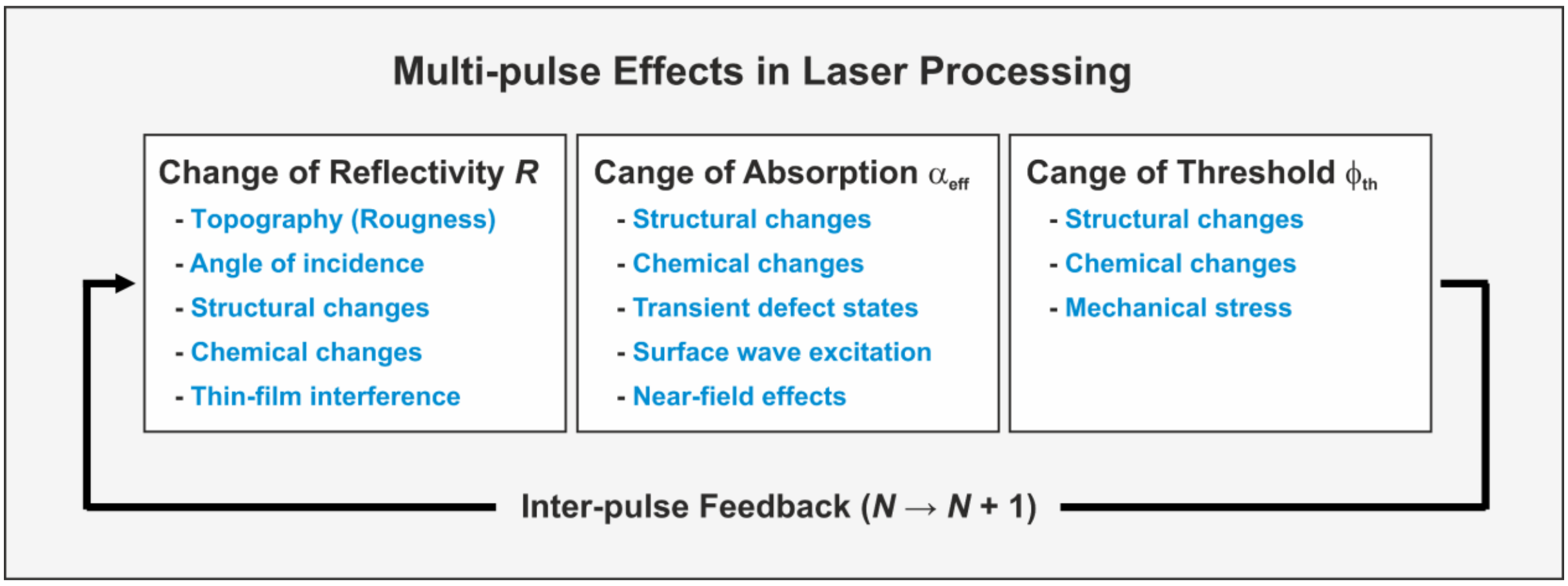

**Fig. 1.21:** Scheme of multi-pulse effects and inter-pulse feedback affecting during laser processing the optical surface reflectivity ($R$), the effective optical absorption ($\alpha_{eff}$) and the intrinsic threshold ($\phi_{th}$) of the irradiated material

## 2.3 Selected Applications

Around the turn of the millennium the first industrial applications taking benefit from the increased laser machining precision addressed the segment of high-priced goods and services. They improved the repair of photolithographic projection masks in semiconductor industry [Haight, 1998 / Haight, 1999 / Lieberman, 1999, Nolte PB, 1999], the cutting of filigree coronary stents in medicine [Momma, 1999 / Korte, 1999], and the deep drilling of fuel injection nozzles in car industry [Korte, 1999]. Since Part III of this book will provide an extensive survey on current applications enabled by ultrashort laser processing, in this sub-section, we will briefly commemorate only a few, mostly early applications of ultrashort laser pulse processing.

### Cutting of Coronary Stents

Coronary stents are specially designed filigree tubular structures with a diameter of approximately 1 – 2 mm, which are inserted in the event of a narrowing of blood vessels (e.g., coronary arteries blocked by plaque) with a special catheter into the affected vessel and then widened by a balloon. Self-fixed in the dilated state, the stent then keeps the blood vessel open.

This procedure is named *angioplasty* and became an established alternative method to bypass operations. With the help of fs-laser processing new materials (replacing the standard stainless steel) have been explored for the fabrication of coronary stents, taking benefit of the increased micromachining precision of ultrashort pulsed laser radiation and its capability to process almost any material. Figure 1.22 exemplifies two different fs-laser processed coronary stents that were manufactured in the metal tantalum (a) or even in a bioresorbable polymer (b), poly-(L-lactide) [Korte, 1999].

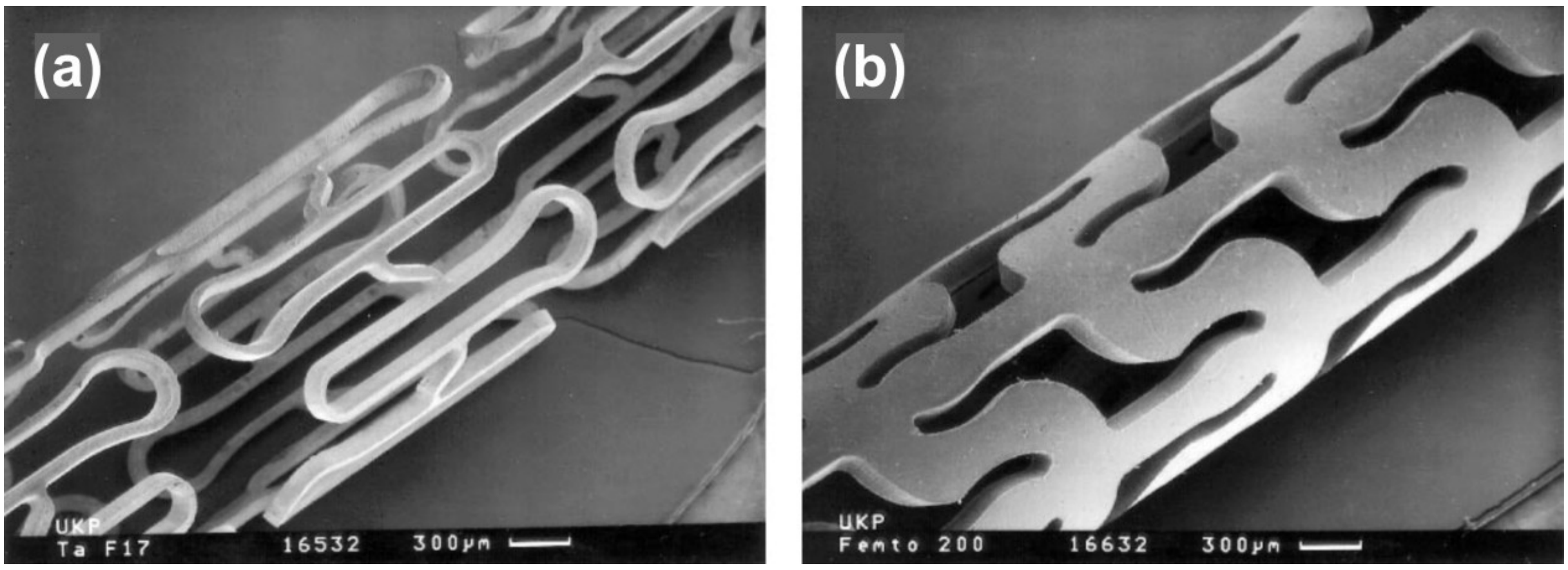

**Fig. 1.22:** Scanning electron micrographs showing two examples for the production of medical implants (stents) made of the metal tantalum **(a)** and of the biologic resorbable polymer Poly-(L-Lactide) **(b)** [Korte, 1999]. (Reprinted from [Korte, 1999], F. Korte et al., Far-field and near-field material processing with femtosecond laser pulses, Appl. Phys. A **69**, S7 – S11 [Suppl.], 1999, Springer Nature)

## Drilling of Injection Nozzles

Another eminent microprocessing application is the drilling of fuel injection nozzles for combustion engines. Improving the geometry and precision of the manufacturing improves the combustion process in the engine, its efficiency and, thus, can save $CO_2$. It was realized early that the requirements — a burr-free bore hole of less than 250 µm in diameter with a high roundness, drilled with a high reproducibility through a metallic material of ~1 mm thickness — can be fulfilled by the ultrashort pulsed laser technology [Korte, 1999]. The contactless laser processing technology even provided the potential of reducing the hole diameters and, thus, was successfully explored together with the German car industry. Figure 1.23a shows the cap of an injection nozzle having two fs-laser drilled bore holes. In Figure 1.23b, a polymer replica of a typical bore hole is presented, demonstrating its high micromachining quality manifesting through very smooth walls of the hole.

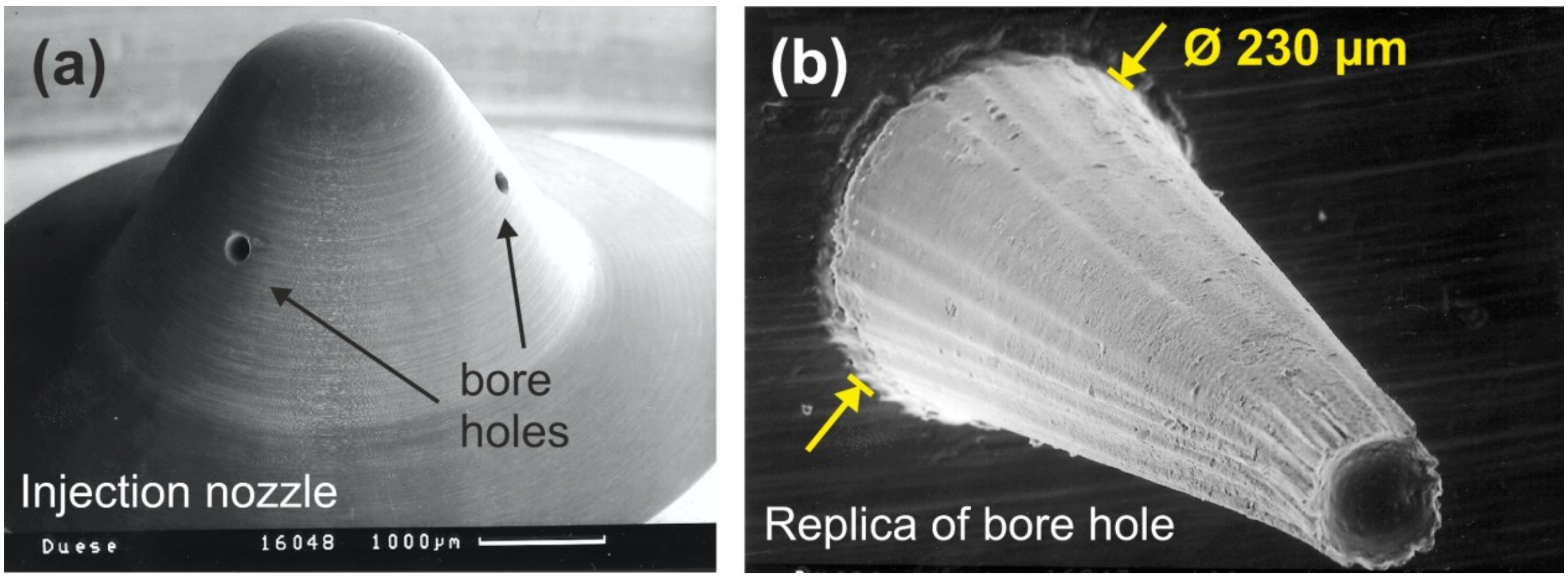

**Fig. 1.23: (a)** Scanning electron micrographs showing the cap of an injection nozzle (steel) with two fs-laser drilled bore holes [Korte, 1999]. **(b)** The micrograph provides a tilted view on a polymer replica of a typical bore hole in the injection nozzle. (Figure (a) adapted from [Korte, 1999], F. Korte et al., Far-field and near-field material processing with femtosecond laser pulses, Appl. Phys. A **69**, S7 – S11 [Suppl.], 1999, Springer Nature)

## Repair of Photolithography Masks

Given the ever-increasing demand of highly integrated semiconductor circuits (chips), suitable tools for photolithographic mask repair are required. Such photolithographic masks typically consist of ~100 nm thick chromium films that are deposited on quartz substrates. Manufacturing-related defects on the photolithographic masks often manifest in the form of excess chromium. A transfer of the pattern to the wafer must be avoided as it would render the integrated circuit inoperative. Thus, a precise defect removal technology is highly preferable and even high related costs can be afforded.

Around 1998, two different concepts for fs-laser-based optical mask-repair systems were proposed and realized, employing either a far-field [Haight, 1998] or a near-field [Lieberman, 1999 / Nolte OL, 1999 / Nolte PB, 1999] based structuring technique. Both methods take benefit from the very different ablation threshold fluences of the Cr-films and the underlying $SiO_2$-substrate ($\phi_{abl}$ (Cr) < $\phi_{abl}$ ($SiO_2$)).

The *far-field structuring* technique was realized by the company IBM, projecting a movable rectangular aperture placed in the beam of a fs-laser oscillator directly to the photomask for ablation purposes [Haight, 1998 / Haight, 1999]. The first generation of mask repair system (MARS I) was installed in April 1998 in IBM's Burlington Mask House for routine manufacturing operation [Haight, 1999], a second-generation system in 2001, which allowed the trimming of mask features to a rms-precision of ~ 5 nm [Wagner, 2002].

The *near-field structuring* technique was realized by Liebermann et al. (Nanonics Lithography Ltd. and Laser Zentrum Hannover e.V.) by injecting the third harmonic generated pulses of a Ti:sapphire femtosecond laser into a light guiding (hollow) chromium-coated micropipette fiber made of fused silica, see Figure 1.24a. The fiber with an aperture diameter of ~400 nm at its end was then used in the scanning near-field mode [Lieberman, 1999 / Nolte OL, 1999 / Nolte PB, 1999]. Laterally protruding excess chrome defects with sizes of 3 µm were carved-off with an accuracy of approximately 50 nm from the chromium layer, see Figure 1.24b.

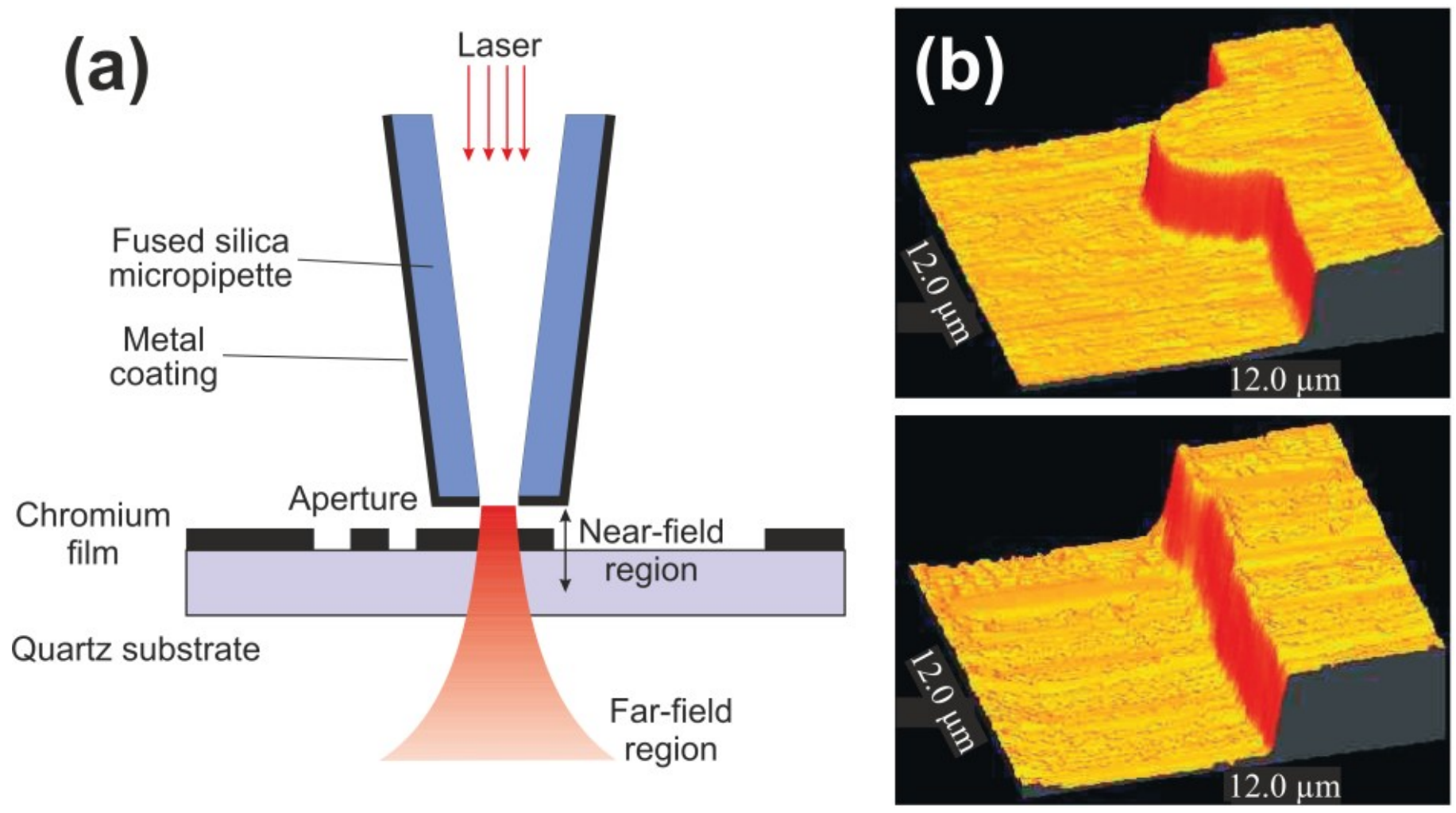

**Fig. 1.24: (a)** Scheme of a near-field optical microscope (SNOM) in illumination mode [Nolte OL, 1999]. Laser light is guided in a tapered and metal-coated silica hollow fiber. The near-field-emitted laser radiation leaving the aperture is capable of structuring a chromium film on a quartz substrate. The structuring resolution is not limited by the optical diffraction limit since the aperture can be smaller than the laser wavelength. (Reprinted (adapted) with permission from [Nolte OL, 1999] © Optical Society of America); **(b)** Example of repairing a lithographic mask. The top image shows an artificially produced defect of the chromium film [Nolte PB, 1999]. This defect was removed through the SNOM-based fs-near-field structuring (bottom image). (Reprinted from [Nolte PB, 1999] by permission from Wiley-VCH-Verlag GmbH: Phys. Bl. **55**, 41 – 44 (Mikrostrukturierung mit ultrakurzen Laserpulsen, S. Nolte et al.), Copyright (1999))

## Repair of Optoelectronic Devices

Another early application of fs-laser processing was proposed by Dupont et al., addressing the laser-based repair of optoelectronic devices, such as light emitting diodes (LEDs) [Dupont, 2000]. Taking benefit of the small HAZ, isolating microtrenches (grooves) were processed with fs-laser pulses for electrically isolating defective regions creating an electric shortcut, see Figure 1.25.

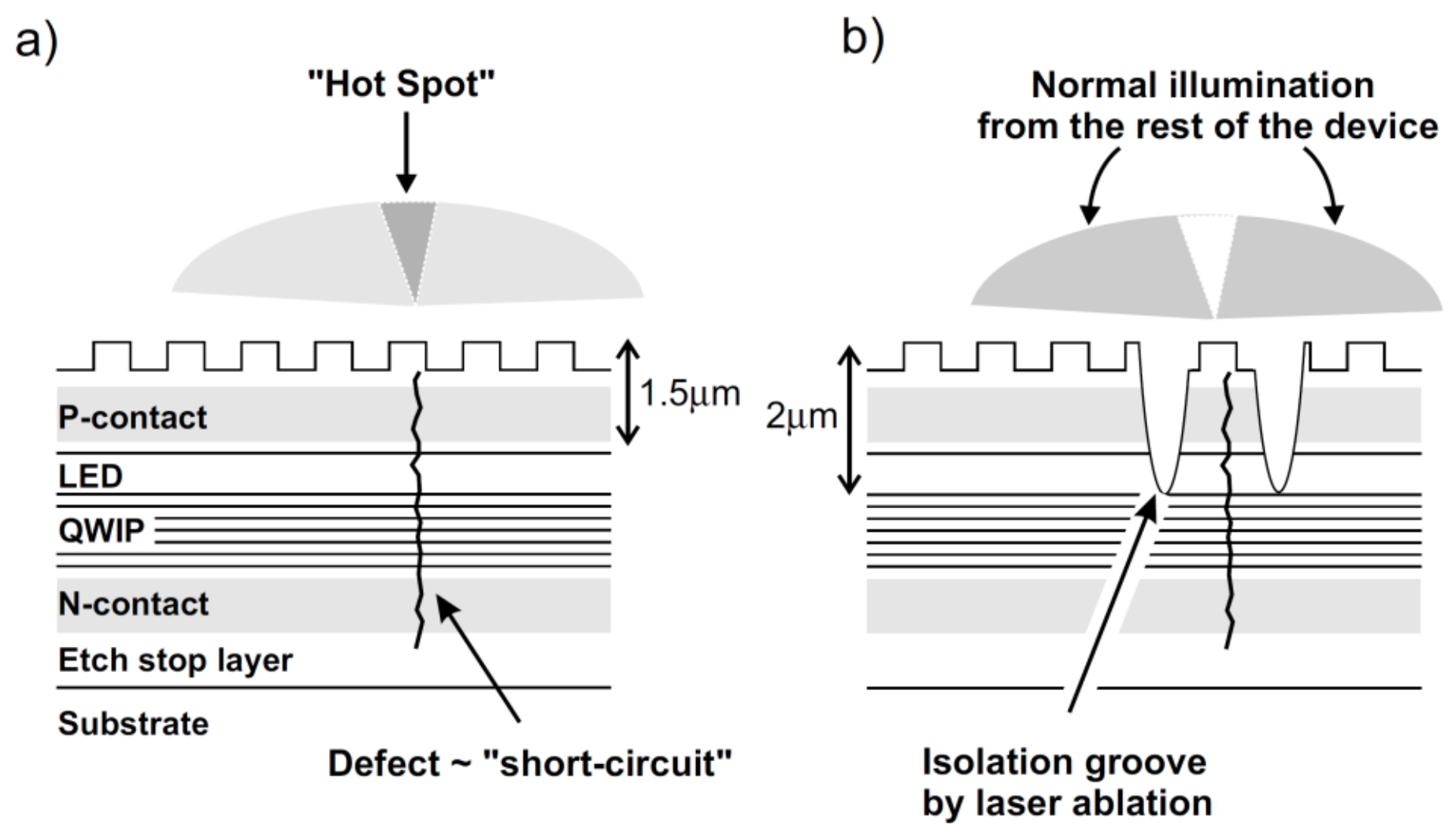

**Fig. 1.25:** Scheme of the principle of laser ablation for isolating a crystallographic defect in a quantum-well infrared photodetector (QWIP) LED device [Dupont, 2000]. (Used with permission of IOP Publishing, Ltd, from [Dupont, 2000], E. Dupont et al., In situ repair of optoelectronic devices with femtosecond laser pulses. Semicond. Sci. Technol. **15**, L15 – L18 (2000); permission conveyed through Copyright Clearance Center, Inc.)

With that approach, the authors demonstrated the successful repair of a single unit in an array of 8×8 mm-sized LEDs that was initially shunted by a crystallographic defect [Dupont, 2000]. Nowadays, the fs-laser-based repair has become a standard procedure in various process steps for manufacturing optoelectronic micro-devices [Zhu, 2022 / Lai, 2023].

## Sub-diffraction Limit Surface Texturing

Another application of ultrashort laser pulses is the possibility of generating "self-ordered" periodic micro- and nanostructures at the surface of almost any material. Self-ordered means here that the periods (Λ) of the surface structures are significantly smaller than the laser beam diameter ($2w_0$) used for their processing and that they emerge upon multi-pulse irradiation without being a simple "imprint" of the laser beam profile. The structures form in the irradiated spot but can be extended to a larger surface area by beam-scanning methods. The approach is very appealing as it enables a large-area surface texturing in a simple processing step.

The phenomenon, known since 1965 [Birnbaum, 1965], is termed *"laser-induced periodic surface structures"* (LIPSS) [van Driel, 1982], and has developed into a scientific evergreen [Bonse, 2017 / Bonse, 2024]. The spatial periods of LIPSS are typically of the order of the laser irradiation wavelength (λ) for the most common type, the so-called *low-spatial frequency LIPSS* (LSFL). But since the mid of the nineties and only upon irradiation with ultrashort laser pulses, another type of LIPSS was discovered (so-called *high spatial frequency LIPSS*, HSFL) that can feature spatial periods of less than 100 nm. This is remarkable since such feature sizes are significantly smaller than the *optical diffraction limit* would allow. For linearly polarized laser radiation, LIPSS are usually oriented either perpendicular or parallel to the polarization direction.

Figure 1.26 presents an example of sub-100 nm HSFL fabricated upon ps-laser processing (925 fs, 1030 nm) at the surface of Ti-6Al-4V titanium alloy. The laser processing was performed in ambient air at a pulse repetition frequency of 400 kHz, using a Galvanometer-scanner and at a laser peak fluence very close to the ablation threshold of the material. The inset in the upper right corner displays the corresponding 2D *Fast Fourier Transform* (2D-FFT) of the scanning electron micrograph, revealing an average spatial period of $\Lambda_{HSFL}$ ~ 81 nm. More detailed morphological and chemical analyses involving different pulse repetition frequencies confirmed that these HSFL are not affected by heat accumulation effects up to 400 kHz [Müller, 2024].

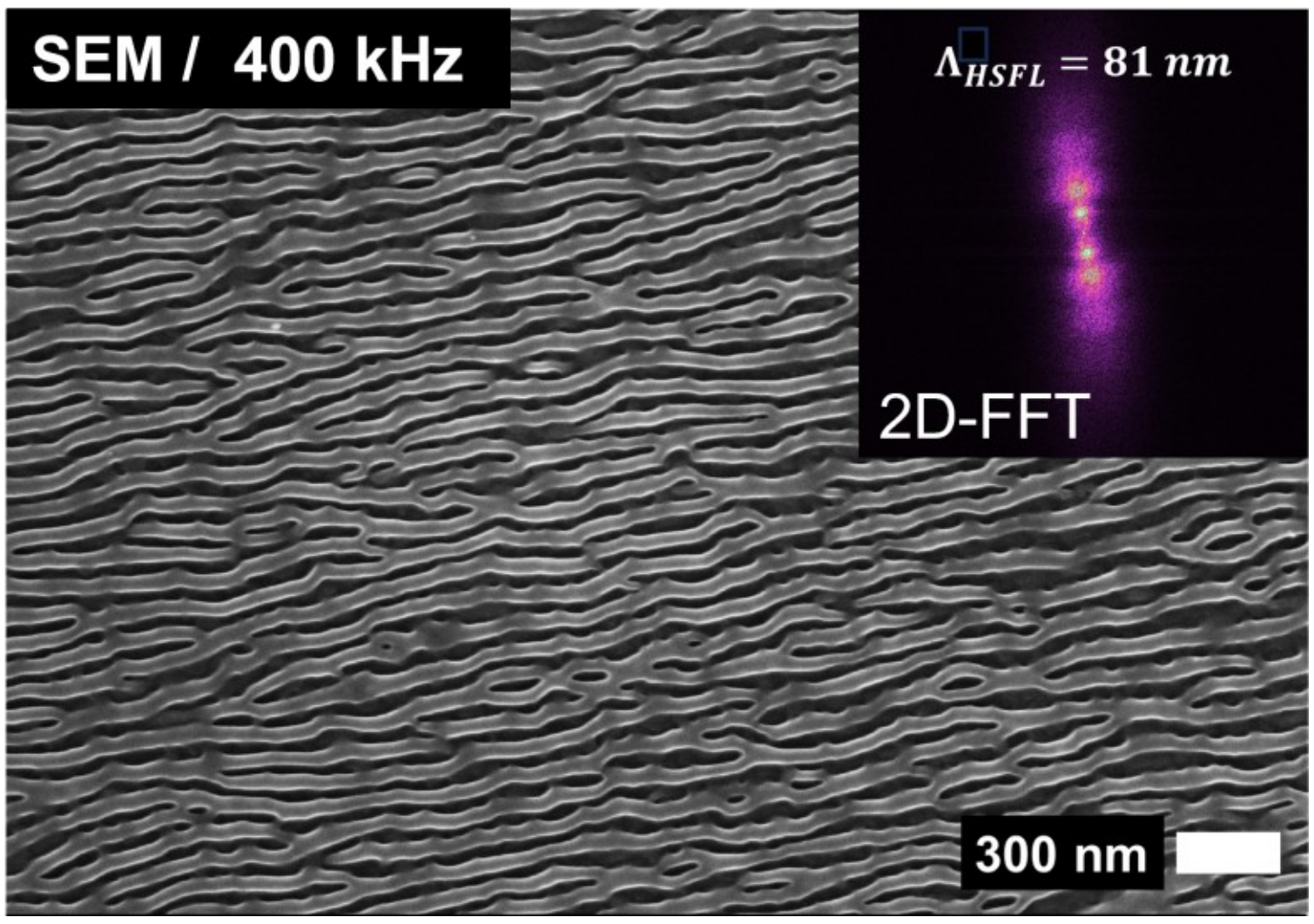

**Fig. 1.26:** Top-view SEM micrograph of ps-laser-generated sub-100 nm HSFL on Ti-6Al-4V titanium alloy processed at 400 kHz pulse repetition frequency [$\phi_0 = 0.26$ J/cm$^2$, $2w_0 = 35.5$ µm, $v_x = 467$ mm/s, $\Delta y = 5$ µm]. The inset represents the 2D-FFT of the micrograph. (Reprinted from [Müller, 2024], K. Müller et al., Chemical and topographical changes upon sub-100-nm laser-induced periodic surface structure formation on titanium alloy: the influence of laser pulse repetition rate and number of over-scans, Phys. Status Solidi A **221**, 2300719 (2024), Copyright 2024 under Creative Commons BY 4.0 license. Retrieved from https://doi.org/10.1002/pssa.202300719 )

It is nowadays clear that both types of nanostructures (LSFL and HSFL) are seeded by coherent scattering and interference effects that occur during the absorption of the laser irradiation at the surface (for times $t < \tau_p$). In a second development stage (at times $t > \tau_{e\text{-}ph}$), this absorption pattern is triggering a plethora of matter-reorganization effects, finally resulting in a rippled surface topography [Sipe, 1983 / Bonse & Gräf, 2020].

The ultimate (minimum) periods of HSFL account to a few tens of nanometers only and are limited by (i) the spatial decay length of optical near fields, and (ii) the requirement of energy confinement at least over the time lapse of electron-lattice temperature equilibration ($\tau_{e\text{-}ph}$) [Bonse & Gräf, 2021]. For both constraints, obviously ultrashort laser pulses are predestinated, which neatly explains the exclusive observation of HSFL with ultrashort laser pulses only. Furthermore, the relevance of near-field-related scattering and interference effects explains non-suited applicability of the diffraction limit as a far-field concept.

LIPSS allow to create a large variety of different surface functionalizations (colorization, wetting, adhesion, etc.) that are enabling numerous applications in optics, electronics, fluidics, tribology, mechanical engineering, and medicine [Vorobyev LPR, 2013 / Reif, 2018 / Florian, 2020 / Mezera Chapter, 2023 / Schwibbert, 2024]. For more details on the formation mechanisms and applications of LIPSS, the reader is referred to review articles [Siegman, 1986 / Buividas, 2014 / Gräf, 2020 / Bonse & Gräf, 2020] and several dedicated chapters in this book, e.g., Chap. 9 (Ancona et al.) focusing on fundamentals of LIPSS, or Chap. 27 (Zielinski et al.) rolling out numerous possibilities of surface functionalization and related applications.

## 2.4 Scaling: Limits Imposed by Plasma Shielding or Heat Accumulation

The development of ultrashort pulse laser with medium to high laser pulse energies has rapidly evolved during the last two decades, manifested through significant advancements in three specific branches of laser technology, i.e. in the fields of *thin disk lasers*, *fiber lasers* and in the *InnoSlab* technology [Jauregui, 2013 / Russbueldt, 2015 / Saraceno, 2019].

Figure 1.27 compiles some data on the average output powers of USP laser systems vs. the past years of the two past decades, currently reaching the 10 kW output power level [Krüger & Bonse, 2023]. Remarkably, the semi-logarithmic plot reveals a Moore-law-like power scaling.

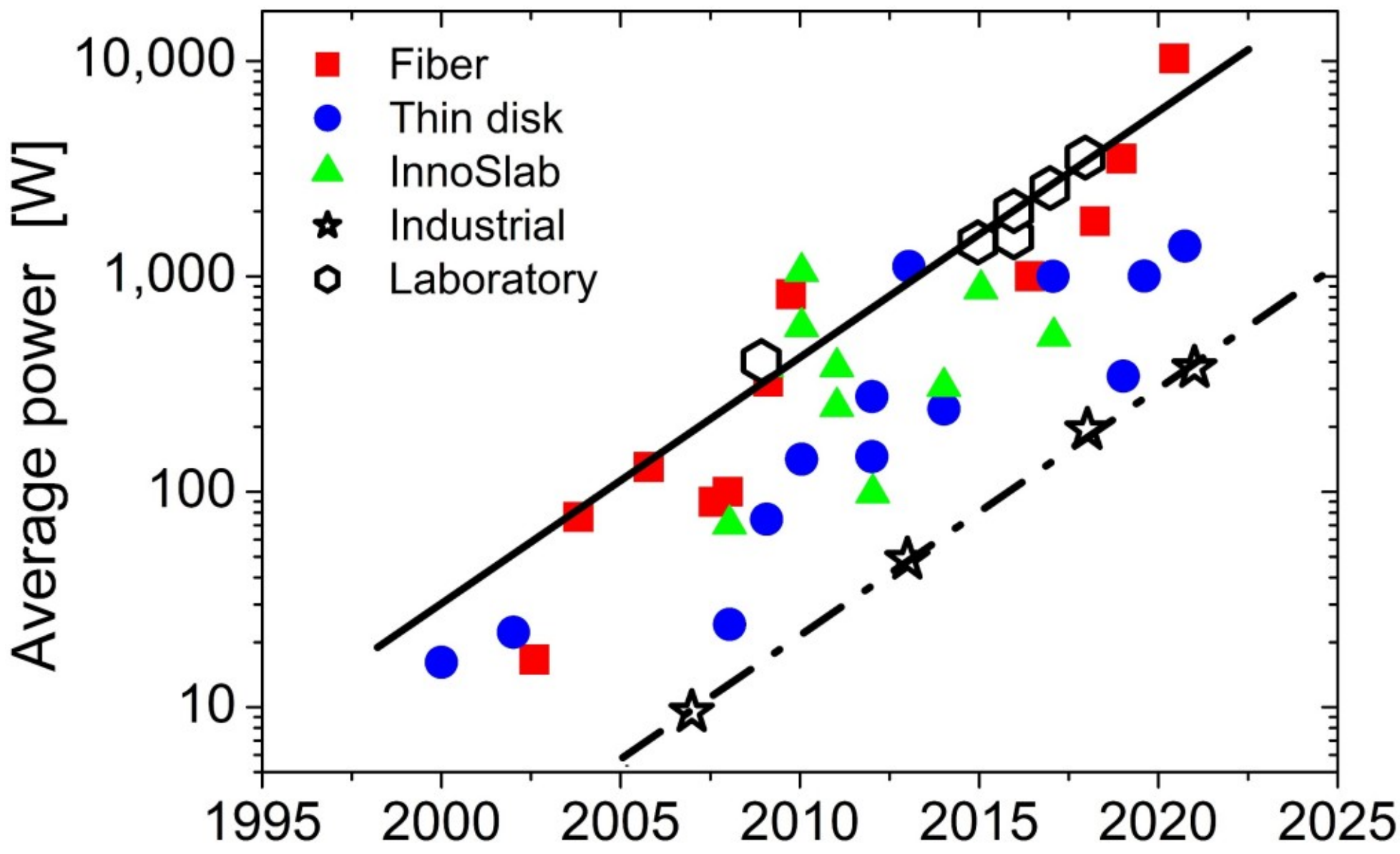

**Fig. 1.27:** Progress in the development of the average output power of ultrafast lasers over the past two decades. (Reprinted from [Krüger & Bonse, 2023], Krüger and Bonse, Special Issue "Advanced Pulse Laser Machining Technology", Materials (Basel, Switzerland) **16**, 819 (2023), Copyright 2023 under Creative Commons BY 4.0 license. Retrieved from https://doi.org/10.3390/ma16020819 )

It can be seen that about every two years, the average power was doubled and that the industrial laser systems lag about 10 years behind the scientific developments (laboratory systems). Note that the up-scaling of the average powers is mainly driven by the increased laser pulse repetition frequencies ($P_{av} = E_p \cdot f_{rep}$). This has important consequences for the laser processing and imposes some limitations.

The most relevant limitations for the up-scaling of laser processing of strong absorbing materials in air with respect to the processing time via an increase of the average laser pulse energy or the pulse repetition frequency are related to the effects of (1) *plasma shielding* through ablating material interacting with the incident laser radiation in front of the sample surface and (2) thermal *heat accumulation* at the sample surface when the irradiated surface spot cannot cool down to room temperature between successive laser pulses (see Sect. 1.2)). While (1) reduces the laser pulse energy that can be effectively used for ablative laser processing, (2) typically limits the precision of the processing results.

Plasma shielding can manifest either as an *intra-pulse effect* or, at high laser pulse repetition frequency, as an *inter-pulse effect*. In contrast, thermal heat accumulation is always an inter-pulse feedback mechanism. In the following section we will first explore in more detail these two effects, before briefly discussing potential solutions and strategies for overcoming the limitations.

## Plasma Shielding

Already at the end of the sixties, Dawson et al. theoretically analyzed the expansion dynamics of the laser-induced plasma and pointed to a strong intra-pulse absorption of laser pulses if the length scale of the laser-induced (overdense) plasma-density fallow $R$ is greater than half of the spatial extent of the laser pulse in propagation direction ($R > c_0 \times \tau_p/2$), with $c_0$ being the speed of light in vacuum [Dawson, 1969]. In other words, strong absorption of the laser radiation may occur in an overdense plasma if the extent of the "slice of light" is smaller than $2R$. At that time, the effect was experimentally widely unexplored since the available laser pulse durations were comparatively long, giving the emerging laser-induced plasmas enough time to expand while becoming optically thin.

In the second half of the Eighties, Eyett et al. reported for ns-laser XeCl excimer laser processing (308 nm wavelength, 11 ns pulse duration) of polished lithium niobate crystals in ambient air a significant dependence of the ablation rate on the laser beam spot diameter, realized through mask projection [Eyett, 1987]. Moreover, at laser fluences largely exceeding the ablation threshold, recordings of the signal being reflected from the surface upon laser irradiation (self-reflectivity) indicated an attenuation of the falling laser pulse flank due to its absorption by the ablating material. In 1990 the research group around Michael Stuke (Max-Planck-Institute, Göttingen, Germany) systematically explored differences between the short-pulsed (ns) and ultrashort-pulsed (ps) ablation of lithium niobate crystals by ultraviolet excimer laser radiation (308 nm wavelength, 1 ps or 16 ns pulse duration) in air [Beuermann, 1990]. They studied the fluence dependence of the ablation rate and the elemental ion yield (time-of-flight mass spectrometry), as well the ablation efficiency for different laser beam spot sizes.

For ns-laser pulses, the authors found a strong dependence of the ablation rate per pulse on the laser beam diameter. For laser beam spot sizes increasing from 30 µm to 206 µm, the ablation rates differed by more than a factor of two at the same fluence. In contrast, for ps-laser irradiation, the ablation rate was independent of the focused laser beam spot sizes and increased when compared to the ns-laser ablation at the same fluence. As another difference, for the ns-laser irradiation, the element-specific atom emission yields differed among the ablated elements lithium and niobium and occurred already at a fluence several tens of percent below the macroscopic ablation threshold, while for the ps-laser irradiation, the fluence threshold for microscopic particle emission coincided with the macroscopic ablation threshold. The authors quantified the ablated lithium niobate material per laser photon and found an optimum ablation efficiency at a fluence about 3-times the ps-ablation threshold and an about 10-times higher maximum ablation efficiency for ps-laser irradiation when compared to ns-laser irradiation [Beuermann, 1990]. This is in line with later publications, where it has been show that for ultrashort laser processing of strong absorbing materials using spatially Gaussian beams the optimum ablation efficiency (removed material volume per incident pulse energy) is obtained at fluences $e^2$-times ($\approx 7.4\times$) the ablation threshold (see Chap. 5 (Neuenschwander and Foerster)).

Importantly, the beam-spot size dependence of the ns-laser ablation rate followed a nearly linear dependence on the reciprocal laser beam spot radius, as predicted by a simple model of plasma plume expansion. At large beam diameters, the ablation rate saturated at a reduced value, limited by the absorption and scattering of the ablating material [Beuermann, 1990]. The latter covers the entire ablation region that expands initially mainly perpendicular to the irradiated surface. For small ablation areas the dilution of the plasma plume through its 3D geometrical expansion leads to a significant increase of the ablation rate then. If the ablation is performed in vacuum, the ablation plume can expand unhindered and the attenuation of the incoming laser radiation occurs faster than in ambient air [Wolf-Rottke, 1995].

Another signature of intra-pulse plasma shielding manifests through a characteristic pulse duration dependence of the ablation rate at high laser fluences. This can be seen in Figure 1.28, which displays the average (multi-pulse) ablation rate per pulse for steel upon irradiation with Ti:sapphire laser pulses (800 nm wavelength) of different pulse durations $\tau_p$ ranging between 200 fs and 5 ns [Momma, 1996]. The data were obtained in spot processing with a focused laser beam by drilling through a 0.5 mm thick steel sheet placed in a vacuum chamber at $10^{-4}$ mbar pressure. The (fluence dependent) number of laser pulses, which were used to drill through the sheet, was limited at $N \leq 3{,}000$.

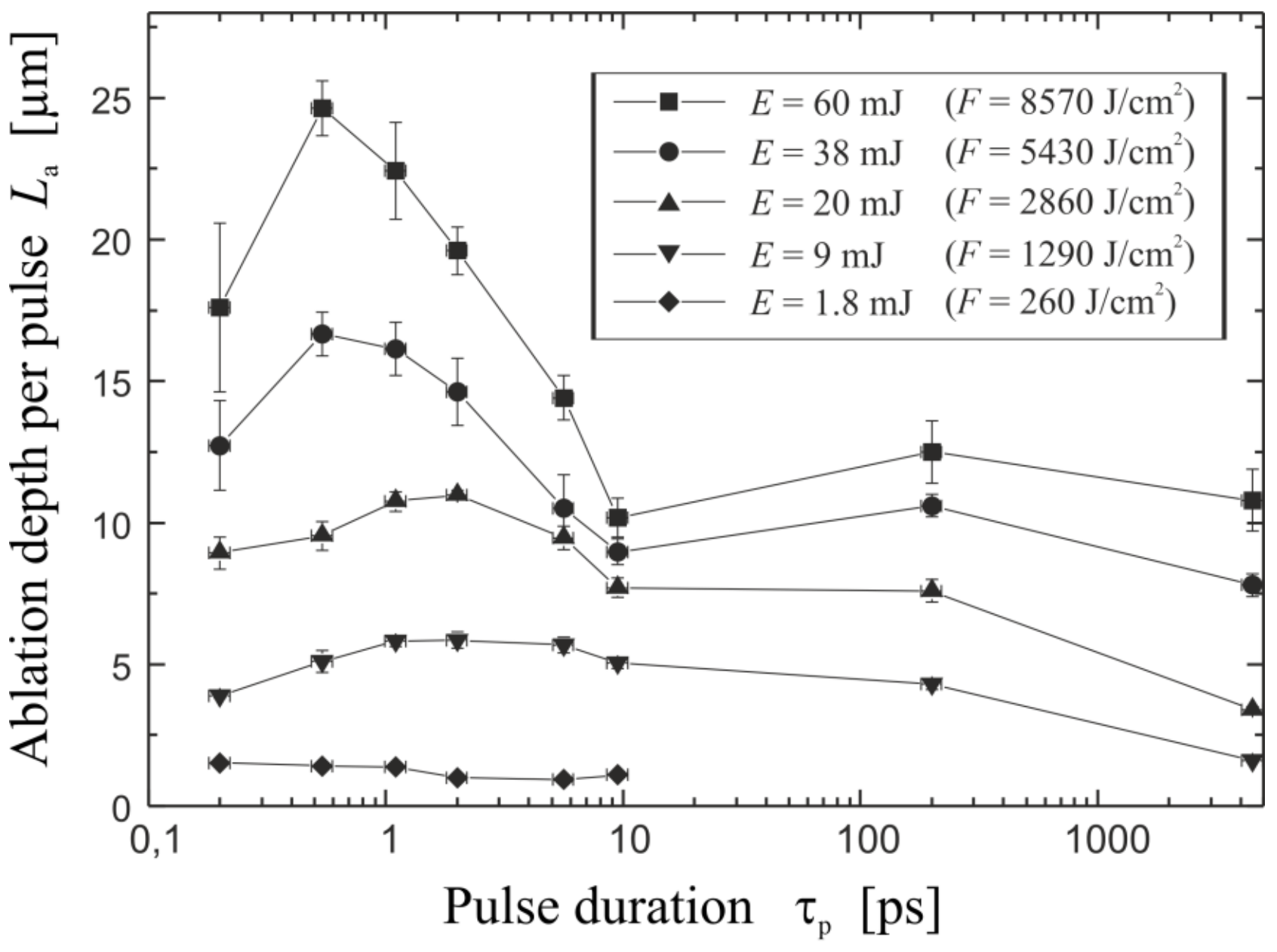

**Fig. 1.28:** Average ablation depth per pulse vs. the laser pulse duration for five different average fluences $F$ in < $10^{-4}$ mbar vacuum [780 nm, 10 Hz] [Momma, 1996]. The target is a 0.5 mm thick steel sheet. (Reprinted from [Momma, 1996], Opt. Comm. **129**, C. Momma et al., Short-pulse laser ablation of solid targets, 134 – 142, Copyright (1996), with permission from Elsevier)

At laser pulse energies $E_p$ < 20 mJ (average fluence $\phi$ = 2860 J/cm$^2$), the ablation rate is practically independent on the pulse duration. At higher laser pulse energies, a maximum ablation rate can be observed for pulse durations $\tau_p$ between 0.5 and 1 ps, reaching a value of up to ~25 µm/pulse here. Particularly for high laser pulse energies exceeding ~10 mJ ($\phi$ = 1430 J/cm$^2$), a characteristic drop of the ablation rate is observed at pulse durations longer than ~10 ps. This is caused by the onset of ablation that is manifesting already during the laser pulse irradiation, see Sect. 2.2 above and Chap. 13 (Bonse). Thus, for avoiding plasma shielding already during a single pulse ablation event, ultrashort laser pulses with durations less than 10 ps should be used.

While the laser-induced plasma characteristics and its evolution dynamics have been studied over several decades already with a focus on the ablating material itself (driven by applications such as spectroscopic material analyses, etc. [Radziemski & Cremers, 1989 / Cremers & Radziemski, 2013]), the implications on laser materials processing have been rather scarcely investigated. This changed with the availability of lasers emitting ultrashort pulses of moderate to high single-pulse energies at high repetition frequencies. For materials processing with such lasers, plasma shielding can also become relevant as an inter-pulse effect. This means that a laser pulse in a train of pulses can interact with the ablation plasma produced by the preceding pulse.

König and Nolte analyzed the plasma evolution during the localized ablation of metals with ultrashort laser pulses, including time-resolved studies of the plasma luminescence, a visualization of the ablated material as well as the optical plasma transmission [König, 2005]. Two ablation fronts were observed that developed at different times after the impact of the ultrashort laser pulse, temporally separated by ~40 ns. They were associated with an early shock wave front (driven by direct ionization, sublimation, photoelectric and thermionic electron emission inducing high pressures and temperatures and further gas- and hydrodynamic effects) and a later vapor front (driven by slower thermal processes like vaporization and boiling). The expansion of the shock wave and the vapor front is symmetric to the laser beam axis, but not completely spherical.

Figure 1.29 provides an analysis of the spatial extent $R(t)$ of the expanding plasma plume of fs-laser irradiated C75 steel (800 nm center wavelength, 200 fs pulse duration, 20 J/cm$^2$ average fluence) by plotting the positions of the shock wave front (blue data points) and the vapor front (red data points) as a function of the delay time $t$.

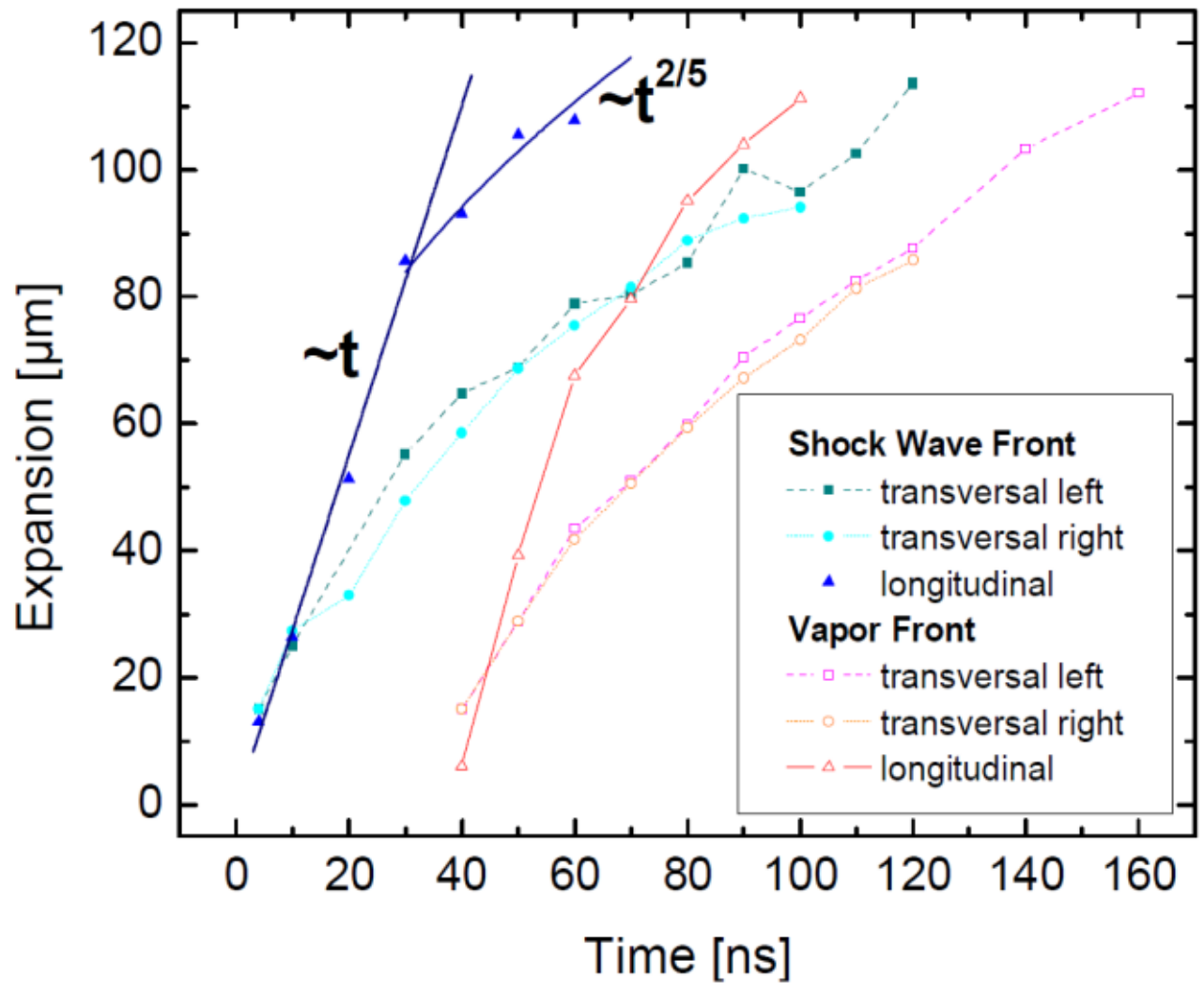

**Fig. 1.29:** Plasma plume expansion position of the shock wave front (blue symbols) and vapor front (red symbols) upon irradiation of C75 steel with a high-fluence fs-laser pulse (800 nm, 200 fs, 20 J/cm$^2$) [König, 2005]. Data are provided for the longitudinal direction (perpendicular to the surface) and for the vertical direction (parallel to the surface). Solid lines represent least-squares fits of the indicated scaling laws to the data. (Reprinted with permission from [König, 2005] © Optical Society of America)

The longitudinal expansion of the plasma plume (perpendicular to the irradiated surface) can be described by a combined propagation model of a shock wave and a vapor plume in gas for nanosecond laser pulses [Bäuerle, 2011]. This model predicts an unhindered linear expansion of the shock wave during the first nanoseconds ($R(t) \sim t$). From the slope of the experimental data in Figure 1.29, an expansion velocity of ~3 km/s is obtained for this early stage of plasma evolution. At a few tens of nanoseconds, a transition from free expansion towards a hindered expansion through a background gas occurs, here being present due to the ambient air. This expansion stage results in a power-law scaling $R(t) \sim t^{2/5}$. The transversal expansion of both ablation fronts (parallel to the irradiated surface) is less linear at the beginning and emerges earlier into that power-law.

Figure 1.30 displays the optical attenuation (transmission) through the ablation plasma plume of various ultrashort pulse irradiated metals (Al, Cu, steel) in a probe pulse delay range between 300 ps and 150 ns. [König, 2005]. The transmission is integrated spatially over the probe region up to 14 µm above the surface, referenced to the case with blocked pump pulse (no ablation). (a) shows the integrated transmission of 400-nm probe pulses upon ablation of aluminum with 200-fs Ti:sapphire laser pulses (800 nm) at an average fluence of 17 J/cm$^2$. (b) visualizes the integrated transmission of 1050-nm probe pulses upon ablation of copper (red data points) and steel (blue data points) with 3.3-ps Ti:sapphire laser pulses at an average fluence of 18 J/cm$^2$.

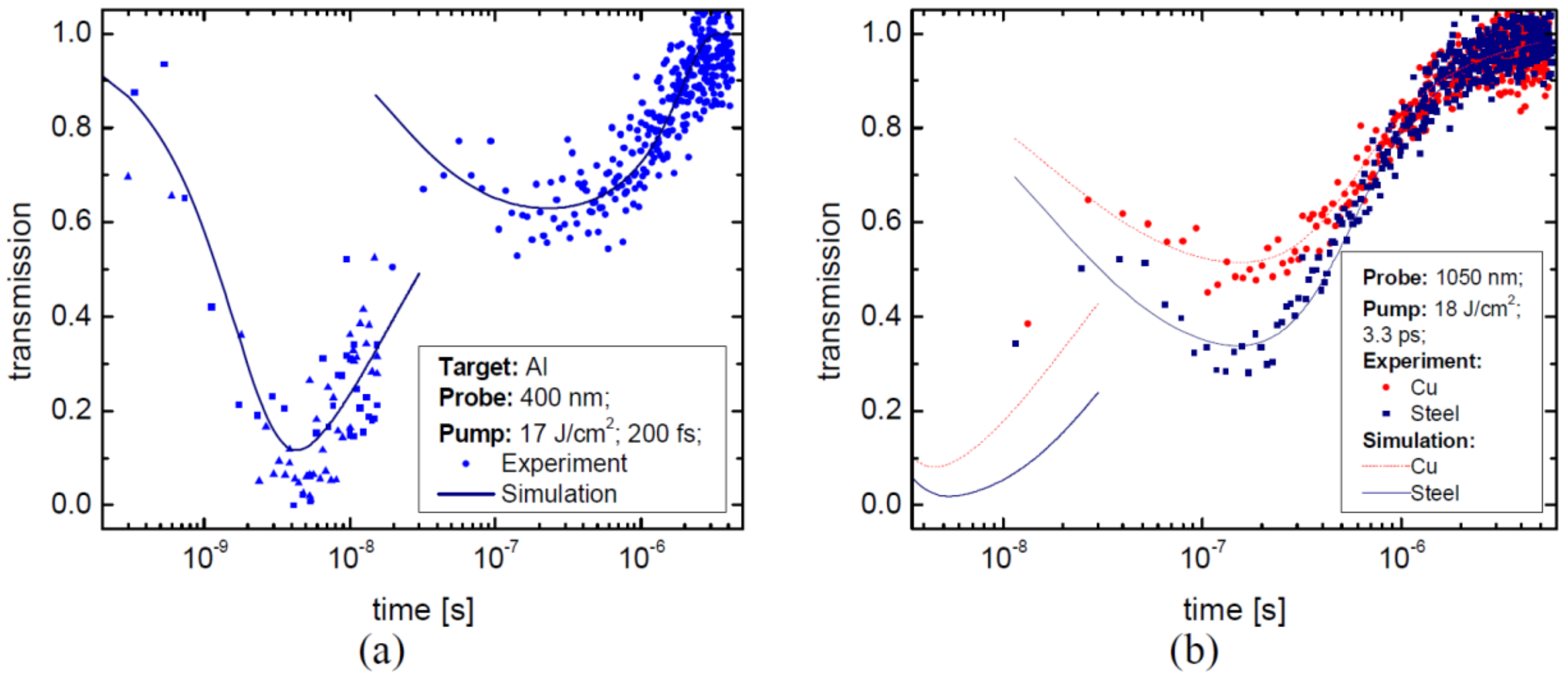

**Fig. 1.30:** (**a**) Transient optical transmission of 400 nm probe pulses during the ablation of aluminum by fs-Ti:sapphire fs-laser pulses [200 fs, 17 J/cm$^2$] [König, 2005]. (**b**) Transient transmission of 1050 nm probe pulses during the ablation of copper and steel by Ti:sapphire ps-laser pulses [3.3 ps, 18 J/cm$^2$]. In both plots the experimental data are plotted as data points, while simulation results are presented as solid lines. (Reprinted with permission from [König, 2005] © Optical Society of America)

The plasma transmission measurements of fs-laser ablated aluminum shows two characteristic minima (Figure 1.30a). The first minimum is reached after ~5 ns and results in almost complete attenuation of the probe beam. The second minimum with a transmission decrease down to 0.6 follows after 100 – 200 ns. Subsequently, the transmission increases within 2 – 3 µs to the initial value. For ps-laser ablated copper and steel, both materials show the same characteristic minima as observed for aluminum but with gradually different minimum transmission values (Figure 1.30b). The second transmission minimum did not show a significant dependence on the probe beam wavelength and was, thus, attributed to Mie scattering from ablated clusters and droplets. In contrast, the first transmission minimum was associated with absorption in an optically dense plasma, evolving from surface electron emission, sublimation, etc. [König, 2005].

These plasma transmission experiments revealed that the probe laser beam is almost completely blocked for up to 5 ns by the laser-induced plasma plume. Based on this result, for high-fluence metal processing by ultrashort laser pulses, an upper limit of the pulse repetition frequency due to plasma shielding was estimated to be around a few hundred kHz, depending on the target material, the laser fluence, and other process parameters. For pulse repetition frequencies higher than this limit, the laser pulse will be significantly absorbed in the ablation plume created by the preceding pulse, preventing that the optical laser pulse energy can be used in an effective way for ablative materials processing [Ancona OPEX, 2008]. Some in-situ visualizations of such effects can manifest at high repetition-rate laser processing be found in Chap. 13 (Bonse).

Förster et al. explored these effects systematically by comparing the fs-laser ablation of copper (515 nm, 10 ps, 50 Hz, 1.1 J/cm$^2$ average single pulse fluence) by singe-, double-, or triple-pulse sequences with inter-pulse delays around 10 ns via shadowgraphic imaging [Förster, 2018]. The fluence value of each pulse in the sequence is about ten times the ablation threshold fluence. The authors identified specific experimental laser-interaction regimes, where the second laser pulse of such a sequence results in a reduced ablation efficiency, a suppression of ablation [Povarnitsin, 2009], or may even lead to a redeposition of previously ablated material still persistent in the atmosphere. Figure 1.31 schematically summarizes the results regarding the ablated volume [Förster, 2021].

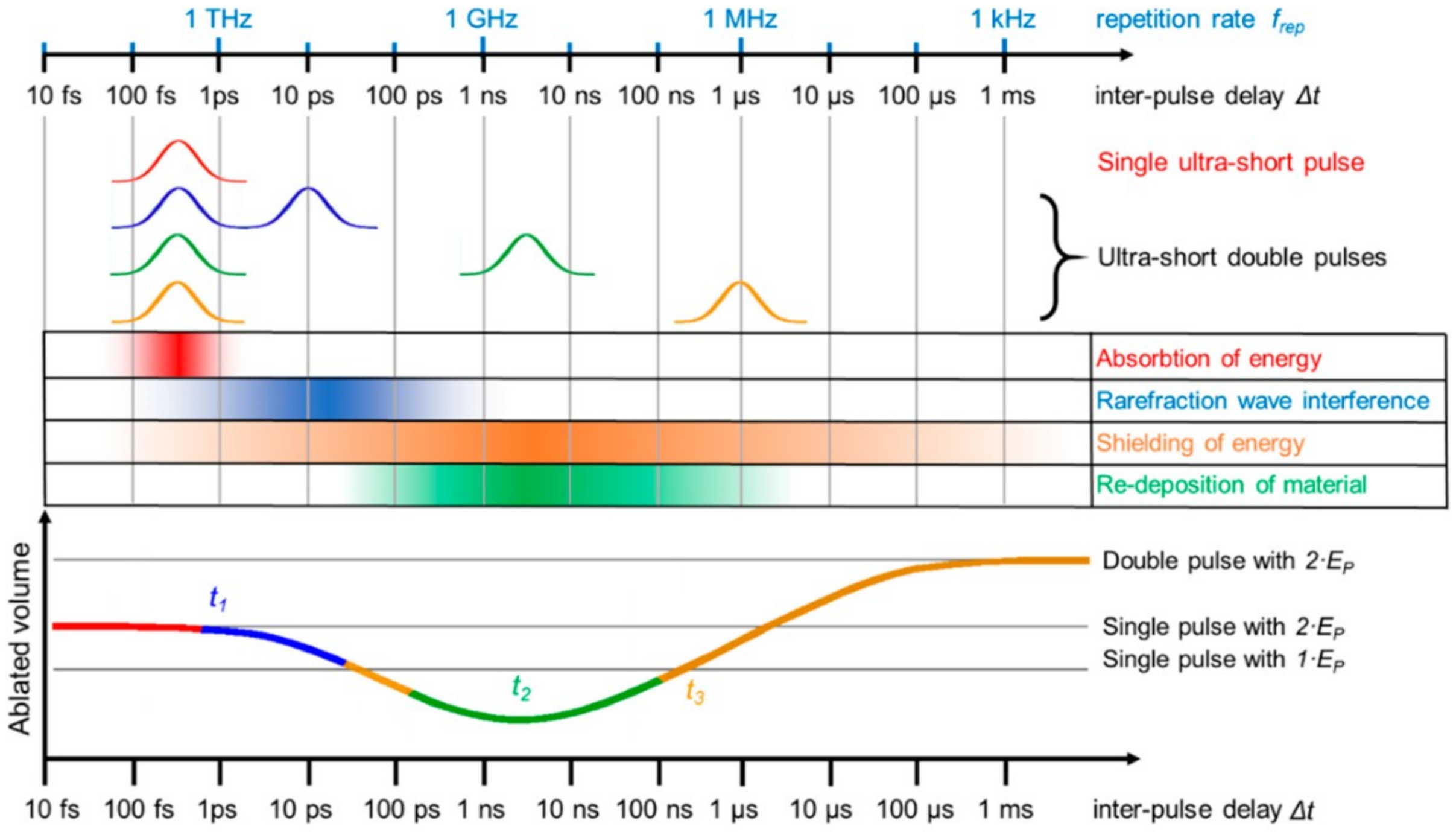

**Fig. 1.31:** Scheme of the interaction mechanisms and the ablated volume of metals upon irradiation with ultra-short single- and double-pulses for different inter-pulse delays. (Reprinted from [Förster, 2021], D.J. Förster et al., Review on experimental and theoretical investigations of ultra-short pulsed laser ablation of metals with burst pulses, Materials (Basel, Switzerland) **14**, 3331 (2021), Copyright 2021 under Creative Commons BY 4.0 license. Retrieved from https://doi.org/10.3390/ma14123331 )

## Heat Accumulation

Another fundamental limitation of laser processing can manifest at high fluences and at high laser pulse repetition frequencies when the laser-induced transient temperature increase at the surface does not return to room temperature between successive laser pulse irradiation events. Obviously, the relevance of such a thermal heat accumulation effect will depend on the two mentioned laser parameters fluence and pulse repetition frequency but also on the thermal properties of the irradiated material (heat conductivity, heat capacity, latent heat of melting), the ambient medium (air, gases, vacuum, liquid) and temperature, the dimensionality of the dominant surface cooling mechanism (1D, 2D, or 3D heat transport in the surrounding), and the selected laser processing strategy (spot processing vs. scanning, scan path and temporal sequence), etc.

Heat accumulation effects are not specific for ultrashort pulse laser irradiation, they can manifest for any pulse duration and were studied in detail in the past already for spot- and scanning laser processing [Carslaw & Jäger, 1959 / Gamaly, 1999 / Ancona OPEX, 2008 / Ancona OL, 2009 / Bäuerle, 2011 / Weber OPEX, 2014 / Bauer OPEX, 2015 / Weber OPEX, 2017]. In the following, we will discuss only the case of heat accumulation effects manifesting during spot processing (sometimes termed percussion drilling in the ablative regime). For the case of scanning laser processing, the reader is referred to the specialized literature [Bauer OPEX, 2015 / Weber OPEX, 2017]. We assume that a train of identical laser pulses (energy $E_\mathrm{p}$) and with a pulse repetition frequency $f_\mathrm{rep}$ is focused on the surface of a selected target material (thermal diffusivity $D_\mathrm{th}$, mass density $\rho$, specific heat capacity $c_\mathrm{p}$). Moreover, we assume that a specific total number of laser pulses ($N_\mathrm{p,tot}$) is required for the desired laser process, being finished after the processing time $t_\mathrm{proc} = N_\mathrm{p,tot} / f_\mathrm{rep}$.

For the above given constraints, Weber et al. derived a *heat accumulation equation* (Equation 1.24) that depends on the dimensionality (1D, 2D or 3D) of the dominant heat flow (cooling) mechanism into the surrounding [Weber OPEX, 2014 / Weber OPEX, 2017]. For ultrashort pulsed laser irradiation at moderate fluences and under loose focusing conditions, 1D heat transport is initially dominant for single-pulse events (see Sect. 2.2), but for long-lasting multi-pulse exposures, 3D cooling can dominate if thermal accumulation effects manifest.

Having a number of $N_\mathrm{p,tot}$ heating laser pulses in a temporal δ-peak at the origin of the heat source $Q_\mathrm{nD}$ and with the laser pulse repetition frequency $f_\mathrm{rep}$, the laser-generated temperature increase $\Delta T_\mathrm{HA,nD}$ due to heat accumulation can be estimated [Weber OPEX, 2014]. Immediately after the laser processing, it accounts to

$$\Delta T_\mathrm{HA,nD}\left(t_\mathrm{proc}\right) = \frac{Q_\mathrm{nD}}{\rho\cdot c_\mathrm{p}\cdot\sqrt{\left(\frac{4\cdot\pi\cdot D_\mathrm{th}}{f_\mathrm{rep}}\right)^{\mathrm{nD}}}}\cdot\sum_{N=1}^{N_\mathrm{p,tot}}\frac{1}{\sqrt{N^{\mathrm{nD}}}}. \qquad (1.24)$$

Herein, "nD" is from the set {1, 2, 3}. The heat source terms $Q_\mathrm{nD}$ can be written as $Q_1 = \sigma\cdot Q_\mathrm{heat}/A$, $Q_2 = \sigma\cdot Q_\mathrm{heat}/\ell$, or $Q_3 = \sigma\cdot Q_\mathrm{heat}$. The parameters $A$ and $\ell$ are the area and the length where the residual laser pulse energy $Q_\mathrm{heat}$ is deposited, and $\sigma = 1$ when the deposited heat can flow in the complete 3D surrounding of the source, and $\sigma = 2$ when the heat can flow in one half space only. The residual heat $Q_\mathrm{heat}$ left in the irradiated sample per incident laser pulse can be written as [Weber OPEX, 2017]

$$Q_\mathrm{heat} = \eta_\mathrm{heat}\cdot\eta_\mathrm{abs}\cdot E_\mathrm{p} = \frac{\eta_\mathrm{heat}\cdot\eta_\mathrm{abs}\cdot P_\mathrm{av}}{f_\mathrm{rep}}. \qquad (1.25)$$

$\eta_{abs}$ represent the fraction of the incident laser pulse energy ($E_p$) that is absorbed in the sample, The fraction of the residual heat $\eta_{heat}$ that is not used for ablation depends on material parameters, laser parameters (such as pulse duration, fluence, intensity distribution, and wavelength), and processing parameters (such as structure size, and structure depth). Usually, it must be determined experimentally.

Approximative solutions of Eq. (1.24) are presented in Sect. 3.5 of Chap. 6 (Lenzner and Bonse) along with some practical approaches for estimating the practical relevance of heat accumulation.

### Potential Solutions for Preventing Heat Accumulation and Plasma Shielding

There are several potential solutions to prevent or reduce heat accumulation effects, such as (i) induce temporal breaks after $N_{p,max}$ to allow the surface to cool down again, (ii) stay below the limiting laser pulse repetition frequency $f_{rep,max}$, (iii) provide active cooling of the target, and (iv) avoid the punctually repetitive laser-induced heat load by moving the focused laser beam across the surface (*scanning*).

If the classical meandering beam scanning approach is still affected by heat accumulation effects, the scanning process can be temporally split and performed in *several passes* over-scans). Alternatively, smart laser beam scanning strategies can be developed that distribute the heat load temporally across the entire scan area, preventing the neighbored sample surface sites from being irradiated shortly after each other.

In a different approach, the optically deposited energy may be *temporally distributed* via tailored temporal pulse shaping or the generation and application of pulse bursts (see Chap. 16 (Förster and Neuenschwander) and Chap. 17 (Manek-Hönninger and Lopez).

If enough laser pulse energy is available, the single laser beam may be split into *multiple beams* of lower pulse energies and the processing can be performed in a parallel way, see Chap. 4 (Holder et al.) and Chap. 15 (Haasler et al.).

For preventing or reducing plasma shielding effects, the inter-pulse separation $\Delta \boldsymbol{t} = 1/f_{rep}$ must be kept larger that the lifetime of the ablation plasma ($\tau_{plasma}$). The latter is defined here as the minimum time until the single laser pulse-induced ablation plasma turns optically thin upon expansion, disintegration, recombination, etc. Its exact value depends strongly on the irradiation conditions (fluence, focus diameter, angle of incidence) and material parameters (threshold fluence, surface reflectivity, etc.) and usually must be determined experimentally. Several of the above-mentioned beam scanning-, temporal pulse shaping-, and parallelization-related strategies to prevent heat accumulation effects may also help to reduce the plasma shielding.

# 3 Volume Processing

This section is devoted to laser-induced modification of material properties by focusing laser beams inside bulk materials which are usually transparent for the applied laser wavelength. Under such conditions, the laser light absorption is governed by nonlinear processes enabling the deposition of laser energy into micro/nanoscale volumes. This technique is already used in a number of smart applications based on 3D photonic and microfluidic structures in bulk optical materials and still has tremendous potential for further developments toward new functionalities and practical utilization. The physics behind laser-induced volumetric modification of materials is very rich and its deep understanding is required for the optimization of 3D laser writing for achieving more localized, controllable, and efficient laser energy deposition. Sect. 3.1, addresses the fundamental processes that are responsible for laser energy absorption in bandgap materials with some implications on how to influence the absorption localization and efficacy by varying the process parameters. In Sect. 3.2, the post-irradiated relaxation of absorbed energy in the bulk material matrix is discussed with focusing on non-thermal, thermal, and mechanical processes. A number of selected examples which are successfully used in scientific and technological applications are described in Sect. 3.3, while Sect. 3.4 summarizes factors, which limit the quality and up-scaling of the 3D material writing technique with analyzing strategies for further enhancing its capabilities.

# 3.1 Optical Energy Deposition: Tight Focusing

The era of volumetric laser processing of transparent materials, which now represents an important field of unique laser applications, started with the publishing of the seminal paper by Davis et al. [Davis, 1996] where the authors demonstrated the possibility to locally induce stable modifications of the refractive index inside bulk glasses by tightly focused ultrashort laser pulses. Since that time, this technique has become widely adopted in a number of scientific and technological applications based mostly on using optical glasses due to their relatively low cost and processability [Gattass & Mazur, 2008 / Sugioka & Cheng, 2013 / Stoian, 2020 / Butkute, 2021]. However, volumetric laser modification of semiconductor materials is also of high interest due to their specific properties and application opportunities [Chambonneau, 2021b / He, 2024], although large efforts are still required for controllable straightforward writing of semiconductor-based 3D photonic structures.

The phenomenon of laser-induced modification of transparent materials by tightly focused laser beams involves the multiplicity of interfering and consecutive physical and chemical processes initiated by radiation absorption during the action of the laser pulse, which can shortly be summarized in a simplified form as following [Itoh, 2006 / Couairon, 2007 / Gattass & Mazur, 2008 / Bulgakova, 2013a]:

- This phenomenon starts from the absorption of photons of the laser beam propagating through a bandgap material. Upon propagation toward its geometrical focus, the beam intensity increases and becomes sufficient to efficiently trigger multiphoton ionization of the initially neutral matrix of transparent dielectrics. The minimum number of photons necessary to photoexcite valence electrons to the conduction band (the order of multiphoton ionization) can be estimated as the ratio of the band gap energy to the photon energy as $K = \lfloor E_g/\hbar\omega \rfloor + 1$. Thus, the material excited by an ultrashort laser pulse becomes locally ionized to the state which is often called "solid plasma" or "free-electron plasma", when the material matrix remains solid and cold while the excited electrons can be energetic.
- At high laser intensities, the potential of the band gap is distorted, and tunneling ionization contributes to free-electron plasma formation. This means that electrons can easier escape from the valence band to the conduction one above the decreased ionization barrier. The transition from the multiphoton ionization regime to tunneling is gradual with increasing electric field strength of the laser wave. The boundary above which the tunneling mechanism starts to dominate is determined by the Keldysh parameter $\gamma = \omega(2m_{\mathrm{eff}}E_{\mathrm{g}})^{0.5}/eE_{\mathrm{L}}$ ($m_{\mathrm{eff}}$ and $e$ are the effective electron mass and the unit charge, respectively; $E_{\mathrm{L}}$ is the electric field of the laser wave; $E_{\mathrm{g}}$ is the bandgap width in dielectric materials [Keldysh, 1965 / Kaiser, 2000 / Derrien Chapter, 2021].

- As soon as the plasma of free electrons becomes dense enough, it starts to efficiently absorb laser light via inverse bremsstrahlung absorption and can create secondary electrons in collisions with neutral atoms called "avalanche ionization".
- Free electrons transfer their energy to the material lattice via collisions and recombine with the creation of numerous defect states. The excitation and recombination processes are schematically presented in Figure 1.32 [Stoian, 2003].
- The electron-lattice thermalization and electron recombination lead to the heating of localized volumes of matter to high temperatures, sufficient for softening or melting in the laser-affected zone and hence high-pressure gradients.
- Consequently, compression waves are generated, leading to material relocation, densification, the formation of rarefaction zones, or even voids, which result in modulation of the optical properties. Additionally, in the laser-modified region, numerous defect states are formed.
- When the next laser pulses couple with the material in multi-pulsed irradiation regimes, they meet a modified material structure. The light absorption process changes from pulse to pulse due to various accumulation effects (density changes, defects, heat) that further complicate the whole phenomenon of writing 3D structures inside bulk materials.

Note that many of these effects triggered by the laser-induced carrier generation in the volume of the material are ruled by the same fundamental physical mechanisms as discussed above in the context of surface processing (see Sect. 2.1). However, the final processing results may be very different due to the altered boundary conditions (material confinement in the bulk, small laser-excited volumes, strongly nonlinear interactions, material-specific energy relaxation and defect formation, etc.).

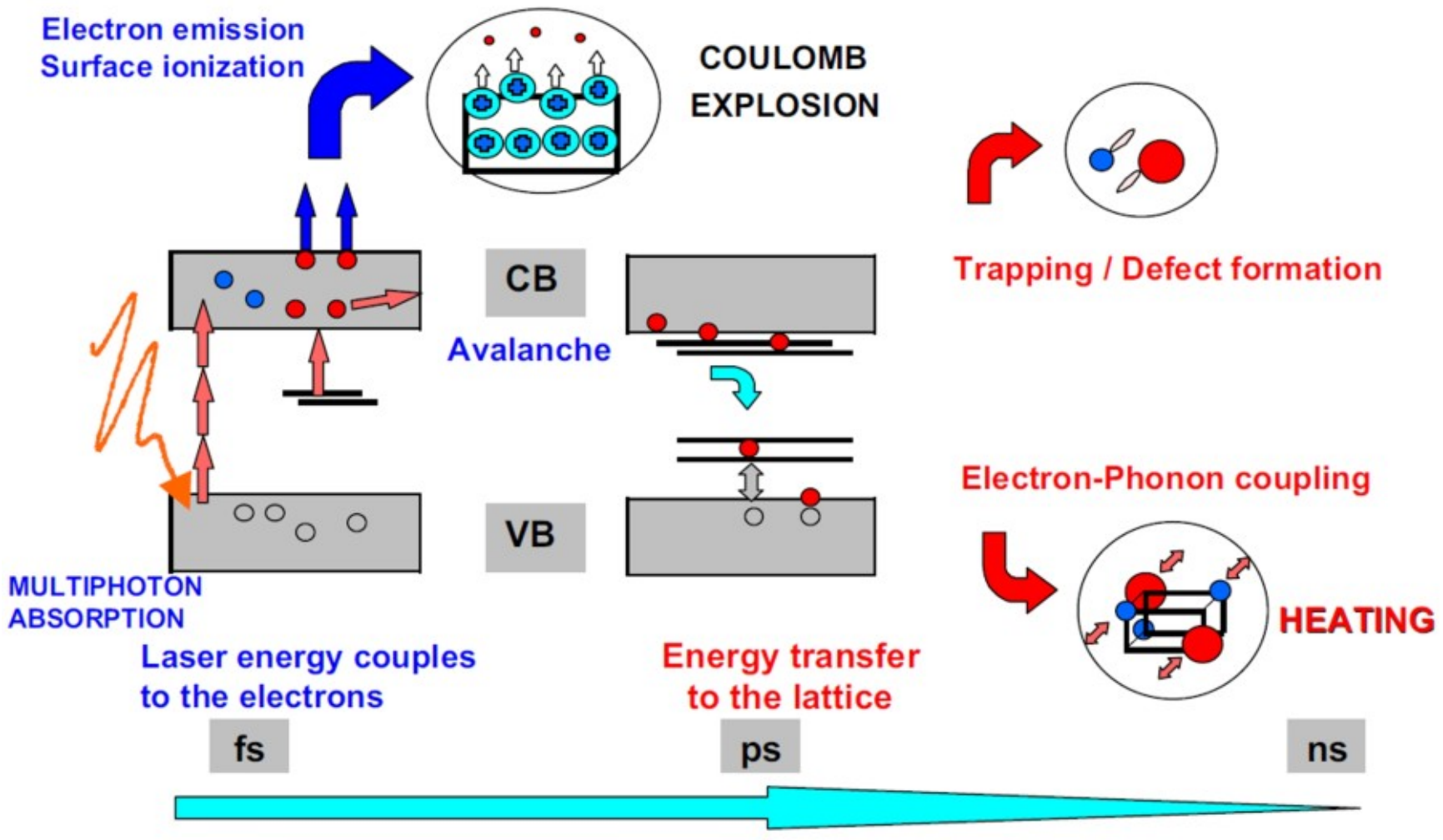

**Fig. 1.32:** Schematic representation of basic physical processes in laser-irradiated dielectric materials. (Reprinted from [Stoian, 2003], R. Stoian et al., Dynamic temporal pulse shaping in advanced ultrafast laser material processing, Appl. Phys. A **77**, 265 – 269, 2003, Springer Nature)

It should be underlined that ultrashort-laser-irradiated material properties can swiftly start to change at femtosecond timescales, exhibiting strong variations of optical response and emerging defect states [Canning, 2011 / Mishchik, 2013 / Shchedrina, 2024]. As a result of this interaction, modification features can be localized at nanometer spatial scales such as nanograting structures [Shimotsuma, 2003 / Bhardwaj, 2006 / Zimmermann, 2014]. The state-of-the-art diagnostic techniques still do not have enough capabilities to directly follow the complexity of the dynamic processes inside the material volume at such temporal and spatial scales [Mermillod, 2011 / Mauclair, 2016 / Hendricks, 2017]. Thus, they give valuable but incomplete information that must be analyzed and properly understood. In such circumstances, theory and especially computer simulations are important means for gaining a comprehensive insight into the multiparametric phenomenon of volumetric modification of bandgap materials via determining the spatiotemporal dynamics of laser-induced material transformation.

Several types of theoretical approaches can be listed that have been developed for the detailed study of the propagation of an electromagnetic wave in transparent materials with accounting laser energy absorption. A simplified semi-analytical model [Rayner, 2005] provides an estimative analysis of the geometry of laser energy absorption regions and light transmission through the sample. The model enables to predict the distributions of laser pulse intensity $I$ and free-electron density $N_\mathrm{e}$ by solving the interrelated equations for $\partial I(z,r,t)/\partial t$ and $\partial N_\mathrm{e}(z,r,t)/\partial t$. Applying this model, the authors have unambiguously demonstrated that, for their specific case of Pyrex glass irradiated by 800-nm, 40-fs laser pulses focused inside the bulk using a lens of NA = 0.25, the laser intensity is strongly clamped by nonlinear absorption to the maximal reached levels of the order of $10^{13}$ W/cm$^2$. The scheme of the clamping effect and comparison of the modeling results of laser energy absorption with the experimental data are shown in Figure 1.33, left and middle panels (a,b), respectively. Another important observation is that the absorbed laser energy is proportional to the pulse duration. It must be underlined that a more rigorous approach based on solving the nonlinear Schrödinger equation (NLSE) supports the latter conclusion, indicating, however, that at longer pulses the laser energy is deposited into a more localized region [Burakov, 2007]. A similar semi-analytical model applied for bulk modification of fused silica with accounting for the generation and re-excitation of self-trapped excitons (STEs) enabled an insight into the roles of defects in multi-pulse volumetric laser processing of transparent materials [Bulgakova, 2016] (Figure 1.33, right panel).

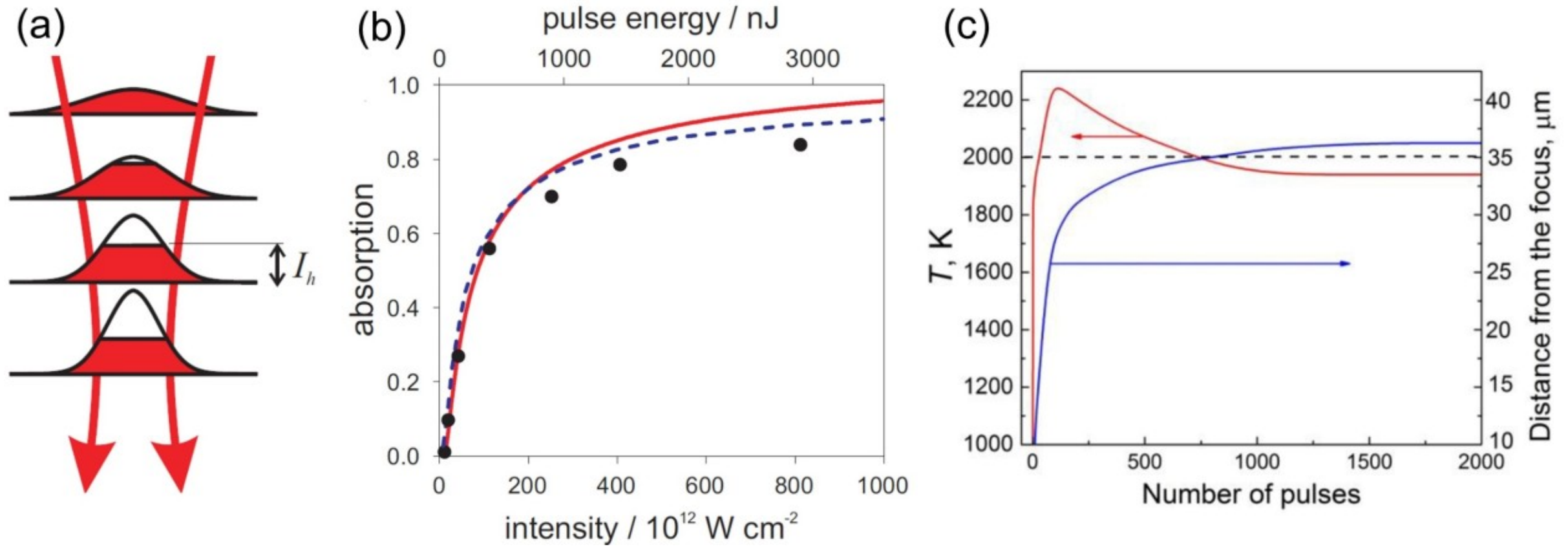

**Fig. 1.33:** (**a**) Scheme of the intensity clamping effect by nonlinear absorption of a focused laser beam in transparent dielectrics [Rayner, 2005]. Far from the geometric focus, absorption is negligible as the intensity is below the level necessary for efficient nonlinear absorption, $I_h$. As the laser beam approaches the focus, the intensity is clamped at $\sim I_h$. (**b**) Absorption in the bulk of Pyrex glass as predicted by the semi-analytical model (solid line) with the dashed curve obtained by numerical analysis based on the Keldysh ionization theory. Dots correspond to experimental measurements. (Figures (a and b) reprinted with permission from [Rayner, 2005] © Optical Society of America); (**c**) Maximum local temperature inside a fused silica sample, predicted by a semi-analytical model for multi-pulse irradiation by a focused laser beam with accounting for the defect accumulation (red line) [Bulgakova, 2016]. The length of the modification structure extends, from pulse to pulse, in the direction toward the irradiating laser (blue line). (Reproduced from [Bulgakova, 2016], with permission from N.M. Bulgakova et al., Ultrashort-pulse laser processing of transparent materials: insight from numerical and semi-analytical models, Proc. SPIE **9735**, 97350N (2016). Copyright 2016, SPIE)

For more predictive descriptions that would be applicable to a wide range of laser irradiation conditions, sophisticated approaches are required that could account for the optical Kerr effect, free-electron plasma defocusing, the beam diffraction in the transverse direction, group velocity dispersion, and other related processes such as avalanche ionization disregarded in the simplified models cited above. One such approach based on the solution of the NLSE is widely utilized for studying the processes of laser excitation of dielectrics in the regimes of modification. The NLSE, which is an asymptotic parabolic approximation of Maxwell's equations, is applicable for describing the unidirectional propagation of slowly varying envelopes of laser pulses. In particular, it describes the self-focusing effect, which manifests itself as a laser beam collapse at beam powers beyond a critical value specific for laser wavelength and material [Couairon, 2007]. Numerical solutions of the NLSE in applications to different irradiation conditions and materials have allowed elucidating important features of laser pulse propagation through transparent solids such as filamentation, intensity clamping, and strong dependence of laser energy deposition geometry on pulse duration [Couairon, 2005 / Couairon, 2007 / Burakov, 2007 / Bogatskaya, 2024]. Figure 1.34 presents the comparison of the experimental and simulation results for fused silica excitation by a Ti:sapphire laser pulse, demonstrating that qualitatively the calculated plasma density (b) and laser intensity (c) contour plots predict the shape of the modification structure (a).

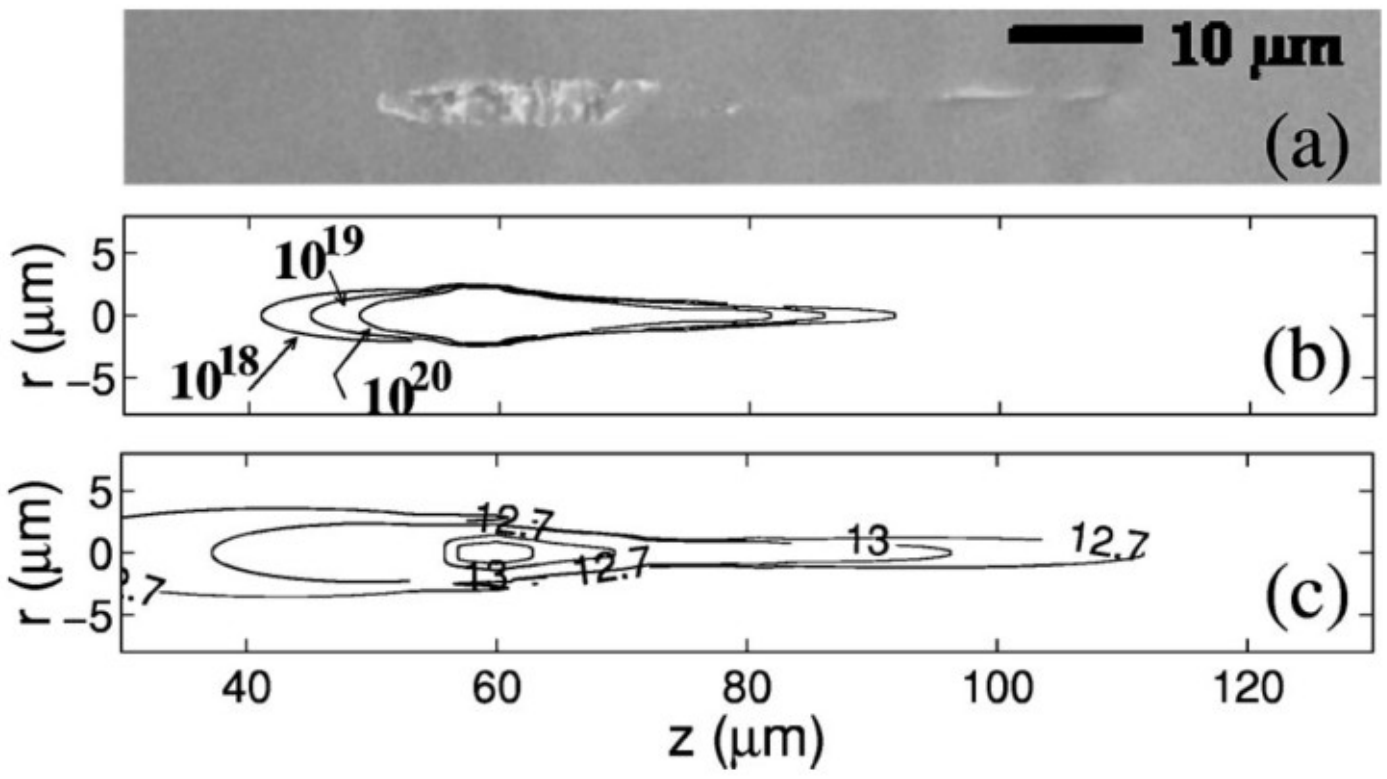

**Fig. 1.34:** Comparison between experiments and simulation [Couairon, 2005]. (**a**) SEM image of a laser-modified region produced by focusing pulses into fused silica [800 nm, 160 fs, 1 mJ, NA = 0.5]. (**b**) Numerically obtained contour plots of the free electron density for the irradiation parameters as in the experiments. (**c**) Computed intensity contour plots. The labels 12.7 and 13 in (c) indicate $5\times10^{12}$ and $10^{13}$ W/cm$^2$. (Reprinted figures with permission from [Couairon, 2005], A. Couairon et al., Phys. Rev. B **71**, 125435, 2005. Copyright (2005) by the American Physical Society)

The NLSE applied to nonlinear Bessel laser beams revealed specific features of such beam propagation in transparent solids [Arnold, 2015]. The authors have shown that the nonlinear Bessel-Gauss vortex beams provide, due to their helicity, an additional control for single-shot long-aspect-ratio microstructuring as compared to Bessel-Gauss beams with zero orbital angular momentum. Thus, for sufficiently large cone angles, a hollow plasma channel is generated whose radius and density increase with helicity and cone angle, respectively. Figure 1.35 presents simulation results for a cone angle of 11 degrees in BK7 glass, demonstrating the plasma and intensity channels through the whole width of the 100 µm sample [Arnold, 2015].

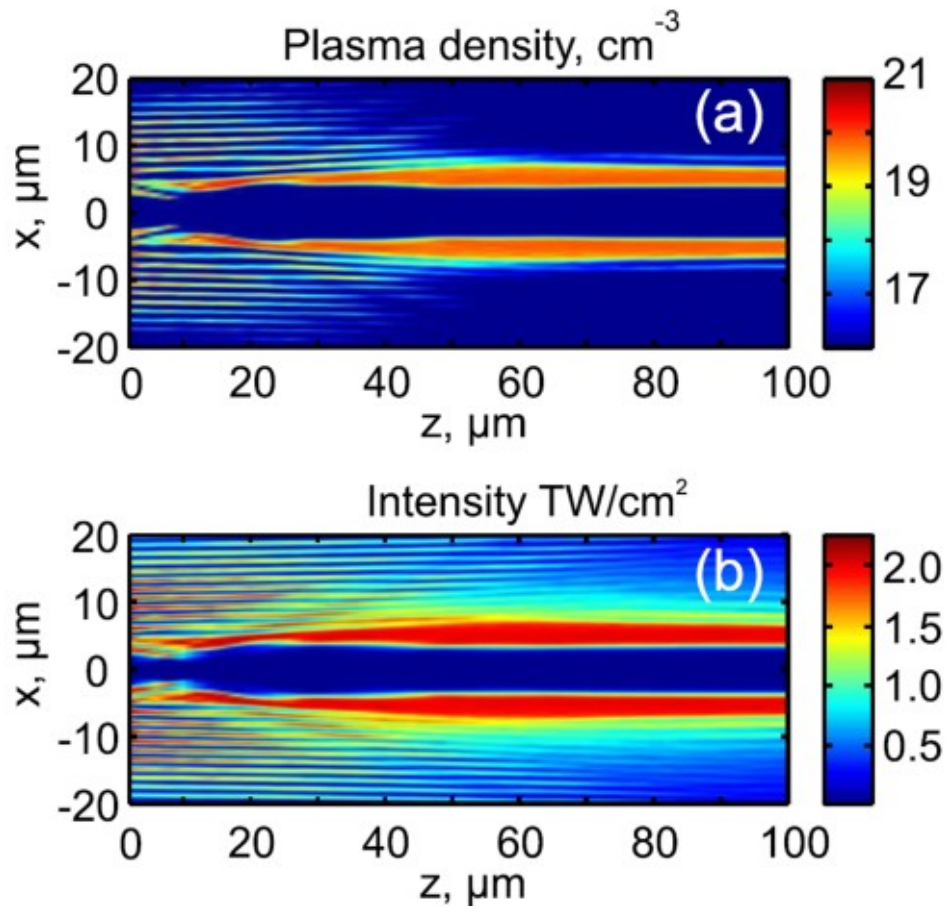

**Fig. 1.35:** Numerically simulated free electron density (**a**) and intensity (**b**) snapshots along the propagation axis for the Bessel-Gauss vortex beam of a large cone angle [Arnold, 2015]. The pulse duration is 1 ps and pulse energy is 37 µJ. The color bar indicates $\log_{10}(N_e)$, with the electron plasma density $N_e$expressed in cm$^{-3}$. (Used with permission of IOP Publishing, Ltd, from [Arnold, 2015], C.L. Arnold et al., Nonlinear Bessel vortex beams for applications. J. Phys. B: At. Mol. Opt. Phys. **48**, 094006 (2015); permission conveyed through Copyright Clearance Center, Inc.)

The validity of the NLSE in application to ultrashort laser beams focused inside transparent crystals and glasses can be subjected to questioning in many situations, that is conditioned by neglecting some terms upon its derivation from Maxwell's equations. For instance, the NSLE condition of a slowly varying envelope limits its applications to relatively long laser pulses, while the requirement of unidirectionality of the light beam makes it impossible to utilize this equation in the cases when a dense electron plasma is generated, causing light scattering to large angles. Although the NLSE can be generalized with additional terms to account for features of extremely short laser pulses, Maxwell's equations are free of the above limitations, however being time- and computational resources-consuming. To describe laser beam propagation through an absorbing ionizable medium, Maxwell's equations are appropriately supplemented to account for photoionization and the corresponding depletion of the laser beam, the Kerr effect, and plasma dispersion while the optical response of the plasma is described in the frames of a plasma fluid model [Popov, 2010]. The application of Maxwell-based models to the problems of volumetric modification of transparent materials gave a fresh impetus to a deeper understanding of this intricate phenomenon and pushed this field toward more quantitative numerical predictions.

For the description of ionizing beam propagation through a transparent medium of plasma clouds, Maxwell's equations and the plasma fluid equations are coupled via the free electron current [Popov, 2010 / Bulgakova, 2013b]. The complete set of equations is solved using a finite-difference time-domain (FDTD) algorithm, which, for large-scale modeling, demands high computer resources. However, at present such models represent the best choice for modeling tightly focused laser beams, which can potentially generate critical and overcritical plasma inside transparent dielectric materials. In 3D geometry, although minimizing the computational domain to relatively small interaction volumes, such a model enabled uncovering the mechanism of volumetric nanograting formation inside glass discovered by Shimotsuma et al. [Shimotsuma, 2003]. Buschlinger et al. [Buschlinger, 2014] introduced tiny inhomogeneities into the glass matrix and, via FDTD simulations, obtained periodic nanoplasma generation whose periodicity is dependent on the laser light polarization (Figure 1.36, left panel). Extending such a grating formation model to the computation of laser-induced hydrodynamics in laser-excited glass advanced the description of grating formation toward higher precision and into multi-pulse irradiation regimes (Figure 1.36, right panel [Rudenko, 2021]).

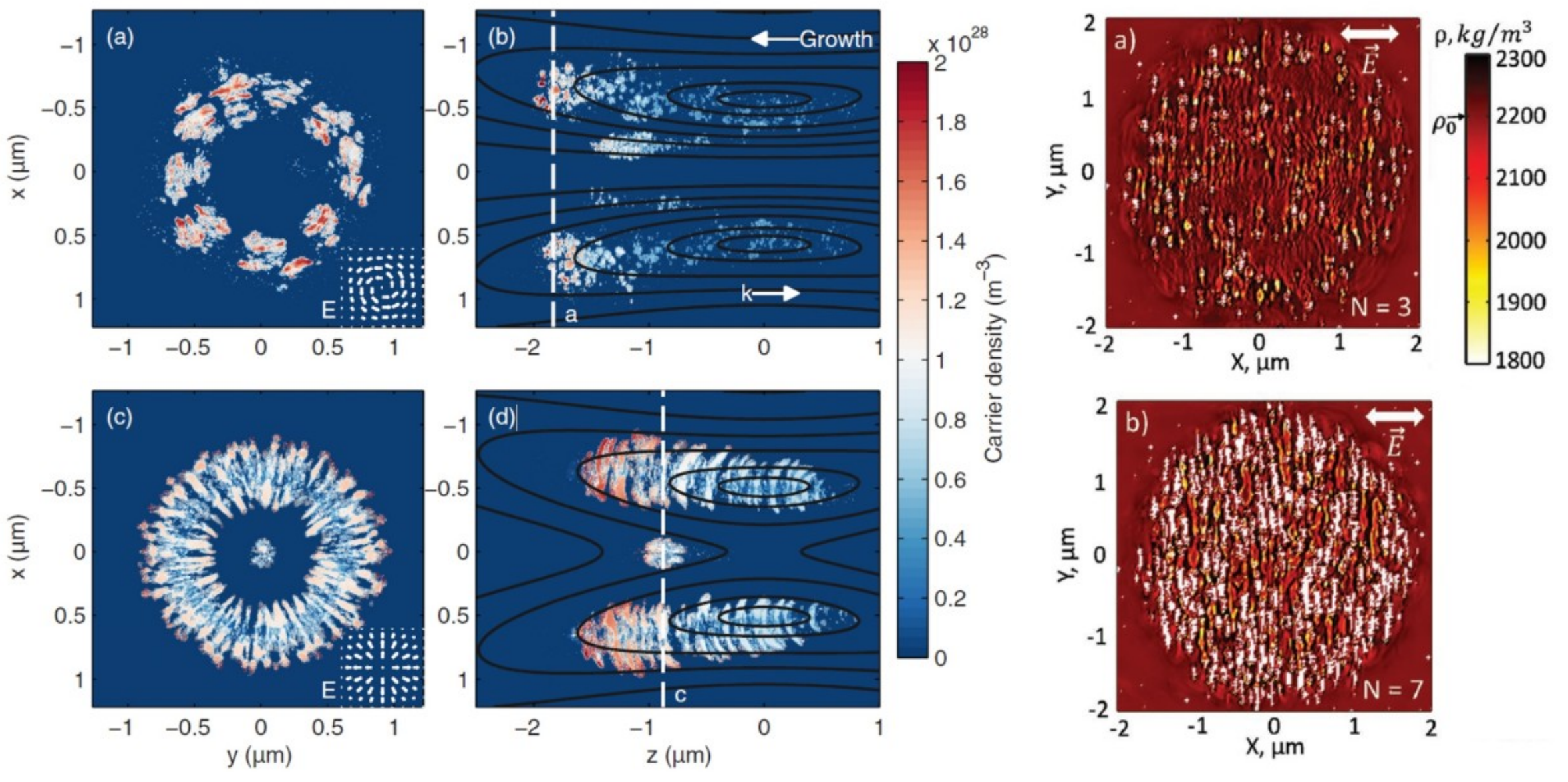

**Fig. 1.36:** (Left panel) Carrier density inside a glass volume with inhomogeneities irradiated with azimuthally and radially polarized laser beams (maximum field strength in the homogeneous case $E_0 = 1.7 \times 10^{10}$ V/m, NA = 0.5) propagating in the *z* direction [Buschlinger, 2014]. (**a**) and (**b**) Azimuthally polarized beam; (**c**) and (**d**) radially polarized beam. Solid lines in (b) and (d) correspond to equal field strength, as expected in the linear case. (Reprinted figure with permission from [Buschlinger, 2014], R. Buschlinger et al., Phys. Rev. B **89**, 184306, 2014. Copyright (2014) by the American Physical Society); (Right panel) Density snapshots in fused silica after irradiation by 3 (**a**) and 7 (**b**) laser pulses, which demonstrate the volumetric formation of periodic patterns [Rudenko, 2021]. The pulse energy is 250 nJ and the pulse duration is 200 fs. (Reprinted from [Rudenko, 2021] by permission from Wiley-VCH-Verlag GmbH: Adv. Opt. Mater. **9**, 2100973 (Genesis of nanogratings in silica bulk via multipulse interplay of ultrafast photo-excitation and hydrodynamics, A. Rudenko et al.), Copyright (2021))

The significance of the Maxwell-based models for investigations of volumetric modification of transparent materials is that such models enable the description of subtle details of laser-matter interaction such as, for example, the effects of the pulse front tilt (PFT). Figure 1.37 presents the extension of the model to multi-pulse irradiation of fused silica with the pulse containing PFT and the so-called lighthouse effect, also taking into account accumulation and re-excitation of defect states [Zhukov, 2019]. Via simulations, it was possible to explain the dependence of the modification structures on the direction of laser scanning and polarization, an effect known as non-reciprocal writing. Hence, the Maxwell-based models are a powerful tool for describing, predicting, and optimizing the conditions for laser direct writing of photonic structures inside optical materials for advanced applications.

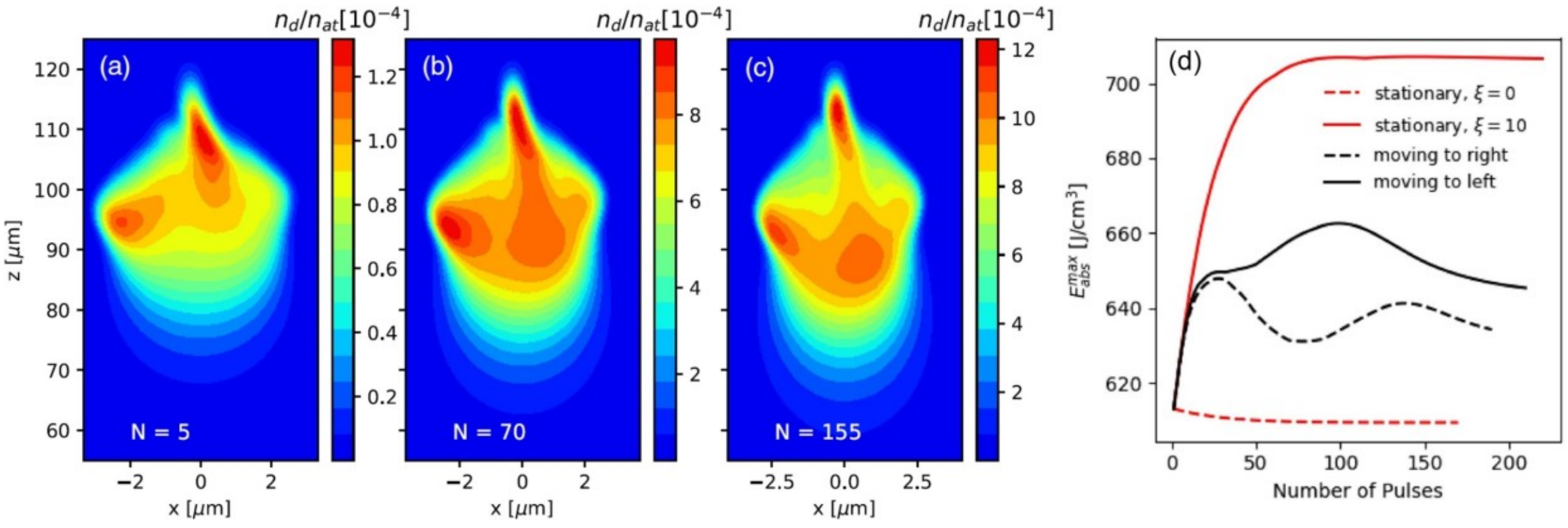

**Fig. 1.37:** (**a**)-(**c**) Numerically obtained distributions of the defect density $n_d$ normalized by the atomic density $n_a$ created by ultrashort laser pulses with PFT to the beginning of (a) 5th, (b) 70th, and (c) 155th pulse [Zhukov, 2019]. (**d**) Maxima of the absorbed laser energy as a function of the number of laser pulses. The red and black lines correspond to stationary and scanning irradiation, respectively. The solid and dashed red lines show the results of simulations with and without introducing into the model a change in refractive index due to defect accumulation with $\xi = 0$ corresponding to the absence of defect accumulation. For both moving cases, $\xi = 10$. (Reprinted with permission from [Zhukov, 2019] © Optical Society of America)

One more mechanism of laser energy absorption inside transparent materials should be mentioned, which was uncovered using particles-in-cell simulations [Ardaneh, 2022]. The authors have demonstrated that, when an overcritical plasma is formed, the resonance of free electron plasma waves at the critical surface plays an important role in the absorption of light from femtosecond Bessel beams inside dielectrics. This collisionless process can be responsible for a significant part of the absorbed laser energy.

In this section, a plethora of the processes connected with ultrafast laser excitation of a free electron plasma has been overviewed with insights, where possible, into their interrelation, based on considerable progress in theoretical modeling of laser-excited processes, mostly during the stage of the laser pulse propagation through a transparent material sample. However, the laser-induced modification of bulk dielectric materials is not finished by this stage, and the post-irradiation effects play a significant role that is considered in the next section.

## 3.2 Energy Relaxation

As shown in the previous section, laser light absorption with generation of a free electron plasma is at the core of volumetric modification of the structure of dielectric materials. However, the free electron plasma is a short-living transient state that is recombining with the transfer of its energy to the material matrix. The absorbed energy relaxation proceeds in several stages, which can coexist in time [Stoian, 2003 / Gattass & Mazur, 2008 / Bulgakova, 2013a]:

- The electrons excited into the conduction band are thermalizing due to collisions between them that, for dielectrics, can take dozens of femtoseconds [Kaiser, 2000].
- In collisions with ions and atoms of the material lattice, the electron plasma transfers its energy to the lattice (electron-phonon coupling). During the laser pulse action, free electrons remain hot, having the source of laser energy obtained by gradual absorption. When the laser pulse leaves the excited plasma region, the electrons cool via collisions with the lattice, moving in the energy scale toward the bottom of the conduction band, and finally recombine.
- Recombination of the laser-excited electrons proceeds, *depending on the material properties*, either by direct recombination to the valence band or via a trapping mechanism when the electrons localize in intra-bandgap states, such as self-trapped excitons (STEs) [Martin, 1997]. STEs are metastable states that further decay toward the valence band with some portion transferring to the long-living electronic defects [Petite, 1999]. In fused silica, for example, the typical long-living defects are color centers and non-bridged oxygen deficiency centers. The characteristic times of electron recombination differ strongly for different materials. For fused silica, it is believed that this time is ~150 fs [Martin, 1997] while, in laser excited $Al_2O_3$, the electrons can survive at the bottom of the conduction band for a hundred picoseconds [Daguzan, 1996].
- Both electron-lattice coupling and electron recombination result in the localized heating of the laser-affected region, considering that the heat transfer is negligible on the picosecond timescale. The steep temperature gradients between the laser-excited zone and surrounding cold material culminate in the emission of stress waves, which dissipate in the nanosecond timescales [Mauclair, 2016].
- Heat dissipation is the slowest process in the chain of the relaxation phenomena listed above. The characteristic time of heat propagation by the distance $x$ can be estimated as $\tau_{\text{heat}} \sim x^2 \rho c/\lambda$ where $\rho$, $c$, and $\lambda$ are the material density, heat capacity, and thermal conductivity, respectively. For fused silica as an example, a heat wave propagates by ~1 μm during 1 μs, while the complete dissipation of heat from the focal volume of a several-micrometer-sized diameter may take up to several dozens of microseconds [Bulgakova, 2013a]. This determines the manifestation of the heat accumulation effect as a function of the pulse repetition rate and the laser scanning velocity.

Thus, the post-irradiation processes are of high importance for the final imprinting of the desired modification structures for technological applications. Returning to the case of fused silica, the final transfer of the absorbed energy into the heat of the material lattice is determined by the decay of the STEs. Grojo et al. [Grojo, 2010] obtained, via fitting the experimental result in the pump-probe configuration by exponential functions, that the STE decay can be described by a biexponential function with the characteristic times of ~34 ps and ~338 ps. Figure 1.38 [Grojo, 2010] presents the measured STE population with accompanying residual plots obtained by nonlinear least-square fitting. Thus, the final heating of the laser-affected region may continue up to several hundreds of picoseconds, when the heat can still be considered as strongly localized in view of the negligible heat diffusion process on such a time scale. We note that the final heating of $Al_2O_3$, although not mediated by trapping, also requires at least 100 ps [Daguzan, 1996].

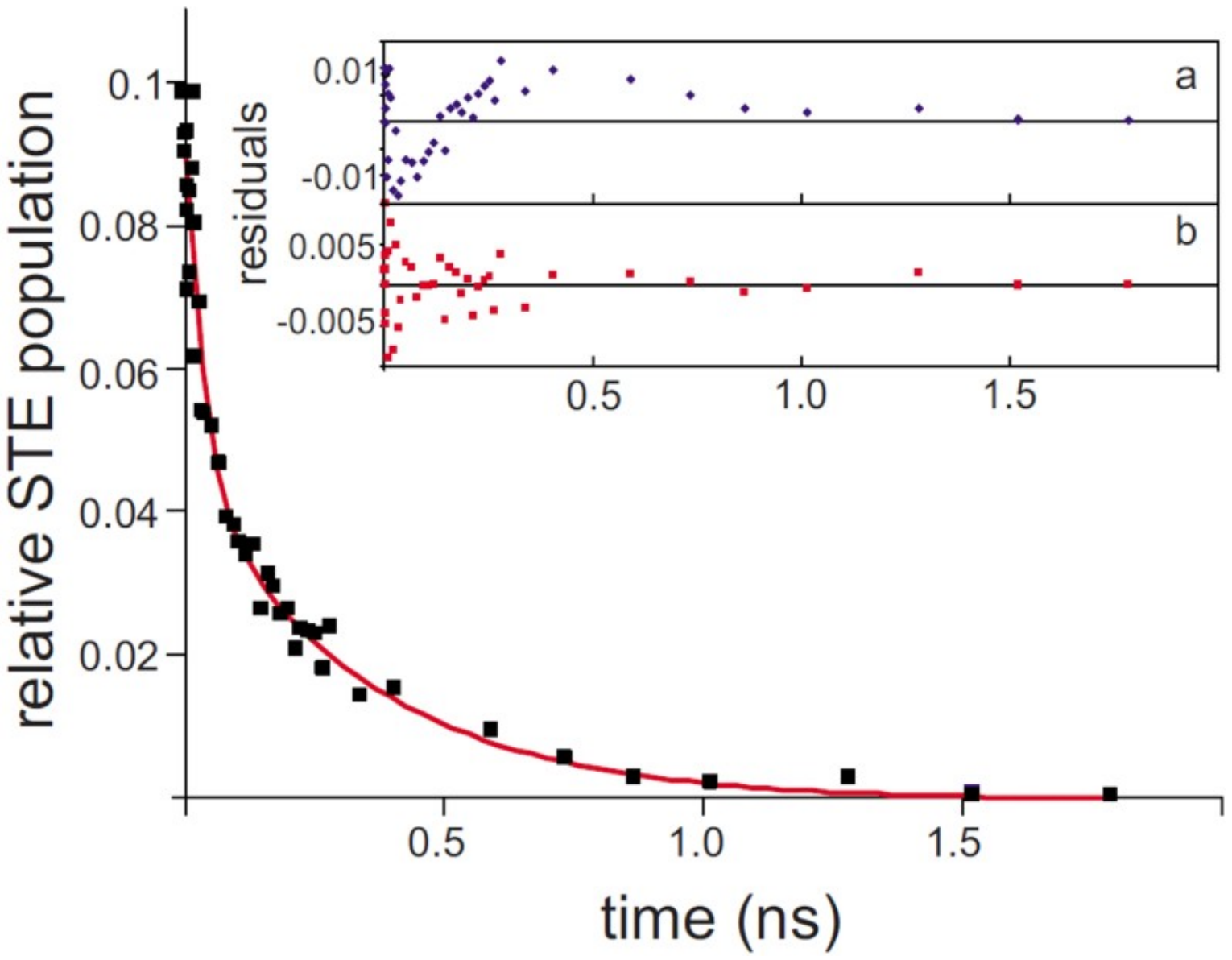

**Fig. 1.38:** Relative STE population, determined from multiphoton-probe absorption as a function of time [Grojo, 2010]. The pump energy was 70 nJ and the probe energy 52 nJ. The inset shows the residuals obtained from (a) single exponential and (b) biexponential nonlinear least-squares fits. The curve through the data points in the main figure is the result of the biexponential fit. (Reprinted figure with permission from [Grojo, 2010], D. Grojo et al., Phys. Rev. B **81**, 212301, 2010. Copyright (2010) by the American Physical Society)

The localized, swiftly heated zone inside a bulk material is first relaxing via the emission of stress waves [Mauclair, 2016 / Sakakura, 2017 / Koritsoglou, 2024], while the heat transfer may take the time up to dozens or hundreds of microseconds. The stress wave evolution under such conditions can be studied by time-resolved phase-contrast microscopy (PCM) and photo-acoustic modelling [Mauclair, 2016], pump-probe polarization microscopy, the Transient Lens (TrL) method [Sakakura, 2017], or time-resolved shadowgraphy [Koritsoglou, 2024]. For examples see Chap. 13 (Bonse). The evolution of the stress wave emitted from the laser-excited region due to high temperature (and, hence, pressure) gradients is shown in Figure 1.39, left panel [Sakakura, 2017]. It is demonstrated that measuring birefringence makes the visualization technique very efficient and precise. Similar to measurements with other visualization methods, the stress relaxation time is reported to be within 10 ns. The last stage, heat dissipation from the laser-affected zone, can also be analyzed using birefringence measurements (Figure 1.39, right panel) [Sakakura, 2017]. As is expected from estimations of heat diffusion, cooling of the laser-heated zone continues to a dozen(s) of microseconds, and, after this final relaxation stage, the modification structure is finally imprinted into the bulk of transparent material. Although the rates of the relaxation/dissipation processes depend on the material properties, the above-described evolution is inherent for inorganic transparent materials whose laser-induced modification is used for practical applications.

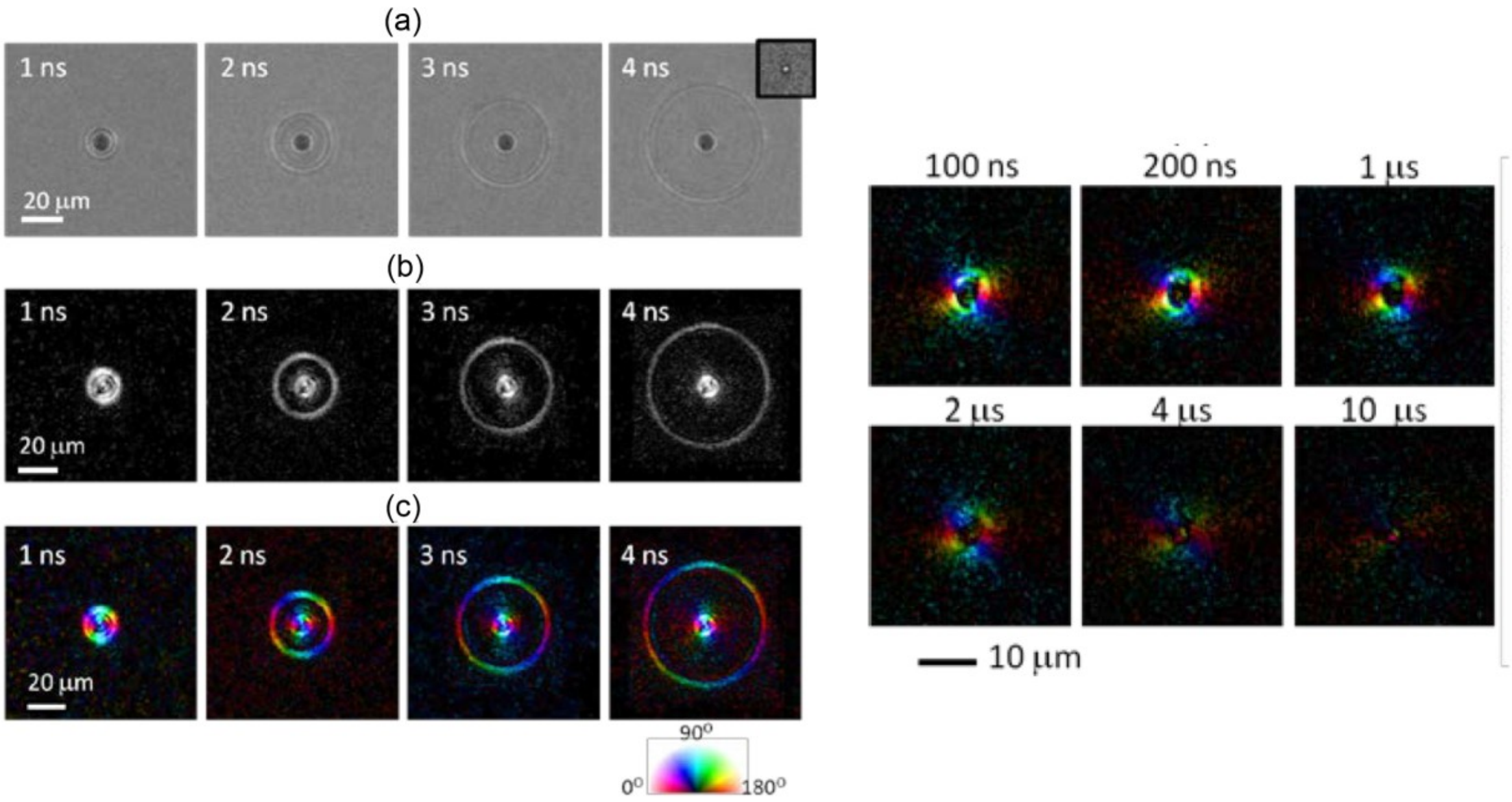

**Fig. 1.39:** (Left panel) Images of fs-laser-excited soda lime glass at different time delays after laser irradiation [Sakakura, 2017]. (**a**) Transmission images. The inset shows the transmission image of the completely relaxed modification. (**b**), (**c**) Distributions of birefringence: (b) for retardance, and (c) for the slow axis of the birefringence. The relation between the slow axis and the color is given by the colored semicircle. (Right panel) Birefringence distributions in a longer time range after photoexcitation inside soda lime glass with a focused fs laser pulse [Sakakura, 2017]. The laser pulse energy was 2 μJ. (Reproduced from Ref. [Sakakura, 2017] with permission from Japan Laser Processing Society)

## 3.3 Selected Applications

### Writing of Optical Waveguides

Optical waveguides are of paramount importance in photonics. Their fabrication has been demonstrated for a wide set of transparent (dielectric) materials, including glasses, crystals, or polymers, always taking benefit from tailoring a locally confined high refractive index region that is surrounded by a lower refractive index. In the *ray-optical model* and for sufficiently sharp interfaces separating the two refractive index regions, total reflection can manifest for light entering the transition region from the higher to the lower refractive index, thus, confining the propagating light into the region of higher refractive index. The *wave-optical model* imposes a lower limit of the extent of the high-refractive index regions (for enabling wave propagation) that crucially depends on the optical wavelength and the amplitude of the refractive index change. It turns out that micrometer sized regions of refractive index changes $\Delta n/n$ of less than a few percent are suitable to ensure efficient low-loss single-mode beam transport. The manufacturing of such light-guiding regions might be obtained either by depositing a high refractive index material on a substrate or by creating a suitable localized refractive index modification of the material, for example via a laser direct writing process in the bulk of dielectrics. The latter can be realized in an elegant way by tightly focusing ultrashort laser pulses into the bulk of the dielectric material. In the vicinity of the focus, nonlinear absorption can trigger structural changes that eventually leads to a locally increased refractive index.

Conceptually, the writing of optical waveguides is very appealing as it allows a flexible integration and inter-connection of optical components on the micrometer dimensions, thus providing the building block technology for extending the well-established microelectronics into three-dimensional all-optical microchips and devices for integrated micro-optics. As of today, the technology is the most efficient method for the 3D fabrication of complex photonic networks. It can be performed as a single-step process, does not require a clean-room environment, and the resulting waveguiding region is neatly sealed and protected in the bulk of the processed samples.

Typically, laser-written optical waveguides are manufactured by focusing a train of ultrashort laser pulses (with sub-ps duration) in the host substrate, while simultaneously translating the substrate in a direction transverse to the propagation direction of the focused laser beam, see Figure 1.40a,b [Mermillod Chapter, 2023]. Nevertheless, the photoinscription of waveguides is also possible in the longitudinal configuration, i.e., with the sample translated in the direction of laser beam propagation, see Figure 1.40c.

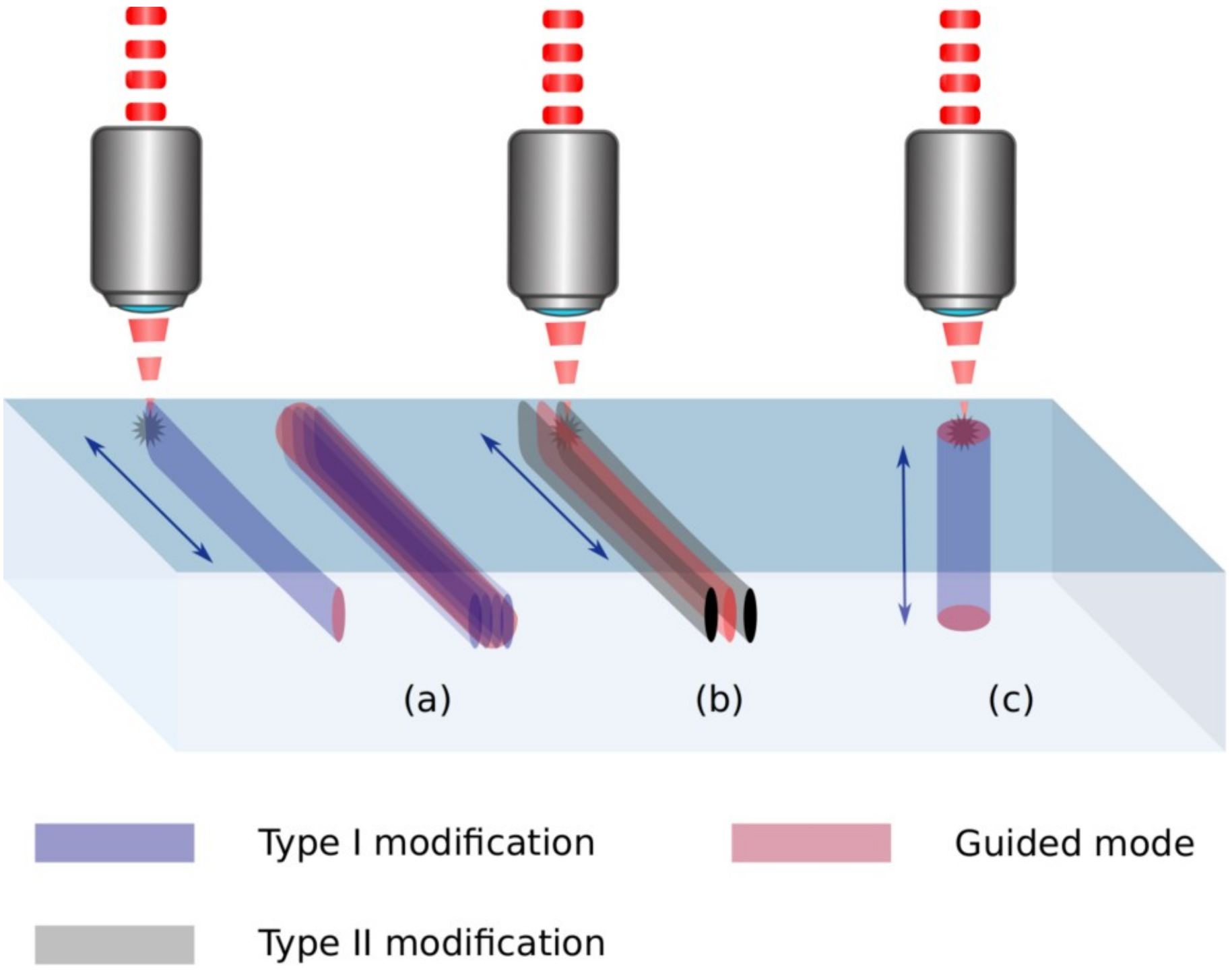

**Fig. 1.40:** Experimental approach used for ultrafast waveguide writing in tight focusing conditions [Mermillod Chapter, 2023]. Schematic representation of a waveguide based on (**a**) type I modifications induced in a transverse configuration, **(b)** type II modifications induced in a transverse configuration, and **(c)** type I modifications generated in a longitudinal writing scheme. The double-arrows indicate the direction of the sample translation. (Reprinted from [Mermillod Chapter, 2023], O. Ghafur et al., in *Ultrafast Laser Nanostructuring*, pp. 759 – 786, 2023, Springer Nature)

The structural response of the laser processed dielectric material and the final refractive index stratigraphy may be different and complex. In the so-called *type I regime*, the high refractive index region manifests in the photoexcited volume, as sketched in Figures. 1.40a and 1.40c. In contrast, in the *type II regime* (see Figure 1.40b), the photoinscription scenario is different [Mermillod Chapter, 2023]: the laser-induced structural modifications lead to material damage [Gross, 2015]. The light-guiding region then appears at some distance from the laser focus, typically formed through a field of mechanical stress [Chen, 2014]. In such a scenario, the type II microstructures play the role of a local “stressor” [Mermillod Chapter, 2023]. In addition, the laser-induced refractive index modifications might be negative. In this case, light can be guided in between such modifications as well, which is the so-called depressed cladding waveguide.

Historically, the first optical waveguide writing using ultrashort laser pulses was demonstrated by Davis et al. in 1996, realizing type I modifications in fused silica [Davis, 1996]. Since that, considerable progress has been made in understanding, optimizing and refining this method of waveguide production. It turned out that the type I modification in fused silica is mainly caused by a laser-induced structural rearrangement of the glass network, where 9-membered rings of $SiO_2$ are transformed into 3-membered rings [Chan APA, 2003], potentially accompanied by the formation of other laser-induced localized electronic defects.

In 2003 Nolte et al. reported the first fs-laser fabrication a three-dimensional 1×3 optical waveguide beam splitter as another important cornerstone towards integrated optics [Nolte APA, 2003], see Figure 1.41. Panel (a) sketches the experimental approach for writing the 1×3 waveguide splitter in a fused silica sample. Panel (b) provides near-field intensity distributions, as simultaneously measured at the exits of all splitters, indicating an almost equal light-splitting ratio among the three branches. Panel (c) provides a quantitative measurement of the local refractive index distribution of one representative splitter branch. The quantification reveals maximum refractive index changes of $\Delta n \sim 1.3\times10^{-3}$ for the type I modification in the silica, here.

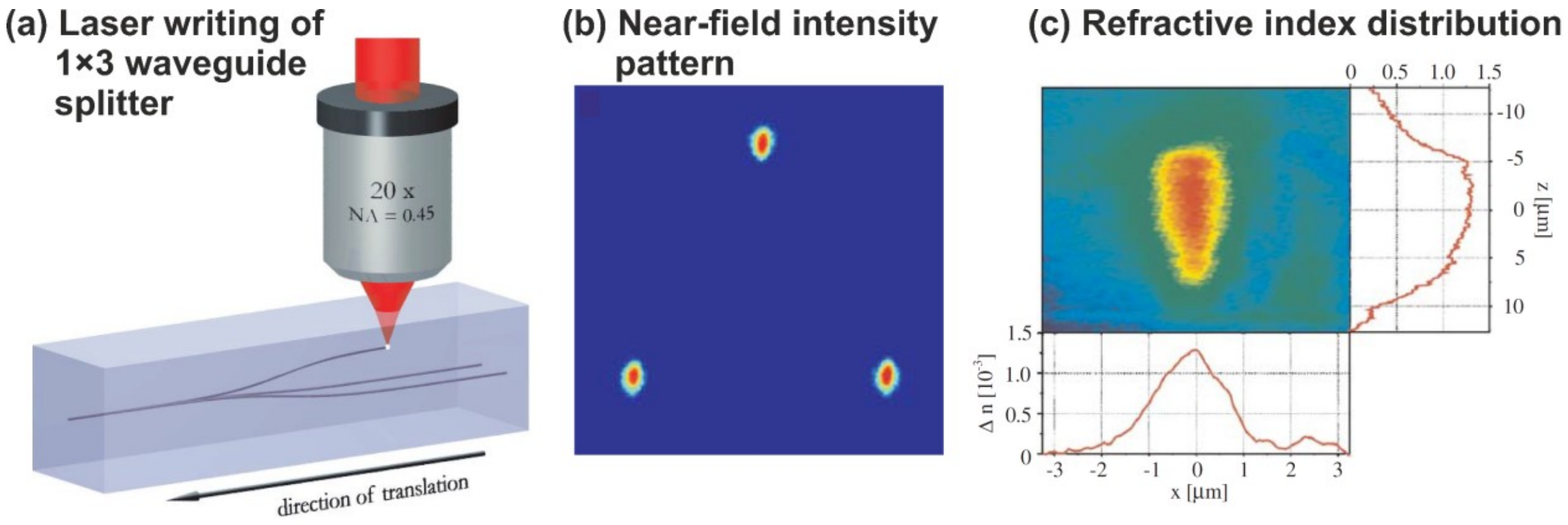

**Fig. 1.41:** **(a)** Scheme of the experimental approach for transversally writing a 1×3 waveguide splitter in a fused silica sample [Nolte APA, 2003]. **(b)** Near-field intensity distribution, as simultaneously measured at all splitter exits. (c) Measured distribution of the local refractive index of one representative splitter branch. (Reprinted from [Nolte APA, 2003], S. Nolte et al., Femtosecond waveguide writing: a new avenue to three-dimensional integrated optics, Appl. Phys. A **77**, 109 – 111, 2003, Springer Nature)

It is important to underline that the sign, magnitude and local distribution of the laser-induced refractive index modification strongly depend on (i) the laser processing parameters and (ii) the specific structural, thermophysical, and thermochemical response of the irradiated dielectric material. Regarding the first aspect, it was found that high repetition rate laser processing, leading to heat-accumulation in the focal region, can be beneficial for the writing of waveguides [Eaton, 2005 / Eaton AO, 2008]. For example, for BK7 glass, low-repetition rate (few kHz or below) written modifications usually show a negative refractive index change that does not support the waveguiding of light. However, with the help of high-repetition rate lasers operated in the few MHz range, low-loss (0.3 dB/cm propagation loss) fs-waveguides were demonstrated in 2008 [Eaton AO, 2008]. These waveguides in BK7 are based on a stress-related type II material modification that increases the refractive index in a pronounced location close to the damage created in the focal region, see the example in Figure 1.42a,b. The complementary optical transmission and phase-contrast microscopy reveal the formation of a ~5 µm diameter core modification zone of increased refractive index, which is embedded within a complex shaped structure of permanent modification. This core modification is able to guide laser ground mode radiation ($\lambda$ = 1.55 µm) having a mode field diameter of ~11 µm, see Figure 1.42c.

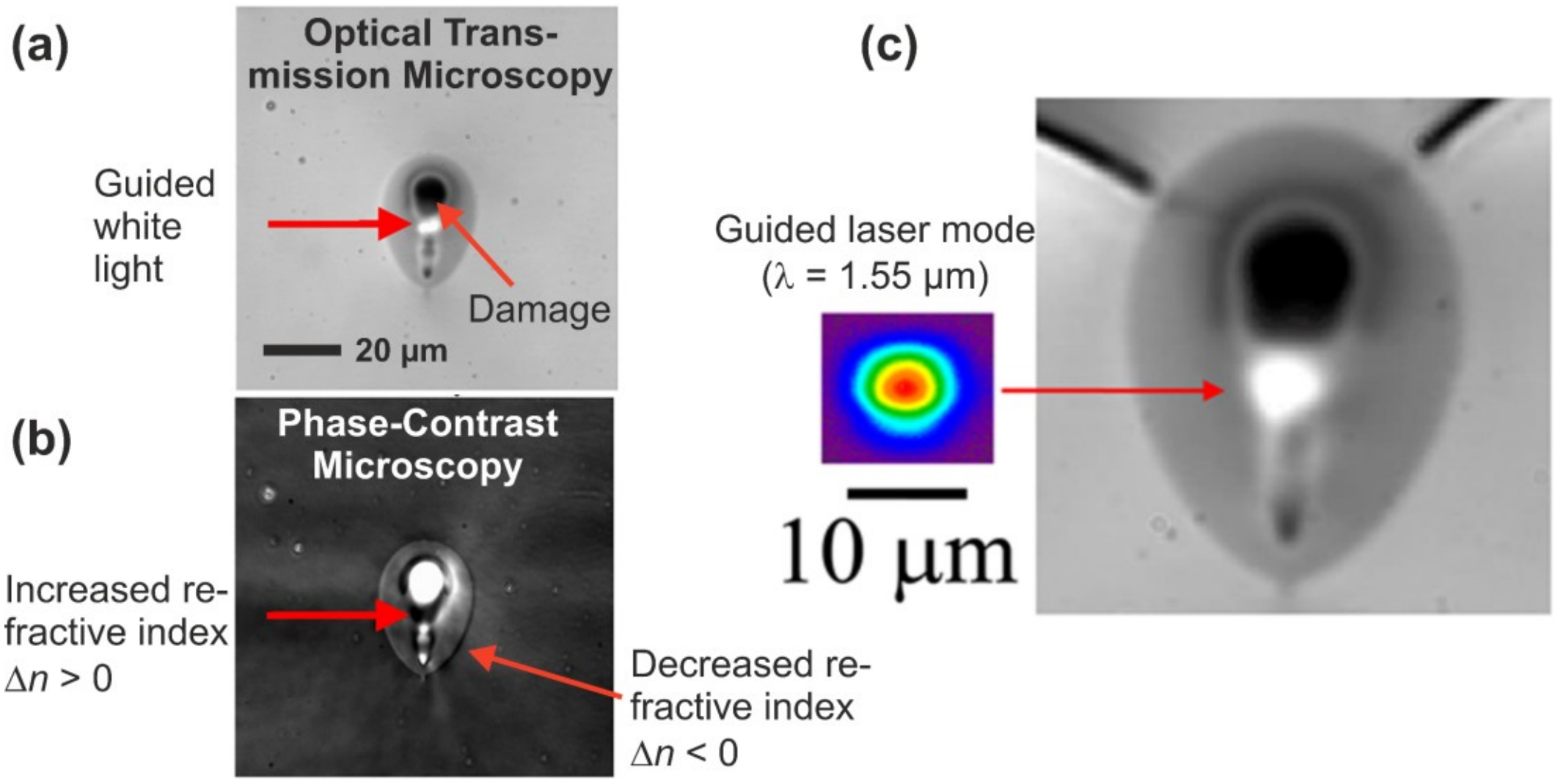

**Fig. 1.42:** Cross-sectional views through transversally fs-laser-written optical waveguides in BK7 glass (Schott AG) [1045 nm, 300 fs, 375 nJ pulse energy, 1 MHz, 1.5 mm/s writing speed] [Eaton AO, 2008]. The fs-laser radiation was incident from the top. **(a)** Optical transmission microscopy (OTM) image [150 µm writing depth]. **(b)** Phase-contrast microscopy (PCM) image [150 µm writing depth]. **(c)** OTM image [75 µm writing depth] and corresponding guided laser mode profile at $\lambda$ = 1.55 µm featuring a mode field diameter of ~11 µm. (Reprinted with permission from [Eaton AO, 2008] © Optical Society of America)

Regarding the relevance of thermophysical and -chemical effects in laser processing of glasses, it was found that heat-accumulative laser processing can facilitate the local crystallization in glasses, featuring the formation nanocrystallites [Miura, 2000 / Cao, 2008]. Similarly, the overall extended heating/melt duration (compared to the non-accumulative regime) allows lighter chemical elements to diffuse out of the laser focal region [Kanehira, 2008 / Fernandez, 2018]. Thus, gradients of the glass composition are generated that are also affecting the local refractive index distribution. The latter effects are particularly relevant for multi-component glasses (such as BK7) and can explain the difficulties for writing optical waveguides in these complex materials.

## Inscription of Bragg Gratings

The precise control of light propagation inside optical fibers is essential for a variety of applications including optical communication, sensing and fiber lasers. Here, in-fiber components are employed for filtering, reflection and mode conversion. The most prominent one is the so-called *fiber Bragg grating* (FBG), a periodic modulation of the refractive index, discovered by Hill et al. [Hill, 1978].

The working principle of a FBG is sketched in Figure 1.43. When a spectrally broad source of light is coupled into a fiber with a FBG, certain wavelengths $\lambda_{Bragg}$ are selectively back-reflected at the FBG's refractive index modulation (representing a phase grating) if they fulfill the (Bragg) condition $\lambda_{Bragg} = 2 \times n_{core} \times \Lambda$, where $n_{core}$ is the effective refractive index of the fiber core and $\Lambda$ is the grating period of the refractive index modulation along the fiber axis. The spectral width $\Delta\lambda_{Bragg}$ of the reflected light depends on the amplitude (depth) of the FBG's refractive index modulation $\delta n$, the length of the refractive index modulation $L_{FBG}$, and the fraction $\eta$ of light

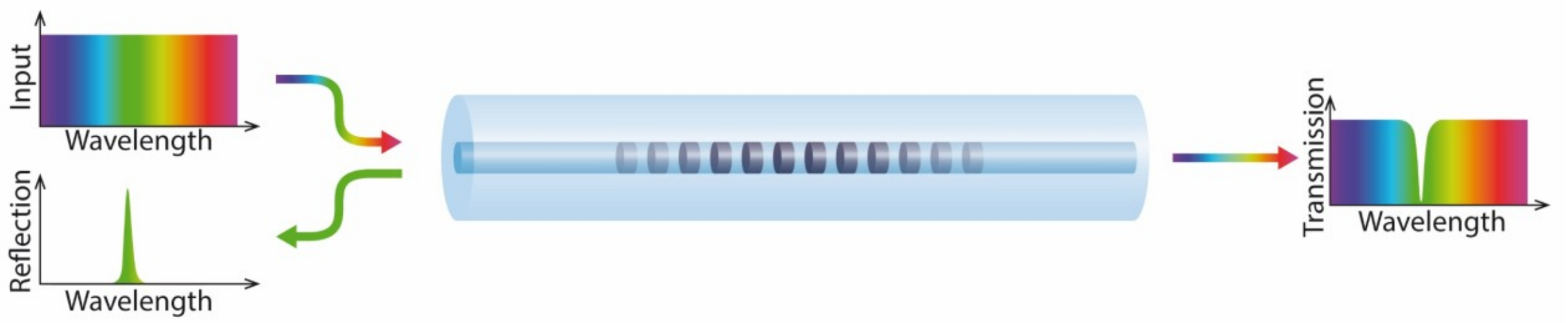

**Fig. 1.43:** Scheme of the working principle of a fiber Bragg grating. (Figure adapted from [Kraemer, 2023])

When the spectrum of the light in a FBG is monitored in transmission, these spectral components are missing, while in reflection a narrow-band spectral feedback can be generated via constructively superimposing the Fresnel reflections from the refractive index (phase) grating. Such an optical Bragg reflector can be used for example as a mirror for building fiber lasers (see the example below), as spectral filter, etc.

The breakthrough of this technology started with the inscription of FBGs by side illumination of the fiber with a two-beam interference pattern of a UV laser [Meltz, 1989]. Despite all the developments and applications realized, one of the bottlenecks of UV inscription techniques is the required photosensitivity, typically addressed by Germanium doping of the fiber core or hydrogen loading [Othonos, 1997]. Nevertheless, this imposes restrictions on the fiber material as well as its geometries.

The localized energy deposition based on nonlinear absorption of ultrashort laser pulses inside transparent materials, as detailed in Sect. 3.1 of this chapter, can be effectively harvested to overcome those restrictions. Mihailov et al. were the first to demonstrate FBG inscription using 800 nm femtosecond laser pulses [Mihailov, 2003]. Based on this approach, different inscription techniques have been developed over the past decades. There are, on the one hand, direct femtosecond inscription techniques, which provide high flexibility. Inscribing the periodic modifications line by line [Zhou, 2010 / Antipov, 2016] allows to individually adapt grating period and strength of each grating plane [Goebel, 2018 / Su, 2024]. The same flexibility is achieved with the point by point or plane by plane inscription techniques [Martinez, 2004 / Marshall, 2010 / Gao, 2024]. While mode coupling and scattering losses might become more critical in this case, this approach is significantly faster and allows e.g. flexible spectral profiles by varying the cross-sectional position of the modification relative to the fiber core [Williams, 2011 / Ioannou, 2023].

In addition to these direct inscription techniques, the phase mask technique can be also applied to imprint the periodic pattern into the fiber core [Mihailov, 2003]. It enables low loss gratings with large cross-sections and offers a highly reproducible and stable inscription process, nevertheless at the expense of flexibility. However, some degree of flexibility can be regained via beam shaping or post-processing steps [Voigtländer, 2011 / Shamir, 2016]. Scanning the laser beam across phase mask and fiber allows to extend the FBG in size and length easily.

There exists a huge variety of different types of fiber gratings, for an overview see e.g. [Canning, 2008]. This is especially important for sensing applications, where the desired type of modification can be employed e.g. for operation in harsh environments [Mihailov, 2017]. More details about fs FBG inscription can be found in previous review papers on this subject [Nikogosyan, 2007 / Thomas, 2012 / Berghmans, 2014 / Zhang, 2024].

Another application we want to highlight here is the use of FBGs as in-fiber cavity mirrors for high-power fiber laser systems. The independence from photosensitivity allows e.g. the inscription directly into the active fiber, thus reducing the number of splices in a monolithic fiber laser system [Wikszak, 2006 / Bernier, 2009a / Becker, 2011]. Based on this approach, a fiber laser with 1.9 kW output power has been demonstrated, proving the high power durability of fs FBGs [Kraemer, 2019]. Even higher output powers of 5 kW have been obtained with FBGs inscribed through the fiber coating into passive fibers with femtosecond laser pulses (see Fig. 1.44) [Kraemer, 2020]. In this case the low reflectivity FBG showed a thermal slope around 12 K/kW, and the high-reflectivity grating only 1 K/kW. The gratings showed excellent high-power stability. No modal instabilities have been observed and further power scaling was only limited by stimulated Raman scattering.

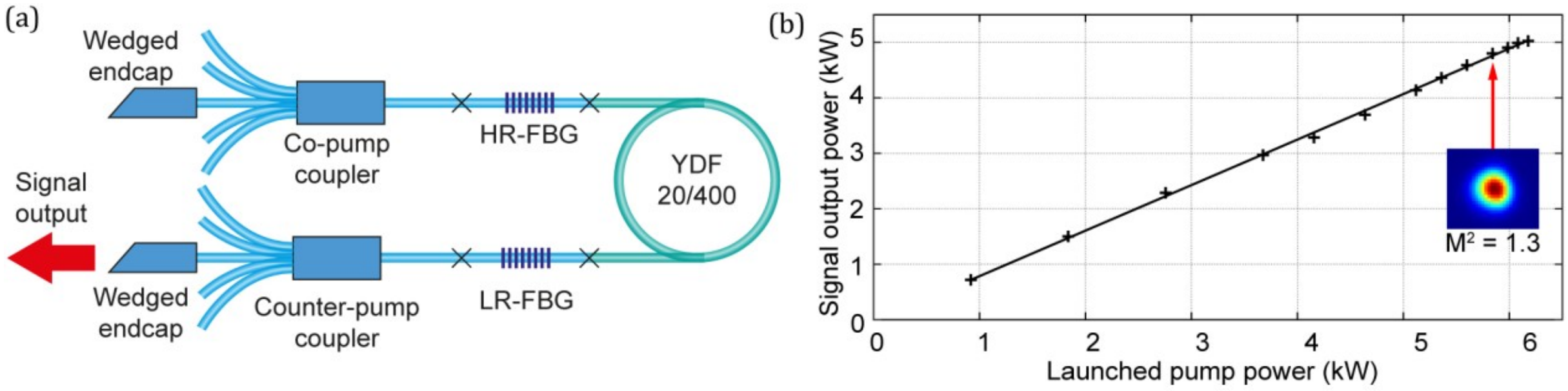

**Fig. 1.44: (a)** Schematic setup of the bidirectionally pumped all-fiber oscillator [Kraemer, 2020]. Both FBGs have been inscribed using the fs phase mask technique through the coating of the fiber. **(b)** Signal output power in dependence of the launched pump power yielding 83% slope efficiency. The mode field image taken at 4.8 kW output power shows single mode operation. (Reprinted with permission from [Kraemer, 2020] © Optical Society of America)

Chirped fiber Bragg gratings exhibiting a varying grating period do not only broaden the spectral response, they can also control the dispersive response of the reflected signal by introducing a time delay between spectral components reflected at different spatial positions along the grating. Fs-chirped FBG are realized by varying the grating period during inscription [Antipov, 2016] or using chirped phase masks [Bernier, 2009b]. Apart from dispersion compensation, they have been successfully implemented e.g. for the chirped pulse excitation of quantum dots [Remesh, 2023].

Apart from Bragg gratings in fibers their counterpart in bulk glass are so-called *volume Bragg gratings* (VBGs). Typical applications include spectral beam combining or separation as well as laser diode stabilization. For the latter, the VBG reflects a small fraction of the emitted laser light at the Bragg resonance wavelength back into the diode [Andrusyak, 2009], setting the laser wavelength. For high-power laser diodes still a sophisticated temperature stabilization of the VBG might be necessary due to residual absorption of the VBG. Again, the independence from photosensitivity enables the use of pure fused silica with low OH content in order to minimize residual absorption. We could demonstrate extremely low residual drifts of only 0.53 pm/W up to output powers of 170 W [Richter, 2017]. In this case the fs phase mask scanning technique has been applied to obtain VBGs with extended size (see Fig. 1.45). The same technique can be applied in other materials like fluoride glasses [Talbot, 2020] or for the inscription of chirped volume Bragg gratings [Perevezentsev, 2023] as required for broad bandwidths or dispersion control.

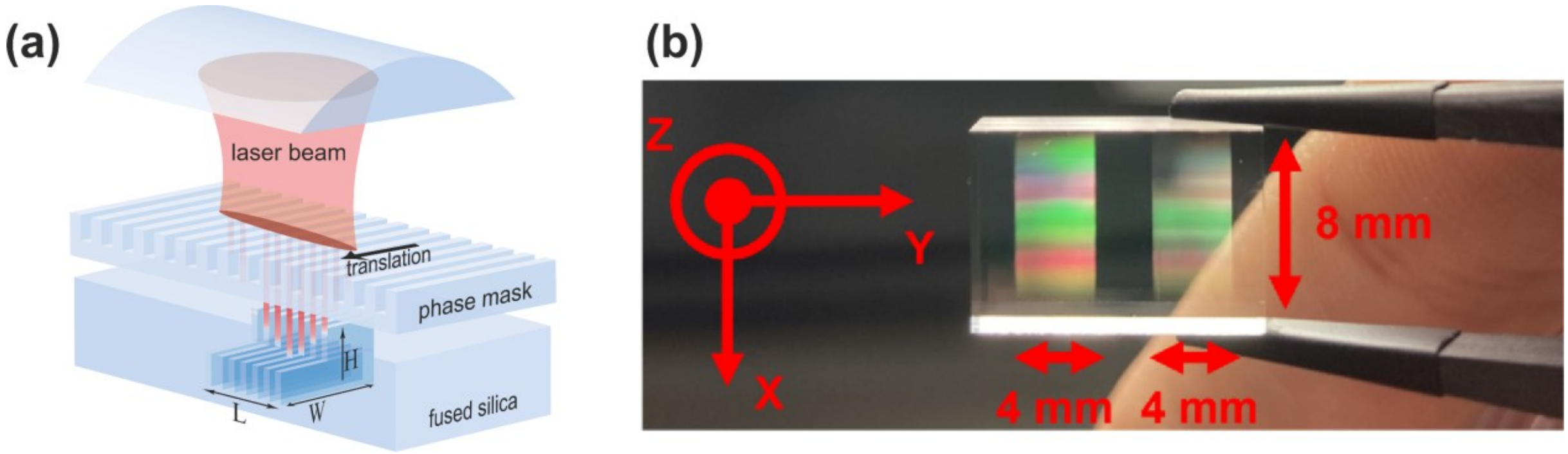

**Fig. 1.45: (a)** Schematic setup to inscribe a VBG into transparent material by applying the phase mask scanning technique and ultrashort laser pulses. (Reprinted with permission from [Richter, 2017] © Optical Society of America); **(b)** Photograph of a VBG diffracting ambient light. The VBG was inscribed into fused silica for mid-IR operation. (Reprinted with permission from [Talbot, 2024] © Optica Publishing Group)

## Microfluidics: Towards a Lab-on-a-Chip

Another direction has emerged around the turn of the millennium in the field of microfluidic "lab-on-a-chip" applications, for example as *micro total analysis systems* (µ-TAS), where miniaturized versions of microfluidic transport channels and reservoirs are combined with chemical or biomedical analytical functions. Such a miniaturization of systems for chemical analysis can lead to an efficient use of resources with respect to the sample size, consumption of reagents, and analytical response times.

Femtosecond lasers can enable such technologies via flexible 3D laser direct writing in the bulk of (photosensitive) glasses [Marcinkevicius, 2001 / Masuda, 2003 / Kim, 2009 / Sugioka, 2014 / Sima, 2018]: in a first manufacturing step (step 1), the focused fs-laser irradiation can locally induce permanent structural change in the material in the shape of a micrometric voxel. That modified region can be spatially extended into the desired microfluidic geometry via 3D scanning processing, including still latently patterned microfluidic channels, reservoirs for reagents, reactor chambers, etc., see Figure 1.46a [Sugioka, 2014]. In a second step (step 2), a heat treatment in a specified temperature regime and for a certain time lapse for further developing the structure, see Figure 1.46b. A third manufacturing step (step 3) is realized via ultrasonic-assisted chemical etching. Since the fs-laser modified and subsequently developed material typically has a higher etch removal rate than the pristine (non-irradiated) glass for HF-based acidic solutions, the latent 3D microfluidic chip geometry can be "photo-chemically sculptured" into the volume of the glass chip material. The ultrasonic supports the liquid transport and reagent exchange processes during the chemical etching. Through this, the final processing result is obtained in the desired lab-on-a-chip geometry, see Figure 1.46c.

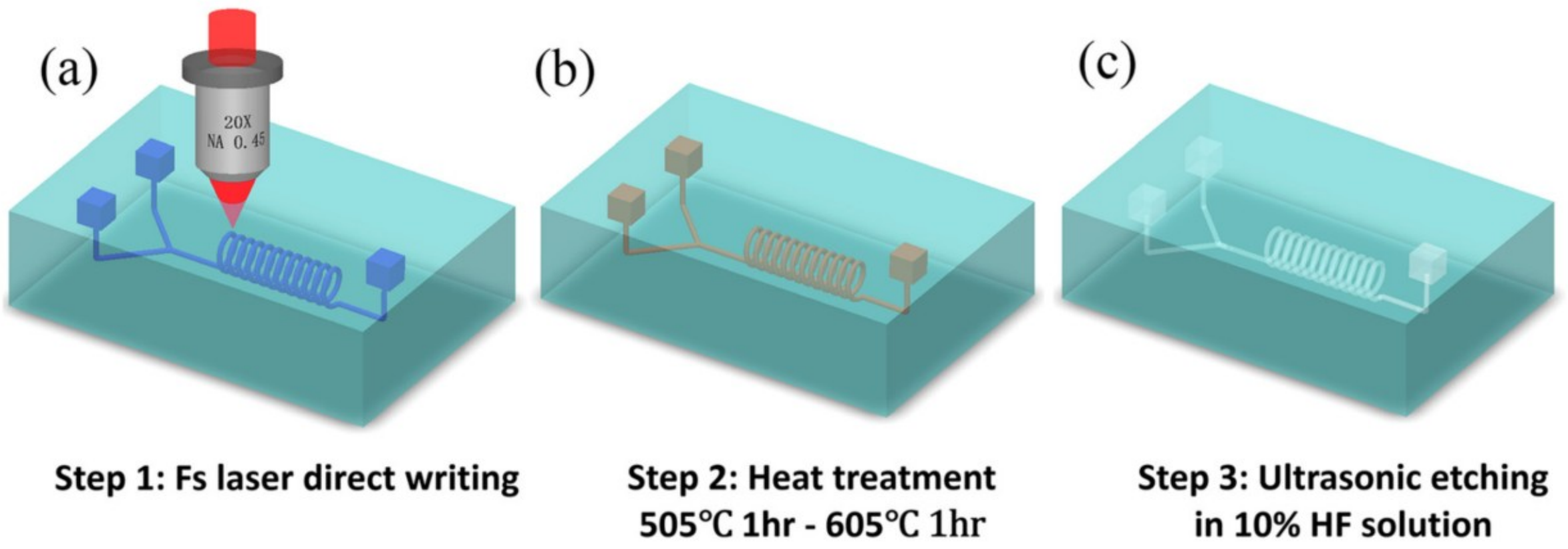

**Fig. 1.46:** Scheme illustrating the process steps for the fabrication of 3D microfluidic structures in a photosensitive glass. (**a**) Step 1: latent images of the desired 3D geometry are created in the photosensitive glass by fs-laser direct writing. (**b**) Step 2: the sample is subjected to a time-programmed heat treatment to develop the laser-modified region. (**c**) Step 3: the sample is embedded in an aqueous solution of hydrofluoric (HF) acid and placed in an ultrasonic bath to selectively etch the laser irradiated regions, leaving behind the desired hollow 3D microfluidic lab-on-a-chip structures in the glass [Sugioka, 2014]. (Reprinted from [Sugioka, 2014].], K. Sugioka et al., Femtosecond laser three-dimensional micro- and nanofabrication, Appl. Phys. Rev. **1**, 041303 (2014), Copyright 2014 under Creative Commons BY 4.0 license. Retrieved from https://doi.org/10.1063/1.4904320 )

Using the above-mentioned approach, Kim et al. fabricated a buried microfluidic channel in a 500 µm thick fused silica wafer, providing an inlet and an outlet at the surface. At the narrowest point (a 5 µm diameter neck) of the hollow microfluidic sub-surface channel, a fs-waveguide was longitudinally written, crossing in a perpendicular direction, see Figure 1.47a [Kim, 2009]. This optical waveguide could be operated either for the detection of transmission changes of a continuous wave Helium-Neon laser, or it could be used for collecting the Argon-laser excited fluorescence of active optical emitters passing the crossing point of the microfluidic channel with the fs-laser written waveguide, see the two illustrative schemes depicted in Figure 1.47b [Kim, 2009].

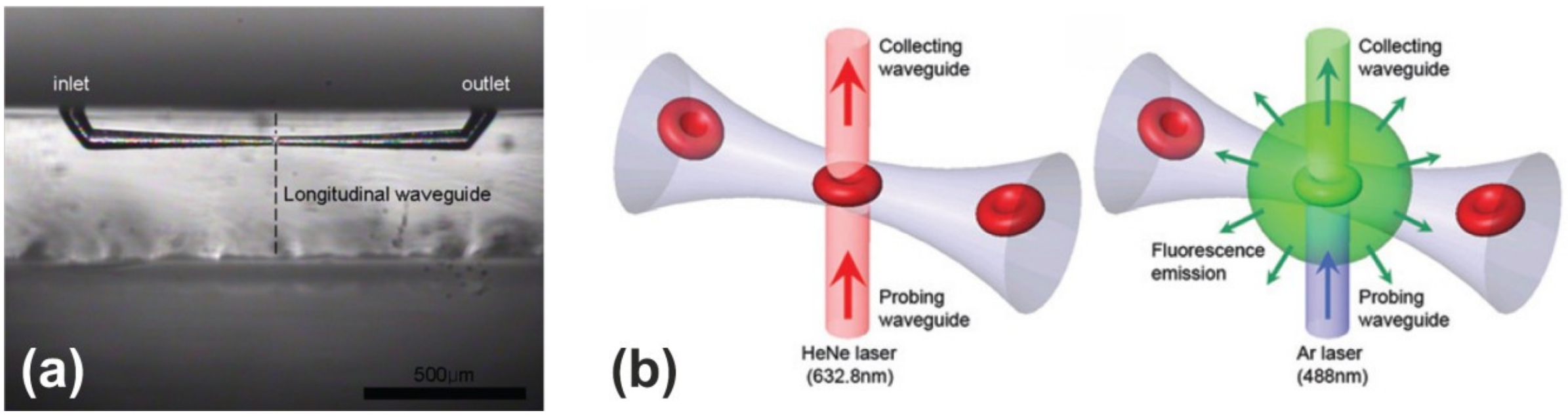

**Fig. 1.47:** Microfluidic lab-on-a-chip device. **(a)** Side-view optical transmission image of a horizontally arranged microchannel fabricated by a fs-laser ~130 μm below the surface of a 500 μm thick fused silica wafer [Kim, 2009]. The black dashed line indicates the location where an integrated optical waveguide (vertically arranged) was written by a focused fs-laser beam in a longitudinal processing geometry. **(b)** Two schemes of cell detection experiment using passive optical transmission changes (left panel) or using active fluorescence emissions (right panel). (Used with permission of Royal Society of Chemistry, from [Kim, 2009], M. Kim et al., Single cell detection using a glass-based optofluidic device fabricated by femtosecond laser pulses. Lab Chip **9**, 311 – 318 (2009); permission conveyed through Copyright Clearance Center, Inc.)

The lab-on-a-chip device presented in Figure 1.47 was used successfully in both detection modes (transmission or fluorescence) with a sensitivity being sufficient for the identification of single human blood cells (data not shown here) [Kim, 2009]. The particle counting efficiency of the device was up to 23 particles per second without clogging. It depends on the flow rate, the particle concentration, the ratio of the particle size to the microchannel neck size, and the geometrical shape of the microchannel. For additional information, the reader is referred to Ref. [Kim, 2009].

## Dicing of Glass

Glass plays an exceptional role for many applications due to its excellent optical, mechanical and chemical properties. In recent years the progress in display technology and especially the advances of smartphones pushed the development of techniques for the precise cleaving of glasses. Traditional approaches like scribing with a diamond tool and subsequent mechanical breaking do not yield the required surface quality, thus requiring elaborate and costly grinding and polishing techniques in addition. Laser-based techniques provide a significant improvement as the process is contact free. However, when the laser is absorbed at the surface, chipping still might be an issue. The use of ultrashort laser pulses offers an interesting alternative, as the nonlinear-based absorption can be localized below the sample surface.

Hosseini and Herman (University of Toronto, Canada) were the first to exploit the use of femtosecond laser filaments [Hosseini, 2012]. In this case the filaments generated during propagation of the ultrashort pulse through the glass sample induce damage tracks weakening the material along the intended cutting plane. For modifications covering a significant part of the material thickness, single pass processing becomes possible, where any lateral position is only irradiated once. After laser processing the work piece can be separated along the weakened plane either by mechanical breaking or by thermal crack propagation with a $CO_2$-laser [Hosseini, 2012].

However, the formation of filaments depends on the focusing conditions and pulse duration. Tight focusing and few picosecond pulse durations result instead in the formation of a so-called moving breakdown, where the absorption and ionization front grows to the direction of the incoming beam [Grossmann, 2016]. A more controlled energy deposition is obtained by applying spatial beam shaping, leading to an elongated laser focal line. This becomes possible with so-called Bessel-Gauss beams, where aspect ratios of focal length to focus diameter exceeding 10,000 have been obtained [Meyer, 2019]. The process results in a kerf-less cut, without the generation of debris, in contrast to, e.g. cutting by rear side ablation [Dudutis, 2020]. With sufficient pulse energy the length of the Bessel beam surpassing the damage threshold can be extended, enabling single-pass cutting of up to 12 mm glass [Bergner, 2018, /Meyer, 2019 / Feuer, 2019 / Jenne, 2020].

Furthermore, breaking the rotational symmetry of the beam leads to preferential crack formation. Thus, lower forces for separation or faster processing speeds become possible [Dudutis, 2016 / Meyer, 2017 / Jenne, 2020]. In any case, using pulse bursts allows to optimize energy deposition and cleaving [Mishchik, 2017], however, the actual laser parameters influence micro crack formation and thus surface quality [Werr, 2022]. In addition to straight modifications, trajectories enabling curved surfaces (C-shaped edges) are highly interesting. This has been investigated by applying Airy beams [Gecevicius, 2014 / Courvoisier, 2016 / Sohr, 2021 / Ungaro, 2021 / Cai, 2024] or a multi-spot approach [Flamm, 2022]. For further details, the reader is referred to Chap. 24 (Thomas) and Chap. 37 (Haupt et al.) of this book.

## Femtosecond Laser Welding

In addition to glass separation as discussed above, the reliable and stable bonding of glasses is important, e.g. for sealing or encapsulating components. Techniques such as gluing, optical contacting, direct or anodic bonding exhibit certain limitations or disadvantages, e.g. aging or outgassing. Thus, welding based on ultrashort laser pulses is an interesting alternative, as demonstrated first by Tamaki et al. in 2005 using 130 fs laser pulses at 1 kHz [Tamaki, 2005]. No absorbing layer is needed at the interface between the glass samples as the intense ultrashort laser pulses trigger nonlinear absorption processes.

Major advances have been obtained with the use of high-repetition rate ultrashort pulse laser systems [Tamaki, 2006 / Miyamoto, 2007]. In this case the high intensity of the ultrashort pulse leads to a localized energy deposition only in the focal volume, too. However, heat is accumulated, when the time required for heat to diffuse out of the focal volume is shorter than the time between successive pulses (see description in Sect. 1.2 of this chapter and Figure 1.4). This timescale obviously depends on the heat diffusivity and thus on the material as well as on the focal volume. For glasses and small focal spots it is typically on the order of few microseconds, thus lasers with repetition rates of several 100 kHz or above have to be used. The use of pulse bursts leads to further improvements [Sugioka, 2011]. In this accumulation regime, the laser acts as a localized heat source, that can be precisely positioned in three dimensions by placing the focus inside the material, i.e., especially at the contact interface between two transparent samples. By scanning the focal spot along a desired track at the interface, a localized weld seam is generated. Optimal welding results are achieved if the two samples are brought into direct contact. Nevertheless, even gaps of few micrometers can be bridged by the generated melt [Chen, 2015 / Cvecek, 2015 / Richter, 2015].

Due to the localized energy deposition, very high temperatures of several 1000 K can be reached in the focal spot, however, the temperature drops to values below the melting temperature of the glass within few micrometers, as proven by simulations and measurements [Miyamoto, 2014 / Hashimoto, 2015]. Therefore, only a tiny volume of the glass is exposed to very high temperatures. Consequently, the total thermal expansion is very small. This allows to weld even glass combinations with different thermal expansion coefficients (see Figure 1.48a) [Richter OPEX, 2013]. Welding strengths of close up to the value of the bulk material can be obtained depending on the material (Figure 1.48b) and the processing strategy [Richter, 2016]. However, the generation of stress and of gas bubbles leading to disruptions after cooling [Richter OPEX, 2013, Cvecek, 2014] should be minimized.

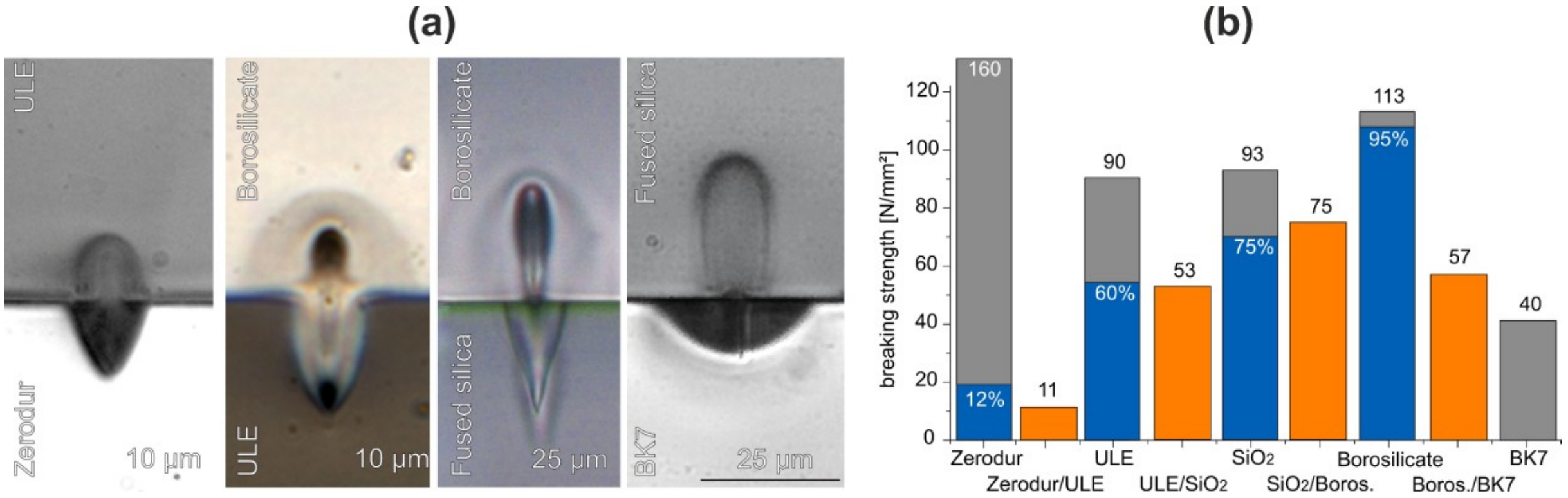

**Fig. 1.48:** **(a)** Welding of different glass combinations [Richter APA, 2013]. The size of the heat-affected zones is in the range of few 10 µm and changes at the interface due to the different thermal coefficients. The focus was placed closely underneath the interface for all experiments. **(b)** Breaking strengths for the different glasses and glass combinations. Gray bars indicate the breaking strength for unmodified bulk materials, blue bars for homogeneous material combinations and orange bars for different material combinations. (Reprinted from [Richter APA, 2013], S. Richter et al., Ultrashort high repetition rate exposure of dielectric materials: laser bonding of glasses analyzed by micro-Raman spectroscopy, Appl. Phys. A **110**, 9 – 15, 2013, Springer Nature)

In addition to welding glass to glass, the small heat affected zones enable also the welding of other material combinations. Technically extremely relevant is the welding of glass to metals [Ozeki, 2008 / Zhang, 2015 / Carter, 2017]. Nevertheless, within the past years various materials and material combinations have been welded using ultrashort laser pulses including glass-semiconductor [Zhang, 2018], polymers [Mingareev, 2016] or ceramics [Penilla, 2019]. As at least one of the two samples to be welded has to be transparent to the laser radiation, for welding silicon to metals [Chambonneau, 202 ] or silicon to silicon [Chambonneau, 2023] an appropriate wavelength of the ultrafast laser has to be chosen (see Chap. 14 (Chambonneau et al.) for further scaling the wavelength). For further details on ultrashort laser pulse welding, the reader is referred to the following review papers [Cvecek, 2019 / Xu, 2023].

## 3.4 Scaling: Limits Imposed by Material Properties or Laser Parameters or Processing Strategies

## Material Properties

Laser processing of bandgap materials (dielectrics and semiconductors) is largely governed by the band gap energy vs. the laser wavelength applied. As analyzed in Sect. 3.1, laser excitation of bandgap materials proceeds via several mechanisms, single- and/or multi-photon ionization, tunneling, and avalanche ionization. However, for avalanche ionization, it is necessary to create seed electrons in the conduction band with a density at which the inverse bremsstrahlung process becomes efficient. Thus, photoionization processes are at the basis of efficient laser processing of bandgap materials. A scheme of photoionization in relation to photon energy vs. material bandgap energy as well as laser light intensity is given in Figure 1.49.

Depending on the band gap, the material can be opaque or transparent at different wavelengths. As an example, silicon ($E_g$ = 1.12 eV), the most demanded material for microelectronics, solar energy harvesting, sensing, as well as for bio-applications due to its biocompatibility and low toxicity, is known to become transparent to light at wavelengths > 1.1 µm. Silicon dioxide, sapphire, and silicon nitride are transparent for visible, near-infrared, and partially for mid-infrared light [Mitchel, 2024], while the transparency ranges of these materials are restricted at wavelengths where vibrational modes become excited. Thus, for optimizing the laser processing of materials, it is necessary to find an optimal balance between the material band gap and the laser wavelength. An illustrative example is the laser crystallization of germanium nanolayers in amorphous Ge/Si multilayer stacks avoiding intermixing with silicon at interfaces and without affecting the a-Si nanolayers [Volodin, 2023]. Successful crystallization was achieved at a laser wavelength of 1.5 µm where silicon is transparent to light, while germanium is efficiently absorbing.

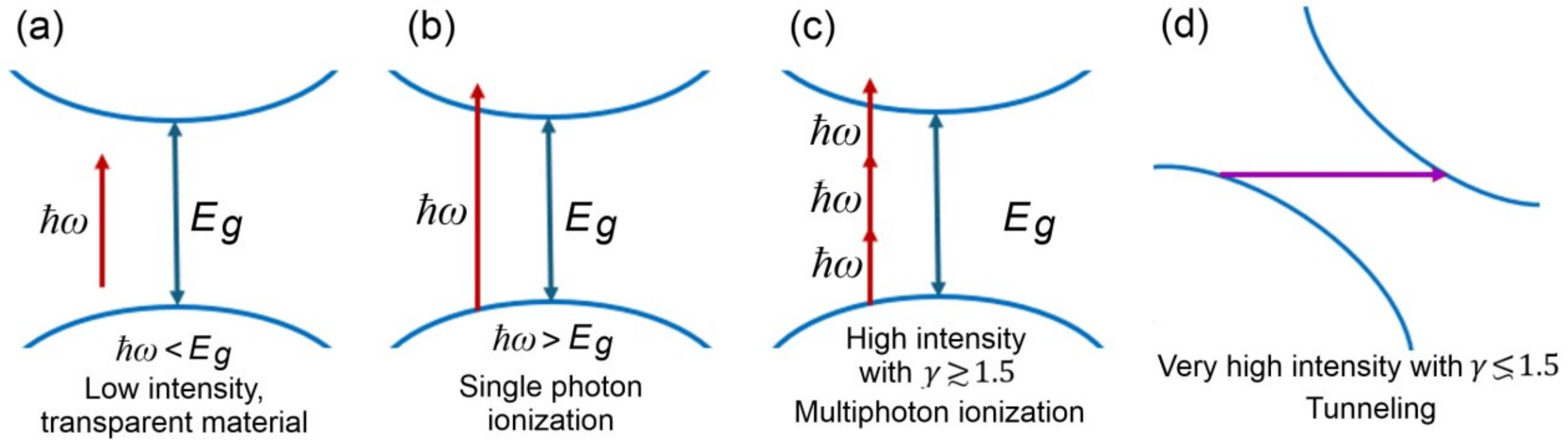

**Fig. 1.49:** Schematics of photoionization of bandgap materials and their linking with material processing. (**a**) Photon energy (ħω) is smaller than the band gap energy $E_g$. The material is transparent for such wavelengths. (**b**) Photon energy is larger than the band gap. The material is opaque to such wavelengths, and absorption proceeds through single-photon ionization even at low laser beam energies. (**c**) Photon energy is smaller than the band gap, but laser intensity is high enough to initiate multi-photon ionization. (**d**) At very high intensities, when the Keldysh parameter γ is below ~1.5, the band gap is distorted by the large magnitude of the laser electric field, and valence-band electrons can penetrate through the potential barrier to the conduction bands. Note that, for 1.5 > γ > 1.5, both multi-photon and tunneling processes contribute to material ionization

Regimes with single-photon and multi-photon absorption are important for different aspects. For the same material, single photon annealing/ablation at short wavelengths, where the photon energy is larger than the material bang gap, is possible at using relatively long laser pulses on large surface areas, while decreasing the photon energy to multiphoton ionization regime and shortening laser pulses to femto-/picosecond duration enables to produce sub-micrometer craters on surfaces [Englert, 2012]. However, by decreasing the laser wavelength and simultaneously increasing pulse duration (and vice versa), it is possible to achieve the same results at very different irradiation conditions as strikingly demonstrated by Herman et al. [Herman, 2000]. On the other hand, the shorter the laser pulse, the more deterministic the material ablation is due to the suppression of the stochastic avalanche process [Lenzner, 1998]. It is necessary to note that, by increasing the laser wavelength to the mid-infrared, new opportunities for processing bandgap materials can be opened [Grojo, 2013]. Indeed, the Keldysh (adiabaticity) parameter is proportional to the square root of the pulse duration and inversely proportional to the wavelength that makes increasing wavelength to be an alternative to laser pulse shortening.

For volumetric modification, e.g. for fabrication of lab-on-a-chip devices [Sugioka & Cheng, 2012], waveguide writing [Davis, 1996 / Chan APA, 2003], or optical memories [Glezer, 1996 / Wang, 2020], multiphoton irradiation regimes using ultrashort laser pulses are typically used, with tight focusing inside transparent materials. The higher the nonlinearity of absorption, the more localized modifications can be achieved in bulk transparent materials.

## Linear and Nonlinear Refractive Index

Particularly for laser processing with ultrashort laser pulses, the pairing of the linear refractive index ($n_0$, not dependent on intensity) and the nonlinear refractive index ($n_2$, intensity dependent) represents an important combination of material properties. It has important consequences for both, the (i) laser beam delivery to the workpiece and (ii) the choice of the focusing element.

The precise knowledge of the refractive index of the corresponding materials (optical component, environment) is required for the design of the laser material processing machine/setup, ruling, for example, intermediate beam diameters and maximum laser intensities in the beam delivery system. For scaling up the average laser power in such a system, care must be taken to not exceed the *critical power of self-focusing* [Marburger, 1975], above which the self-focusing through the material's intensity dependent refractive index according to Equation (1.1) overcomes diffraction. The values of $n_2$ crucially depend on the material, the laser wavelength, and the pulse duration, but not on the laser focus spot size. Note that its contribution to the total refractive index is additionally multiplied by the intensity profile $I = I(x,y,t)$ of the laser beam, i.e., transient lensing effects may occur in the optical components of the beam delivery system. Such transient lensing effects may prevent that the complete laser pulse energy can efficiently be used for precise laser processing. For experimentally assessing the relevance of such transient effects in the beam delivery system, the linearity of the energy/power input-output characteristics can be checked in the desired scaling regime with a combination of a partly laser beam-clipping transmissive aperture placed directly in front of the energy/power detector (in analogy to the method referred to as "z-scan" for the assessment of nonlinear optical properties).

For ultrashort laser pulses, an additional comment must be spent on the selection of the suitable focusing element and the processing environment. The influence of the ambient air (causing transient optical phase distortions that may be quantified via the B-integral) was already discussed in Sect. 1.3 above in the context of nonlinear effects limiting the precision of *surface processing* (see Figure 1.5). Additional effects can manifest at the sample side and may also limit the scaling of the laser processing: Ashkenasi et al. found in experiments with ultrashort (fs – ps) laser pulses being loosely focused to the front surface of dielectrics that the rear surface of the sample may get damaged first due to self-focusing effects within the dielectric material [Ashkenasi, 1997 / Ashkenasi, 1998]. This potentially detrimental effect depends on the type of material and the sample thickness.

Moreover, for the *volume processing* of transparent materials, early works raised doubts that, due to self-focusing and plasma defocusing, it is impossible to generate damage in the bulk of dielectrics just by scaling up the pulse energy of the fs-laser beam. While this claim may be true for loose focusing conditions, it can be rebutted for tight focusing conditions, where the large geometrical beam divergence prevents that the B-integral can reach sufficiently large values for manifesting significant self-focusing effects. Hence, care must be taken for a proper choice of the numerical aperture of the focusing optics, particularly for scaling the processing of transparent materials by ultrashort pulsed lasers. Note that this can become extremely critical and may require extreme focusing conditions for materials with high nonlinearities as e.g. silicon (see Chap. 14 (Chambonneau et al.) and [Chanal, 2017]).

## Defect Accumulation

It is important to note that any initial defects that are present in materials may considerably influence laser-material processing. This especially concerns glass materials, where tiny bubbles or metal impurities strongly affect the process of laser light absorption. Furthermore, mechanical processing of optics leads to the appearance of micrometer-sized defects such as scratches, cracks, and pits [Sun, 2022]. All these defects result in the reduction of the laser-induced damage threshold. At the defect states, initial or induced by previous laser irradiation, atomic- or micrometer-scale, the incident laser energy is redistributed and unevenly absorbed, thus creating temperature and stress gradients [Liu, 2024]. The generation of nanograting structures inside glasses was, in particular, attributed to tiny nanovoids [Buschlinger, 2014 / Rudenko, 2017], which are intrinsically present inside amorphous solids.

For bulk laser processing of transparent materials by multiple laser pulses, the defect accumulation or resonant absorption phenomena (such as Plasmon Polaritons) can play an essential role in laser light absorption for glass and crystalline bandgap materials [Ashkenasi, 1999], or polymers [Baudach, 2001]. For waveguide writing inside glass materials, it was also found that the higher repetition rates enabled reducing waveguide losses, which was explained by the heat accumulation [Eaton, 2005].

It can be concluded that, similarly to the laser surface processing (see the subsection Material Incubation in Sect. 2.1) in multi-pulse irradiation regimes, each subsequent laser pulse couples with material whose properties are modified by the previous pulses. A gradual accumulation of defect states with larger absorption cross-sections facilitates excitation for the following pulses. Hence, particularly for bandgap materials, the dominant absorption mechanism can change from nonlinear towards linear material interactions – eventually ruling the resulting localized energy deposition. As a consequence, the refractive index varies due to defect generation, change of material density, formation of porosity, and stress accumulation that strongly affects the laser processing of materials and should be considered for optimizing processing quality.

## Processing Strategies

When a high-intensity laser pulse propagates through a radiation-absorbing medium, it can undergo nonlinear effects such as self-focusing, followed by a beam collapse and the formation of plasma that, in turn, leads to plasma defocusing. Upon laser ablation in air or liquid environment, this can lead to an attenuation of the laser beam whose intensity is limited to a certain value along the beam propagation path due to energy ionization losses and plasma scattering [Becker 2001 / Liu, 2002 / Couairon, 2007]. A laser beam focused inside transparent material can also be subjected to self-focusing, which is manifested in shifting the geometrical focus position toward irradiating laser, as well as in plasma defocusing with light displacement from the excited plasma region [Couairon, 2007 / Popov, 2010 / Bulgakova, 2013b]. The beam self-focusing is governed by the change of the refractive index of the propagation medium with the light intensity according to Equation (1.1). The threshold power for the self-focusing can be evaluated as $P_{cr} \cong 3.72\lambda_0^2/(8\pi n_0 n_2)$ [Marburger, 1975 / Couairon, 2007] that yields ~3.2 GW and ~120 MW for air at wavelengths of 800 nm and 248 nm, respectively [Couairon, 2007], while in fused silica at 800 nm the critical power for self-focusing can be evaluated as ~1.98 MW [Couairon, 2005]. Thus, for the surface processing of materials in air, this effect is initiated only at rather high laser powers, while the self-focusing effect at volumetric processing of solid dielectrics can represent a challenge.

A vital feature of the intensity clamping effect is that it prevents the beam collapse upon self-focusing due to swiftly generated free-electron plasma. It is believed that the clamping intensity is of the order of $2\times10^{13}$ W/cm$^2$ for a pulse duration of 100 fs at 800 nm wavelength [Couairon, 2007]. Depending on the laser beam power, the focused light can undergo either a single self-focusing and, being attenuated, leaves behind in the focal zone a tightly localized modification region or, at high powers, it can refocus and defocus several times, thus creating long filaments that can be more visible at multi-shot irradiation regimes [Liu, 2002 / Couairon, 2005] as demonstrated in Figure 1.50 [Couairon, 2005].

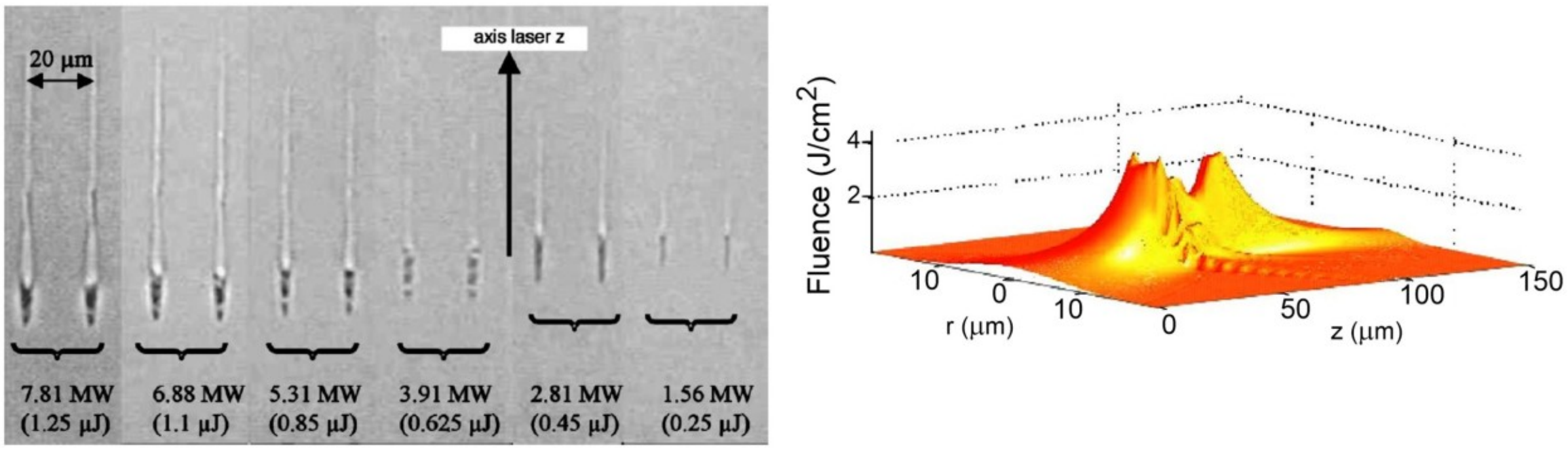

**Fig. 1.50:** Left: Observation of the shift of damage tracks in the bulk of a silica sample as a function of the laser energy [Couairon, 2005]. The pulse duration and laser wavelength were 160 fs and 800 nm, respectively. The laser was operated at a pulse repetition frequency of 200 kHz. The irradiation time was 2 s with the objective of numerical aperture NA = 0.5. Right: Numerically computed fluence as a function of the propagation distance, demonstrating laser beam refocusing. Laser pulse energy was 1.1 µJ at a wavelength of 800 nm and pulse duration of 160 fs, NA = 0.5. (Reprinted figures with permission from [Couairon, 2005], A. Couairon et al., Phys. Rev. B **71**, 125435, 2005. Copyright (2005) by the American Physical Society)

The phenomena of self-focusing, beam intensity clamping, and filamentation are tightly connected. When carefully controlled via the focusing conditions (NA of the focusing optics), the laser parameters (pulse energy, duration, and wavelength) and adjusted to the materials parameters ($n_0$, $n_2$) of the propagation medium, very long self-guided beam filaments with extreme length-to-diameter aspect ratios can be generated. An impressive example of the time-resolved visualization of fs-laser beam filament propagating at the speed of light in the bulk of fused silica glass was provided by Wang et al. in 2023 using compressed ultrafast spectral photography (CUSP) [Wang, 2023], see Figure 13.42 in Chap. 13 (Bonse).

In atmospheric air as a propagation medium, such ultrashort-pulse-generated self-guided filaments can reach lengths in the kilometer range. The related potential was recognized and used by the Franco-German collaborative research project "Teramobile" [Wille, 2002]. Within the frame of this project, the world's first mobile terawatt fs-laser was developed and used for atmospheric studies. The self-guided filaments in air emit a spectral white light continuum from the ultraviolet to the infrared. This white light additionally exhibits a directional behavior with enhanced backward scattering. It was detected from altitudes up to 20 km and allows remote identification of aerosols or atmospheric trace-gas remote sensing or remote through *Light Detection and Ranging* (LIDAR) [Kasparian, 2003]. Some years later, the atmospheric lightening control through fs-laser generated filamentary plasma channels was demonstrated [Graydon, 2009] and has been proven in a campaign conducted during summer 2021 to be even capable to contact-less protect human infrastructure on the Säntis mountain in north-eastern Switzerland [Houard, 2023].

Moreover, atmospheric self-guided fs-laser beam filaments bear the potential of a long-range transport of optical energy being sufficiently intense for laser material processing and allowing a scaling of the remote processing position over large distances. This was already demonstrated over distances of tens to hundreds of meters, for example, for the analysis of the chemical composition of copper, aluminum, and steel samples via *Laser-Induced Breakdown Spectroscopy* (LIBS) [Rohwetter, 2004 / Stelmaszczyk, 2004]. The method has the potential for various contact-free long-distance laser processing, e.g. in applications in steel production, for nuclear waste management, or in cultural heritage preservation [Salle, 2007].

While the remote laser processing with guided fs-filaments has the unique advantage of being not affected by any "depth of focus limitations", i.e., being suitable even for curved sample surfaces, the processing at long distances may suffer from some beam-pointing instabilities. However, by using a ~100 µm diameter self-guided Ti:sapphire fs-laser-induced filament in air ($\lambda$ = 800 nm, $\tau_p$ = 120 fs, $E_p$ = 5.2 mJ, $N \leq 150$) Schille et al. demonstrated successfully the processing of self-ordered sub-micrometric laser-induced periodic surface structures (LSFL-I type) over a spat diameter of less than 1.5 mm on AISI 304 stainless steel at remote distances exceeding 50 m from the laser source [Schille, 2022]. Recently, it has been demonstrated that "stitching" filaments together by using GHz bursts is beneficial for filament control and ablation at long distances [Kerrigan, 2024]

# 4 Outlook: Pushing the Limits through Tailored Ultrashort Laser Processing

**Is there an optimum laser pulse duration for laser processing?**

From a technological and practical point of view, the duration of ultrashort laser pulses in the visible to infrared spectral range can be kept short with affordable means in beam delivery systems if the pulse duration is not significantly smaller than ~100 fs. This duration is shorter than the electron-phonon coupling time $\tau_{e\text{-}ph}$ of most materials of interest, allowing to take benefit of the energy confinement during the laser ultrashort pulses for improving the machining precision. Hence, for the laser processing of strong absorbing materials (metals, semiconductors), a pulse duration between ~0.1 ps and a few ps usually will be a good choice when traded off against the costs of the laser processing system. Pulse durations towards the nanosecond range, as compared the picosecond and femtosecond regimes, result in higher ablation efficiency but always at the cost of processed surface quality [Mustafa, 2020].

In this context it is important to recall the lesser-known fact that even for normal ns-laser processing, the spectrum of longitudinal laser modes being present in the pulse is very relevant for the determinism of laser damage and the materials processing quality. Vogel et al. demonstrated that for the laser processing of transparent materials a ns-pulsed laser seeded at a single longitudinal resonator mode (single-frequency laser operation) enables superior results compared to the unseeded (common free-running) longitudinal multi-mode laser operation [Vogel, 2008 / Linz, 2025].

For laser processing of the surface or in the volume of transparent materials (dielectric crystals, glasses, polymers), where the peak intensity and temporal shape of the excitation sequences are key, the situation must be assessed individually and specifically optimized for the individual material. Here, with additional efforts, extremely short few-cycle optical femtosecond laser pulses can be generated by propagating sub-ps ultrashort laser pulses through a low pressure noble gas-filled hollow fiber [Nisoli, 1997] or through a gas-filled photonic fiber [Markos, 2017] for inducing a coherent spectral broadening via the nonlinear effect of *self-phase modulation* (Kerr effect). With a subsequent additional spectral chirp compensation and pulse compression [Nagy, 2020], energetic sub-10-fs laser pulses were generated for the Ti:sapphire laser platform that allowed to remarkably improve the laser machining precision even of wide band gap materials, such as fused silica [Lenzner, 1998 / Lenzner, 1999 / Lebugle, 2014].

**Is attosecond pulse machining possible and advantageous?**

Although attosecond (as) laser processing was already proposed and theoretically addressed around the turn of the millennium [Temkin, 1999], so far, the necessary as-pulse energies are currently only available at large scale *Free-Electron Laser* facilities [Yan, 2024]. Since the pulse duration has to be longer than the period of the optical field, attosecond pulses can only have a carrier frequency in the XUV region and below. The main advantage (precision) of few-cycle optical femtosecond laser pulses has so far only been shown for transparent materials [Lenzner, 1999 / Lebugle, 2014], which do not exist in the XUV region. Hence, the question arises if the instrumental effort for as-pulsed material processing would be justified.

**Are Carrier-Envelope Phase (CEP)-stabilized lasers beneficial for material processing?**

The technique of *Carrier-Envelope Phase* (CEP)-stabilization represents an important step towards attosecond science [Krausz, 2009]. The control over the CEP allows to precisely set and fix the temporal positions of the maxima of the electromagnetic field (oscillating at the light carrier frequency) under the overall laser pulse envelope (accounting for its duration). With the help of CEP-stabilized lasers, extremely fast physical processes can be deterministically triggered on the quantum level. This enables the investigation of fundamental processes in atomic or molecular systems (awarded the Nobel Prize in 2023 [Krausz, 2023]). Even if technologically demanding, the generation of CEP-controlled optical laser pulses may be interesting for the laser processing of solids for "coherently driven optical excitations". That could improve the deterministic nature of the laser-matter interaction and, thus, further enhance the laser machining precision.

**Is temporal laser beam shaping the solution?**

One approach for improving laser processing may be temporal beam shaping. This approach may help, particularly for band-gap materials showing a nonlinear interaction, in order to temporally tailor the localized absorption in the material. Upon shaping, the train of single pulses delivered directly from the laser system is redistributed in time into generalized temporal sequences, e.g. pulse bursts. Practically this can be realized by individually controlling the spectral phases and amplitudes through spatial light modulators (SLMs) that are implemented to manipulate ultrashort laser pulses. In this way, apart from the time-envelope of the laser radiation, polarization states, or wavelengths can be manipulated.

The potential of this SLM-technology was demonstrated already successfully for tailoring ultrafast pulse double-pulse sequences that optimize the generation of specific ionic species from the laser irradiated silicon surfaces [Spyridaki, 2003]. Similarly, the approach was used to optimize the localized refractive index modification obtained under tight focusing conditions within the bulk of dielectrics [Mermillod, 2008].

In addition to pure temporal or spatial shaping of the laser beam, the simultaneous spatio-temporal shaping (SSTF) has attracted interest in the past years. The basic concept behind is to spectrally disperse the pulse, e.g. using a grating and then to focus the spatially separated spectral components, ensuring their overlap in space and time at the focus. Originally developed for applications in multiphoton microscopy [Zhu, 2005 / Oron, 2005], this concept soon found applications in confining the 3D modification in glasses e.g. for generating microfluidic channels with circular cross section [He, 2010]. In addition, as the intensity is significantly reduced before the focal region, nonlinear propagation effects can be significantly reduced [Kammel, 2014]. Although the spatio-temporal evolution of the pulse is complex, one can gain an intuitive understanding based on an ABCD matrix approach [Durfee, 2012]. Recently, the concept of SSTF has been combined with SLM shaping in order to achieve holographic control over multiple foci in the SSTF geometry [Sun, 2018]. This opens additional degrees of freedom and is a pathway to increase processing speed.

While in early experiments *genetic algorithms* were used for adaptive control in laser processing [Lim, 1994], nowadays artificial intelligence (AI) and *machine learning* based on neuronal networks will be feasible and the optimization method of choice [Dubey, 2008], see also Chap. 19 (Zwahr et al.). The great advantage of such "parallelized" optimization strategies is that good processing solutions are obtained automatically without the need of gaining additional knowledge on the specific material system or the process parameter space. The only requirement is to have a reliable method for uniquely quantifying the quality of the laser processing results, along with available computing power and data storage capabilities. The ease of such approaches sometimes comes at the cost that the identified optimized solutions cannot be rationalized anymore.

**How to take benefit of the ever higher average output powers of ultrashort pulses lasers in surface processing?**

With the ever-increasing higher average output powers of ultrafast lasers – currently driven mainly by increasing pulse repetition rates – limitations through heat accumulation at the laser processed sample surface and through plasma shielding of ablation products become relevant (see Sect. 2.4).

For avoiding or at least minimizing such effects, the laser processing strategy may be adapted. For example, the complete processing could be divided in several steps and executed via different processing passes (at reduced fluence / ablation rate or at increased scanning velocity for a spatial separation of consecutively irradiated surface spot), with temporal breaks in between, allowing the sample surface to cool down again between successive passes and to allow the exhaust suction system to carry away the ablation products. Certainly, a good thermal coupling of the sample to a heat sink or even actively cooled sample holders may be beneficial for reducing heat accumulation effects. A theoretical analysis of the laser-induced thermal heat loads and maximum possible heat extraction can help exploring the ultimate physical limits. Sect. 3.5 in Chap. 6 (Lenzner and Bonse) may help to identify limits imposed by heat accumulation in the standard laser processing strategies.

Optimized scanning approaches can deviate from the most common line-wise meandering areal scanning, where within such a scanned line, there is a spatial overlap of consecutively irradiated surface spots, thus promoting heat-accumulation via the release of residual heat into the neighbored sample regions. This can be avoided by not irradiating neighboring surface spots by consecutive pulses. Examples for such approaches can be found in eye surgery, where sequences of spiral scan patterns are used. "Non-neighbored" spot processing may be effectively realized with tailored scanning patterns in combination with fast electronic scanning devices. Spatial beam modulators (SLMs) can be implemented as mass-less scanners and additionally allow a spatial beam shaping in the focal region. Digital-controlled micro-mirror arrays (DMAs) can be used as beam shaping and focusing elements, taking benefit of the reduced movable masses of the individual mirrors for lowering their response and settling times.

If high energetic laser pulses are available at high pulse repetition rates, the laser processing may be *parallelized* to speed up the total processing times (at the cost of additional resources). This may be implemented either through different independent laser processing devices, or by splitting the high-energy pulsed beam into several lower-energetic identical beamlets that are spatially separated at the sample surface, e.g., via tailored diffractive optical elements (DOEs) or an SLM. For additional details on the mentioned concepts, smart scanning strategies, and laser process optimizations for high repetition rates, the reader is referred here to Chap. 3 (Gräf and Müller), Chap. 4 (Holder et al.), and Chap. 5 (Neuenschwander and Förster).